\documentclass[preprint,aps,showpacs]{revtex4}
\usepackage{graphicx}
\usepackage{bm}
\usepackage{amsmath}

\begin{document}

\title{Coherent Collective Excitations in a Superfluid:
Spontaneously Broken Symmetries and Fluctuation-Dissipation}
\author{S. J. Han}

\affiliation{P.O. Box 4684, Los Alamos, NM 87544-4684\\
email:sjhan@cybermesa.com}

\begin{abstract}

 This paper presents an elementary theory of the phase-coherent
collective excitations in both He II in the gravitational field
and atomic Bose-Einstein condensation in a trap. The theory is
based on the concept of off-diagonal long-range order by Penrose
and Onsager and the quantum theory of Bohm, with emphasis on the
broken symmetry in a Bose-Einstein gas with repulsive
interactions. It is shown that a spontaneously broken symmetry
that accompanies a phonon (Nambu-Goldstone mode) takes place at
the surface layer of an inhomogeneous Bose system in the presence
of an external field. The spontaneously broken symmetry in a Bose
system is described and is shown to manifest itself in both He II
and the Bose-Einstein condensation of an imperfect Bose gas - a
shell-like structure of the Bose-Einstein condensation in a trap.
The broken symmetry gives a coherent explanation for a number of
long standing puzzles in He II.

The collective excitations in a liquid helium are essential for a
complete understanding of certain properties of the liquid helium
at low temperature, particularly its interaction with the external
gravitational field, which is similar to that of the Meissner
effect in a superconductor in an external magnetic field except
for a characteristic difference arising from the statistics. We
present a detailed derivation of dispersion relations for the
various collective excitations in a surface layer in He II, in
which Landau's two-fluid model breaks down. It is shown in the
surface layer that, in addition to the usual surface waves (a
gravity and a capillary waves), there is also a transverse
collective excitation with entirely different behavior from the
phase-coherent longitudinal excitation (phonon). It is further
shown that the spontaneously broken symmetry due to fluctuation
and dissipation of the system is both a necessary and a sufficient
condition for the conservation of energy in an isolated system.

\end{abstract}


\maketitle

\section{\label{sec:level1} Introduction}

The two-fluid model as a phenomenological theory of superfluidity
for He II was a major conceptual breakthrough in quantum
hydrodynamics since it gives an excellent account of the data in
those cases to which the model is applicable, namely a spatially
uniform fluid in a Hilbert space, but it has long been known as an
incomplete theory to describe many spatially inhomogeneous
problems \cite{Landau41,Landau47,Tisza38,Tisza47,London45}.

Specifically, the remarkable hydrodynamic properties of He II have
been explained by the two-fluid model in which He II is considered
a mixture of two interpenetrating components: the superfluid with
$\bm{\nabla}\times\bm{v}_{s}=0$ and the normal fluid with
viscosity; the two-fluid picture has been confirmed in an
experiment with a torsion pendulum \cite{Andronika46} and also the
theory predicted the correct energy spectra of collective
excitations (phonons and rotons) below a $\lambda$ point
\cite{Landau41,Henshaw61,Yarnell59}. However, it cannot explain
satisfactorily the peculiarly normal-fluid like behavior of the
surface curvature of a rotating He II to this date
\cite{Osborne50,Meservey64}.

Over half of a century the dynamics of quantized vortices have
dominated the hydrodynamics of a superfluid. And yet, the
mechanism, by which vortices are created and nucleated in a
superfluid, is still poorly understood. In particular, the
breakdown of superfluidity at the vortex core, which is closely
related to the normal fluid-like behavior of the surface of a
rotating He II, has remained a mystery in low-temperature physics.
In spite of the wealth of experimental data on the apparent
breakdown of superfluidity
\cite{Vinen61,Reif64,Glaberson68,Hendry88}, the clue for this
long-standing riddle has remained elusive. Before proceeding to
the explicit discussion of this problem, we mention some
qualitative features of the two-fluid model by Landau
\cite{Landau41} that is valid in a Hilbert space. In the absence
of an external field, there was no need to address the boundary
conditions. Consequently, the model assumes that the fluid cannot
actually separated into two parts (superfluid and normal parts).
This assumption is not in accord with Mott's analysis on the
surface energy due to velocity discontinuity of a superfluid
\cite{Mott49,Anderson66}. The presence of the surface energy
implies that the particles in the layer are in excited states with
finite energy gap from the ground state \cite{Mott49,Landau80} and
behave like a normal fluid - breakdown of the two-fluid model. We
will show this in more detail with explicit calculations on the
surface waves in the surface layer.

A degenerate Bose gas exhibits a peculiar condensation phenomenon
that is now known as the Bose-Einstein condensation (BEC) which is
often described as the momentum condensation; this implies that
the condensation is due to the symmetry of the wave function with
respect to the interchange of any pair of particles in the
degenerate Bose system \cite{London38,Daunt39,London54}. The BEC
shows a remarkable feature that for $T<T_{0}$ the pressure of the
condensed state is independent of density, where $T_{0}=3.31
(\hbar^{2}/m)(N/V)$ \cite{Landau80}; this is the consequence of
the degenerate state of the Bose particles with $\varepsilon=0$
(and $\bm{p}=0$). Since all real gases are condensed differently
by the presence of molecular forces, this mechanism appeared to be
devoid of any real physical significance.

The similar phase transition ($\lambda$-transition) in liquid
helium may be regarded as the condensation phenomenon of a
degenerate Bose-Einstein gas modified by the presence of molecular
forces. This was first recognized by London who has proposed that
the transition between liquid He I and liquid He II - $\lambda$
transition - is the same process that causes the degenerate Bose
gas to condense. In support of London's view, Feynman
\cite{Feynman53} argued that, in a liquid-like quantum mechanical
system, strongly interacting atoms behave in some respects very
much like free particles. In other words, the strong interaction
forces do not prevent these particles from behaving as if they are
free particles. Hence the inclusion of strong pair-interactions
will not alter London's view on the $\lambda$ transition as the
same process that causes the condensation of an ideal
Bose-Einstein gas, so long as there is a pair-interaction however
weak the interaction might be \cite{Bog47}.

The work of London \cite{London38,London54,Schafroth54} and the
seminal paper by Bogoliubov \cite{Bog47,Beliaev58} on a
microscopic model of an imperfect Bose gas have laid the basis for
much of our theoretical understanding of liquid helium and the
Bose-Einstein condensation (BEC). Penrose and Onsager have
extended the concept of the Bose-Einstein condensation due to
London \cite{London38} to a strongly interacting Bose system by
introducing the concept of off-diagonal long-range order (ODLRO)
\cite{Penrose51,Penrose56,Yang62,Schafroth54,Beliaev58,Anderson66}
and suggested that superfluidity be described by the one particle
ground state wave function $\psi$ which has a macroscopic mean
value in the thermodynamic, quasi-equilibrium sense, the simplest
example of which is an atomic Bose-Einstein condensation in a
trap. Thus the Bose-Einstein condensation in a trap can be defined
by ODLRO \cite{Penrose56,Schafroth54,Yang62}.

In 1942, Daunt and Mendelssohn \cite{Daunt42} presented
interesting experimental evidence for a striking similarity
between the surface flow of He II and the electric surface current
in a superconductor, which goes much farther than a superficial
similarity. They have noticed that in both cases there are upper
limits for the current densities which depend on temperature.
London studied a theoretical model to account for the similarity
with the critical surface transfer rate $R_{c} \approx n\hbar$
where $n$ is the density of superfluid particles per $cm^{3}$
\cite{London45}, but was not able to carry through due to the
insufficient experimental data; the measurements of Daunt and
Mendelssohn with mobile helium films were the only data of which
he could make use \cite{Daunt39}. He did, however, correctly point
out the need to find a quantum mechanical mean field that
represents the ground state of a many-particle system, and which
can also describe the hydrodynamics of He II. What's more, he
emphasized that the peculiar low temperature transfer processes in
both a superconductor and He II should be explained on the basis
of that familiar mixture of quantum statistics and classical
mechanics \cite{London35,London38}.

But up to now it has been customary to neglect the gravitational
field in the study of quantum fluid for mathematical convenience,
\cite{Landau41,Tisza38,Tisza49,Feynman53}, which was justified by
the extreme weak gravitational force compared to the
pair-interaction force between He atoms. However, it is essential
to include the external field in the discussion of a surface layer
which is defined by the gravitational field and in which the
two-fluid model of Landau \cite{Landau41,Feynman53} breaks down.
In particular, the BEC in a trap shows this peculiar shell-like
structure \cite{Lamb41,Mott49}. This is precisely the reason why
the two-fluid model by Landau \cite{Landau41} cannot explain the
surface phenomena, since the two-fluid model does not allow a
separation of the superfluid and normal parts of the He II. More
specifically, one of the problems we would like to address in this
paper is to show that there is a striking similarity between the
Meissner effect \cite{London35} in a superconductor in an external
field and the surface phenomena in He II in the gravitational
field \cite{Daunt42}. The question naturally arises what physical
significance has to be given to this analogy. It will be shown
here that the surface layer is composed of the excited atoms with
the energy gap from the ground state \cite{Mott49} and behaves
like a normal fluid with viscosity. The particles in the layer do
not interact with the bulk He II. This separation under the
gravitational field has been observed by Lamb and Nordsieck
\cite{Lamb41} and supports Mott's analysis \cite{Mott49}.

In recent years there has been considerable interest in certain
basic properties of the atomic Bose condensation in connection
with trapped alkali-metal gas at extreme low temperature
\cite{Stringary99,Leggett01,Fetter01}. In a magnetic trap
experiment, Anderson {\it et al.,} \cite{Davis95} have observed a
rapid narrowing of the velocity distribution and density profile
of trapped $^{87}$Rb gas (and also sodium atoms) at low
temperatures. This has been interpreted as evidence for a Bose
condensate in the trap. It would be, however, difficult to make
any quantitative statement, based on first principles
\cite{Schafroth54,Penrose56}, about an actual realization of the
Bose-Einstein condensation in a trap. Before making such a
dramatic claim, we must be sure that the data which appear to
support the BEC in a trap are consistent with the quantum
many-body theory, that is, the data on the collective excitations
must confirm a phonon spectrum as in Henshaw and Wood's experiment
\cite{Henshaw61,Yarnell59}, since the ground state wave function
is given by the mean field defined in ODLRO. In a curious
circumstance, this pertinent question \cite{SJHan1,SJHan2,SJHan3}
was never taken seriously in the physics community
\cite{Stringary99,Leggett01,Fetter01}.

Thus the preliminary study, for its final quantitative
description, must wait for the correct theoretical analysis of a
dynamical proof of the BEC to confirm the realization of a BEC in
a trap. Because the weak interaction between the atomic particles
in the BEC can be described by the hard sphere approximation as in
Bogoliubov's theory of superfluidity, it is essential to
demonstrate the Bogoliubov's phonon spectrum $\epsilon = \hbar ck$
\cite{Bog47} in the the study of collective excitations in the
BEC. It is also the purpose of this paper to discuss a variety of
additional conceptual details of the broken symmetry
\cite{SJHan1,SJHan2,SJHan3} in an imperfect Bose gas which
completes the theory of the Bose system.

It is extremely challenging to carry out an experimental study of
(first) sound wave propagation in a trap to confirm the presence
of BEC \cite{Jin96,Mews96}, because 1) the speed of sound wave
$c=[4\pi a \rho(r)\hbar^{2}]^{1/2}/M$ depends on both the local
density (and the wavelength of a traveling sound wave, which is
$\lambda \sim c/\omega=1/k$); 2) it is impossible to study the
quantum effects of the BEC by the de Broglie wave
$\lambda=2\pi\hbar/p$ of a particle in the trap whose wavelength
is equal in order of magnitude to the macroscopic size of the trap
as $p\rightarrow 0$ in condensation. It should be emphasized,
however, that, in liquid helium, the quantum effects become
important when the de Broglie wavelength $\lambda=2\pi\hbar/p$
corresponding to the thermal motion of atoms becomes comparable
with the distances between atoms at about the $\lambda$ point
\cite{Landau80}. The above discussion applies entirely to the
atomic Bose-Einstein condensation in a trap and so remains valid
in the liquid helium below the $\lambda$-point as well
\cite{Feynman53}.

It will be shown, therefore, that an experimental confirmation of
the Bogoliubov spectrum of a spherical sound wave \cite{Bog47} is
both a necessary and a sufficient condition to observe the Bose
condensation in a trap. Hence the study of collective excitations
occupies a unique place in an interacting Bose system
\cite{SJHan1,SJHan2,SJHan3,Jin96,Mews96}. Just as the phonons that
mediate strong attractive electron-pair interactions in a
superconductor brought about Cooper pairs which obey Bose
statistics and become a superfluid by condensation - the
super-conducting transition \cite{Schafroth54,BCS57,Rick59}, a
pair-interaction between Bose particles which brings about
$\lambda$ transition in liquid helium also ensures that, in the
absence of the long-range order interaction \cite{Penrose56}, the
system possess certain certain collective excitations. More
specifically, we study the collective excitations in a BEC to
probe the dynamics of interacting Bose particles and thereby to
provide a course of experiments to confirm the BEC in a trap
\cite{Schafroth54,Penrose56,Anderson66,SJHan2,SJHan3}. Hence this
discussion is not merely a question of the mathematical technique
to obtain a correct dispersion for a sound wave leading to the
Bogoliubov's phonon spectrum $\epsilon = \hbar ck$, but is the
question of a scientific merit of the Bose-Einstein condensation
in a trap \cite{Bog47}.

The present research was undertaken in the hope that a
comprehensive investigation of phase-coherent collective
excitations in He II and the atomic BEC in a trap would bring a
unified picture to a number of long standing problems in low
temperature physics based on first principles. A new perturbation
method similar to that of Feynman \cite{Feynman53} in concept but
mathematically more precise technique \cite{BFKK58}, is developed
in section III to deal with the region of low energy excitation
(phonon) in both He II and BEC in a trap where the conventional
perturbation method of quantum field theory fails \cite{Bog47}.
Along with the concept of ODLRO of Penrose and Onsager for both He
II and BEC as a superfluid \cite{Penrose51,Penrose56,Beliaev58}, a
detailed mathematical analysis is carried out based on Bohm's
theory of quantum mechanics \cite{Bohm52,Aharonov63} and the
semi-classical perturbation method of the Lagrangian displacement
vectors that overcome a major stumbling block in a finite space
problem in the study of a quantum fluid \cite{BFKK58,SJHan91}.

\section{\label{sec:level1} Brief Outline of the Rational}

Since the mathematical development is fairly complex and involves
the various types of lengthy algebras for each point in the
development of the theory, we wish to outline the rational behind
the theory, leaving the detailed algebras to later sections.

 At low temperature ($T \ll T_{\lambda}$), Bogoliubov's theory of
superfluidity gives the dispersion relation for the collective
excitations of quasi-particles in the form of $\omega = ck$ in the
long wavelength limit (a longitudinal sound wave), where $c=[4\pi
a \rho\hbar^{2}]^{1/2}/M$ and k is the wave number \cite{Bog47}.
However, Bogoliubov assumed in his derivation of the dispersion
relation for a phonon that the Bose quantum fluid is homogenous as
well as isotropic, and of unlimited extent - of a Hilbert space.

In order to incorporate a well-known theorem (Goldstone) in
quantum field theory into our study of an imperfect Bose gas at
low temperature, we review its implications in a many-body system
of bosons. The theorem states: the spontaneously broken gauge
symmetry in a nuclear many-body system always accompanies a
longitudinal phonon, that is, a zero-mass, zero-spin
Nambu-Goldstone boson (or the Goldstone mode of symmetry in
quantum field theory) \cite{Anderson07,Anderson58,Nambu60,Nambu61,
Goldstone61,Levy60,BenLee73,Weinberg96}. The reason why the
spontaneously broken symmetry may play a fundamental role is due
in part to its practical usefulness as a basic mechanism by which
we can explain the apparent breakdown of superfluidity at a nodal
surface for a study of surface phenomena in both He II and BEC
\cite{Meservey64,Reif64,Glaberson68}, for which a standard
perturbation method fails due to the singularity in the mean
field.

A proper treatment of the boundary conditions is especially
important in a finite space problem; we have developed the method
that replaces the physical surface by boundary conditions from
which we can separate the surface layer from the bulk fluids by
introducing the Lagrangian displacement vectors
\cite{BFKK58,Low58,SJHan91}. Moreover, we have established the
similarities between the Meissner effect in a superconductor in an
external field and the characteristics of the surface layer in He
II under the gravitational field. In fact a dominant feature of
the similarities is so striking that it is quite reasonable to
inquire whether the Goldstone theorem can be equally applied to
the imperfect Bose system.

It will be shown, with a specific example, that the broken
symmetry is both a necessary and a sufficient condition for the
conservation of energy in an isolated Bose system. Thus the
spontaneously broken symmetry in a Bose quantum fluid is a
fundamental phenomenon that may explain the peculiar normal
fluid-like behavior of the surface layer of a rotating He II under
the gravitational field; similarly it can also explain the
breakdown of superfluidity at a vortex core. This peculiar
behavior of a surface layer of a rotating He II has troubled the
intuition of many experimental physicists since the its first
observation by Osborne in 1950
\cite{Osborne50,Meservey64,Packard92}.

The new BEC in a trap is a small droplet with tens of microns in
radius and has a low density, that is to say, the mean particle
density $\bar{\rho}$ satisfies $\bar{\rho}|a|^{3}\ll 1$, where $a$
is an $s$ wave scattering length. Thus we can adequately describe
the pair-interaction by the hard sphere approximation. A simplest
procedure we take is to assume the realization of BEC in a trap
and inquire how we may prove it. Of course, one may try to confirm
a phonon spectrum in experiments as shown in Bogoliubov's theory
of superfluid \cite{Bog47}. There are numerous papers on the
theories of a Bose quantum fluid
\cite{Bog47,Landau41,Mott49,Feynman53,Brueckner57}; they are,
however, developed for a spatially uniform Bose system in a
Hilbert space for mathematical convenience. But anyone who is
familiar with modern quantum field theories \cite{Bog59,note1}
should see at once that the conventional perturbation theories
based on the field theories ({\it i.e.,} the pair interactions
through a quantized field) may fail entirely in our problem of
atomic BEC in a trap.

In a finite, spatially, inhomogeneous many-body system, it was
demonstrated \cite{SJHan2} that the most serious difficulty of
applying these theories is that the broken symmetry at the nodal
surface yields a new set of dispersion relations for the
collective excitations, so that the picture of the collective
excitations in the previous theories
\cite{Bog47,Landau41,Feynman53} is significantly altered due to
the presence of a nodal surface. Similarly, the boundary
conditions play a peculiar but a decisive role in a finite space
problem. Both for this reason, and because the quantum field
theoretic (microscopic) approach cannot be applied to a spatially
inhomogeneous problem with a boundary \cite{note1}, it is
therefore essential to find {\it a new mathematical method} by
which we may implement the necessary boundary conditions that are
consistent with the modified two-fluid model of Landau
\cite{Landau41,Mott49}. This is most simply carried out by
introducing the semi-classical perturbation method of Lagrangian
displacement vectors \cite{BFKK58,SJHan91}.

Our discussion of these problems is based on the following
physical picture: an incompressible surface excitation involves a
breakup of phase coherence over the surface of BEC droplet,
whereas longitudinal excitations, however, such as those generated
by compression, do not break up the phase coherence in a
superfluid and yield the Bogoliubov spectrum for a phonon. Thus
the longitudinal and surface excitations in turn identify the type
of fluids that support these excitations. The surface waves are
essentially transverse in nature. In addition to the usual surface
waves, we will show that, in the surface layer, there is also a
transverse wave that is driven by entropy change in the layer,
which is different from that of temperature waves (second sound)
\cite{Landau41,Tisza47}.

One of the essential new steps in our study of the broken symmetry
is to adopt Mott's suggestion \cite{Mott49} on the extension of
the two-fluid model of Landau \cite{Landau41}. Based on London's
interpretation of the $\lambda$-transition \cite{London38} and the
explanation of energy gap by Bijl, et al., \cite{Bijl41}, Mott
suggested that, to describe the behavior of helium in Rollin films
and to derive the critical velocity of a superfluid flow in a
capillary, it is necessary to extend the two-fluid model by
identifying the superfluid with the atoms in the ground state and
the normal fluid with the atoms in the excited states with a
finite energy gap (see Figs. 3, 5 and 6 of Ref. \cite{Mott49}).
Particularly remarkable, however, is the anticipation of the
Onsager-Feynman quantization of superfluid velocity in a rotating
He II as a mechanism to give rise the surface energy by velocity
discontinuity \cite{Onsager49,Feynman55}. Thus we here define the
normal fluid as composed of phonons (and rotons) and the excited
atoms with finite energy gap from the ground state as \textit{an
extended two-fluid model of Landau} \cite{Landau41,Landau80}.

It will be shown in this paper that the above characteristic
difference between the two waves (the sound wave and the surface
wave) provides the reason why the spontaneously broken symmetry
must take place at the nodal surface - a breakdown of
superfluidity. A proof of the broken symmetry lies at the core of
many of unresolved problems in low temperature physics, especially
since there is convincing evidence for the break-down of
superfluidity at the free surface of rotating He II. Perhaps a
more fundamental explanation for the broken symmetry in He II is
that one cannot hold the law of conservation of energy in a
finite, isolated system by dissipation process unless the symmetry
is spontaneously broken at the nodal surface. This is because a
phonon cannot interact with a superfluid component, but with the
aid of Mott's extension of the two-fluid model \cite{Mott49}, it
does, however, interact with a normal fluid, giving rise to a
surface energy - a capillary wave by the surface tension.

The above discussion is by no means a complete answer to the
surface layer problem. By the well-known arguments of
spontaneously broken symmetries
\cite{Goldstone61,Nambu61,Anderson63,BenLee73,Weinberg96}, this
peculiarly universal behavior of a surface layer of a superfluid
in a gravitational field cannot be explained by the Goldstone
theorem alone, since the degree of broken symmetry depends on the
curvature of the free surface and the energy of a superfluid in
the system. For example, the breakdown of superfluidity at the
vortex core accompanies a roton whose effective mass is
$\mu_{ro}=0.16m_{He}$ (a pseudo-Goldstone boson)
\cite{Levy60,Weinberg72} but not a massless phonon (a Goldstone
boson). Since the non-trivial irreducible linear representation of
the group $U(1)$ is a real two-vector $\psi=\psi_{1}+i\psi_{2}$,
we write the ground state function in ODLRO as
$\psi(r,t)=f(\bm{r},t)exp[\frac{i}{\hbar}S(r,t)]$, where $S(r,t)$
is the action and is a solution of the quantum Hamilton-Jacobi
equation \cite{Bohm52}. Following Anderson \cite{Anderson66} we
also interpret the particle field operator $\psi(r,t)$ as our
definition of a superfluid which has a macroscopic mean value
$\langle\psi(r,t)\rangle=f(r,t)exp[i\phi(r,t)]$ with a definite
phase $\phi(r,t)$ \cite{Aharonov63}.  Thus the collective
excitations (a gravity wave, a phonon, and a roton) and Bohm's
quantum theory \cite{Bohm52} are essential for a complete
understanding of properties of a superfluid in a gravitational
field, particularly its interaction with the gravitational field
and the spontaneously broken symmetry at the nodal surface.

We proceed further in two stages: we first derive collective
excitations in a spherically symmetric BEC droplet to show the
reason why the hidden symmetry is broken on the nodal surface of
the BEC in a trap - a shell-like structure and then extend the
calculations to He II. In the second stage of calculation, we
discuss the mechanism by which the symmetry is broken by
fluctuation-dissipation in terms of the action and the effective
quantum mechanical potential \cite{Bohm52,Aharonov63}.

With the spontaneously broken symmetry, I will then discuss the
three outstanding problems: a) a breakdown of superfluidity at the
vortex core in He II for which there is still no proven mechanism
that explains the breakdown in superfluidity at the core
\cite{note2}, and also resolves a long-standing controversy over
the form of the Magnus force that has troubled many theoreticians
\cite{Thouless99} since the first experiment on a vortex
quantization by Vinen in 1961 \cite{Vinen61}; b) we discuss the
Onsager-Feynman quantization of circulation in He II which is an
extension of the Bohr-Sommerfeld condition in phase-space for
which we show the identifiable direct cause for it to take place
in He II with Planck's quantum condition; c) a longstanding riddle
on the curvature of a rotating He II under the gravitational field
- the theoretical curvature based on Landau's two-fluid model is
incompatible with that of observed in the experiments
\cite{Osborne50,Meservey64}. The phenomenon of broken symmetries
confined, as it is in both cases, to the geometrical surface,
might be explained by the fact that the particles in the surface
layer (normal fluid) are free from from the interaction with the
bulk superfluid, they can be accelerated only by external forces.
With these qualitative remarks as an introduction, we proceed to
the development of phenomenological theory of a quantum Bose
liquid.

\section{\label{sec:level1} Phenomenological Theory of a Bose Quantum Fluid}

The broken symmetries in a superconductor have been studied
extensively since the remarkable discovery of Josephson junctions
\cite{Josephson62,Anderson63-1,Josephson65,Anderson84}. And yet
there is no systematic study of the broken symmetry in He II and
an imperfect Bose gas in a trap (BEC) to date. Moreover, there is
actually a fundamental difference between the ways in which the
broken symmetry is realized in a superconductor and in a Bose
quantum fluid (He II and BEC). The problem is then to define in a
mathematically precise way the broken symmetries in an interacting
Bose system in which both the law of conservation of energy and
the number density must be preserved.

The basic idea of spontaneously broken symmetry in a superfluid
can be explained by the Goldstone theorem alone
\cite{Goldstone61,Goldstone62,BenLee73}. The physical principles
underlying mathematical analysis in the derivation of the
Goldstone theorem are so clear that it is easy to see how the
theorem can be applied to a number of problems in low temperature
physics. However, we would first study the broken symmetry with
the dynamical calculations based on the ground state wave function
defined in ODLRO \cite{Penrose56} and Bohm's quantum theory
\cite{Bohm52}. The fist step is therefore to define the ODLRO and
thereby the ground state wave function - mean field. For this
purpose we follow Anderson's analysis \cite{Anderson66}.

\subsection{\label{sec:level2} Basic Equations}

The idea of off-diagonal long-range order (ODLRO) by Penrose and
Onsager \cite{Penrose56,Schafroth54,Yang62} may be stated
precisely as the following: the superfluidity be described as a
state in which the reduced density matrix of the condensed Bose
system can be factorized,
\begin{equation}
\rho(r,r^{\prime})=\psi^{\dag}(r)\psi(r^{\prime})+
\gamma(|r-r^{\prime}| \label{Odlro}.
\end{equation}
where $\gamma\rightarrow 0$ as $|r-r^{\prime}| \rightarrow
\infty$. Here the single particle wave function $\psi(r)$
represents a Bose condensed ground state wave function in ODLRO
and describes a superfluid in terms of the single particle wave
function. We take $\psi(r)$ as the mean value of the quantum
particle field or simply the mean field and consider it as the
definition of a superfluid in this paper
\cite{Penrose51,Penrose56,Yang62}. More important is the
requirement that the wave function be single valued in this
measurable part of the density matrix is what leads to the
quantization of vorticity in He II.

With the the hard sphere approximation for the repulsive pair
interactions for Bose particles \cite{Lee57,Brueckner57}, one can
show that the ground state wave function satisfies the nonlinear
Schr\"{o}dinger equation (Gross-Pitaevskii) in the
self-consistent Hartree approximation \cite{Fetter71},
\begin{equation}
i\hbar\frac{\partial\psi}{\partial t}=-\\
\frac{\hbar^{2}}{2M}\nabla^{2}\psi + [V(\bm{x})_{ext} + g_{1}|\psi|^{2}] \psi,\\
\label{Cat2}
\end{equation}
where $g_{1}=4\pi\hbar^{2}a/M$ and $a$ is an s-wave scattering
length \cite{Lee57,Fetter71}. Eq.~(\ref{Cat2}) is a
self-consistent Hartree equation for the Bose condensed wave
function. It should be emphasized that the nonlinear term
$|\psi|^{2}$ is invariant under a $U(1)$ group transformation. The
Lagrangian from which we can derive Eq.~(\ref{Cat2}) is therefore
invariant under the $U(1)$ group provided that $V(\bm{x})_{ext}$
is invariant, which we assume to be the case [See also Eq. (5a)].
It is also well worth to point out that we can still apply Bohm's
interpretation of quantum theory to Eq.~(\ref{Cat2}) with the
understanding that the self-interacting term be a part of the
potential defined in his quantum theory \cite{Bohm52,Beliaev58}.
Furthermore, the nonlinear term in the nonlinear Schr\"{o}dinger
equation is responsible for an emission of a phonon in a Bose
liquid just like the nonlinear Maxwell equations describe a photon
emission \cite{Bohm52}.

Since the mathematical analysis of the quantum phenomena with the
ground state bounded by a free surface presents much greater
difficulties due to a inhomogeneous spatial density, the question
naturally arises: what is the critical parameter that can assure
us a semi-classical mathematical method as an adequate
perturbation method within given limits of accuracy? The answer to
this question is obviously Plank's constant $\hbar$, since quantum
effects become important as the temperature of the system
approaches to absolute zero and thus the pair-interaction
potential is calculated by the hard-sphere approximation. In the
classical limit ($\hbar\rightarrow 0$), the action $S(\bm x)$ is a
solution of of the Hamilton-Jacobi equation as is shown below.
This implies that if a solution is independent of $\hbar$, it is
then a classical part of the solution. Thus $\hbar$ gives the
fundamental difference between a normal fluid and a superfluid,
since the pair-interaction, which is derived from a hard sphere
approximation \cite{Lee57}, brings about the $\lambda$ transition
in liquid helium as in a superconductor ({\it i.e.,} Cooper
pairs). We also notice that the dispersion relation for the
collective excitations in a superfluid is given by $\omega = ck$,
where $c$ is speed of (first) sound $c=[4\pi a
\rho(r)\hbar^{2}]^{1/2}/M$ and $k$ is the wave number in the
phonon regime in He II ($T \ll T_{\lambda}$). In order to
understand what is involved in such an analysis, we first consider
the problem of collective excitations in an atomic BEC in a trap
and derive the Bogoliubov spectrum to show that the perturbation
method is indeed correct to the first-order.

Thus the whole problem of superfluid dynamics depends on the
physical meaning that is ascribed to the particle field operator
$\psi$ in the equation Eq.~(\ref{Odlro}) as the macroscopic mean
value in a Bose system. Let us now take up the main problem of
superfluid dynamics and ask the question of how to apply the mean
field $\psi$ to the solution of the broken symmetry in He II. It
is important to keep in mind that $\psi$ is also a complex
order-disorder parameter in the sense of Landau
\cite{Landau41,Ginzburg50}. The first step is, therefore, to
express the wave function in terms of an action and amplitude that
yield the equations of motion which are consistent with the
modified two-fluid model of Landau
\cite{Landau41,Mott49,Landau80}. The action satisfies the quantum
Hamilton-Jacobi equation that gives an explanation for a
spontaneously broken symmetry in the BEC in a trap. This is
precisely the reason why we adopt Bohm's interpretation of quantum
mechanics \cite{Bohm52} in our study.

The basic ideas we apply in our discussion is best understood in
terms of a more familiar example. We thus begin our discussion by
recapitulating the basic concepts of Bohm's quantum theory. The
Bohm's interpretation of the quantum theory is based on the three
special assumptions:

a) the single particle field $\psi$ satisfies Schr\"{o}dinger's
   equation;

b) if we write $\psi(r,t)=f(\bm{r},t)exp[\frac{i}{\hbar}S(r,t)]$,
   then the particle momentum is restricted to $\bm
   {p}=\bm{\nabla} S(\bm {x},t)$ in the classical limit ($\hbar\rightarrow 0$);

c) we have a statistical ensemble of particle
   positions, with the probability, $Pr=|\psi|^{2}=f(\bm{r},t)^{2}$.

Plank's constant $\hbar=1.054\times 10^{-27}$erg sec here plays a
critical role in our entire discussion on He II and the BEC; the
transition from quantum mechanics to classical mechanics can be
formally described as a passage to the limit $\hbar \rightarrow 0$
for which the effective quantum mechanical potential becomes
$U_{eqmp}=0$ (see below). As discussed above, we have modified the
first assumption by including the nonlinear term for the
pair-interaction of Bose particles, but it is still invariant
under the $U(1)$ group transformation and becomes a part of the
potential in the wave equation. Thus the wave field $\psi$
satisfies a Schr\"{o}dinger equation Eq.~(\ref{Cat2}) with the
nonlinear interaction term (or simply the nonlinear
Schr\"{o}dinger equation). By deriving the well-known Bogolibov
spectrum in quantum liquids \cite{Bog47} from the nonlinear
Schr\"{o}dinger equation, we demonstrate that our modification of
the Bohm's quantum theory does not affect his interpretation. The
essential point is to write the wave function in the form
$\psi(r,t)=f(\bm{r},t)exp[\frac{i}{\hbar}S(r,t)]$, where $S(r,t)$
is an action and $\rho=|f(r,t)|^{2}$ may be interpreted as a
number density for a system of $N$ Bose particles \cite{Bohm52}.
In the quantum many-body theory \cite{Anderson84}, $\psi(r,t)$ has
been interpreted as a complex order parameter; it has both an
amplitude $f(\bm{r},t)$ and a phase $\phi=\frac{i}{\hbar}S(r,t)$
which is coupled to external forces, whereas $f(\bm{r},t)$ is
merely an internal order parameter in the sense of Landau's order
parameter of the systems \cite{Landau41,Ginzburg50}. Thus Bohm's
quantum theory does not create any difficulty in the conventional
many-body theory and can be viewed as an extension of it
\cite{Anderson84}.

Furthermore, it should be emphasized that Bohm's quantum theory is
essentially equivalent to Feynman's space-time approach to quantum
mechanics \cite{Feynman42}. Bohm's approach is, however, better
suited especially to boundary value problems, because the action
$S(r,t)$ satisfies the quantum Hamilton-Jacobi equation (QHJE)
exhibiting the broken symmetry by the effective quantum mechanical
potential at the nodal surface in the limit $\hbar\rightarrow 0$,
whereas $\rho=|f(r,t)|^{2}$ satisfies the equation of continuity
from which we can derive boundary conditions at the nodal surface
\cite{Lamb45}. Both $S(r,t)$ and $\rho=|f(r,t)|^{2}$ require the
boundary conditions for their solutions and are not quite
independent from each other \cite{Bohm52,Aharonov63}.

We now derive the following basic equations from the wave equation
Eq.~(\ref{Cat2}) with
$\psi(r,t)=f(\bm{r},t)exp[\frac{i}{\hbar}S(r,t)]$,
\begin{subequations}
\label{allequations} \label{Mean}
\begin{equation}
\frac{\partial\rho}{\partial t}+\bm{\nabla}\cdot(\rho\frac{\bm\nabla S}{M})=0\\
\label{subeq:1}
\end{equation}
\begin{eqnarray}
 \frac{\partial S}{\partial t} + \frac{(\bm\nabla S)^{2}}{2M} +
 V(\bm{x}) - \frac{\hbar^{2}}{4M}[\frac{\nabla^{2}\rho}{\rho}-
\frac{1}{2}\frac{(\nabla\rho)^{2}}{\rho^{2}}]=0, \label{subeq:2}
\end{eqnarray}
\end{subequations}
where $\rho=f(\bm {r},t)^{2}$, $V(\bm{x})\equiv V_{ext} +
(4\pi\hbar^{2}a/M)\, \rho$. Here $(4\pi\hbar^{2}a/M)\, \rho$ is
taken as an external potential \cite{Feynman53,Beliaev58}.
Eq.~(\ref{subeq:2}) is the quantum Hamilton-Jacobi equation (QHJE)
and reduces to the Hamilton-Jacobi equation in the limit of $\hbar
\rightarrow 0$.

It may be worth to note the dependence of the potential energy on
$(4\pi\hbar^{2}a/M)\, \rho$ that arises from the pair-interaction
of an imperfect Bose gas \cite{Lee57}. In the model we study in
this paper, the dependence of $\hbar$ in the potential arises from
a hard sphere approximation of two-particle scattering through a
repulsive potential of the hard-core range $a$; therefore, the
speed of the (first) sound $c=[4\pi a \rho\hbar^{2}]^{1/2}/M$
should depend on $\hbar$ in the analysis of collective excitations
in the atomic BEC in a trap and He II.

By the Hamilton-Jacobi theory with Eq.~(\ref{subeq:2}), it
follows that the second assumption of Bohm's quantum theory $\bm
{p}=\bm{\nabla} S(\bm x)$ is consistent with the usual
interpretation of the quantum theory, in the sense that if it
holds initially, it will hold for all time.  From
Eq.~(\ref{subeq:1}), it follows that if $\bm {v}=\bm {\nabla}
S/M$ and $\psi$ satisfies Schr\"{o}dinger's equation, then the
third assumption of Bohm's interpretation implies that the
probability $P=|\psi|^{2}=f(\bm{r},t)^{2}$ is conserved. Thus if
one interprets $\rho=f(\bm{r},t)^{2}$ as a number density, it is
also conserved by Eq.~(\ref{subeq:1}). Moreover,
$\rho=f(\bm{r},t)^{2}$ manifests itself as a useful function of
hidden variables by which we describe the collective excitations
in a Bose condensate. The last term in Eq.~(\ref{subeq:2}) is the
effective quantum mechanical potential (EQMP) defined by
\begin {equation}
U_{eqmp}=-\frac{\hbar^{2}}{4M}[\frac{\nabla^{2}\rho}{\rho}-
\frac{1}{2}\frac{(\nabla\rho)^{2}}{\rho^{2}}]=
-\frac{\hbar^{2}}{M}\frac{\nabla^{2}f}{f}. \label{Eqmp}
\end{equation}

In the limit ${\bm\nabla}\rho=0$ for a homogeneous system,
$U_{eqmp}=0$ and thus $S(r,t)$ becomes a solution of the
Hamilton-Jacobi equation which is consistent with the usual
definition of a phase $\phi=S(r,t)/\hbar$.

In general, Eq.~(\ref{subeq:2}) implies, however, that the
particle moves under the action of the force that is not entirely
derivable from the potential $V(\bm{x})$, but which also obtains a
contribution from the EQMP. The important aspect of
Eq.~(\ref{Eqmp}) is that it fluctuates near the nodal surface as
$\nabla\rho\rightarrow \infty$ and drives the system to undergo
the spontaneously broken symmetry at the nodal surface. The
concept of EQMP is the main point of Bohm's quantum theory that
shows how {\it Bohr's correspondence principle} breaks down at the
nodal surface. Thus we see, in the language of quantum field
theory \cite{Weinberg96}, the spontaneously broken local gauge
symmetry in He II and the atomic BEC in a trap
\cite{Anderson07,Anderson58,Nambu60,Nambu61,
Goldstone61,Levy60,BenLee73}. As will be shown below, it should be
emphasized that there is no unique way of making transition from
classical to quantum mechanics other than taking the classical or
correspondence-limit $\hbar\rightarrow 0$. However a simple
quantum system in which the density is uniform ${\bm\nabla}\rho=0$
in space yields the correct classical limit.

Next we write the equations for the ensemble average energy in the
usual quantum theory \cite{Bohm52,Aharonov63}:
\begin{subequations}
\label{allequations} \label{ground}
\begin{eqnarray}
{\cal H} =\int \psi^{\dagger}\left(-\frac{\hbar^{2}}{2M}\nabla^{2}
+
V(\bm{x})_{ext} +\frac{g}{2}|\psi|^{2}\right)\psi d\bm{x}, \label{grounda}\\
{\cal E}_{ave} =\int\left(\frac{\hbar^{2}}{2M}|\nabla\psi|^{2} +
V(\bm {x})_{ext}|\psi|^{2}+\frac{g}{2}|\psi|^{4}\right) d\bm{x},
\label{groundb}\\
\int\psi^{\dagger}\psi d\bm{x}=N.\label{groundc}
\end{eqnarray}
\end{subequations}

 At this point it is possible to show that the ground state density
profile is given in terms of an external potential and the
chemical potential upon minimizing the energy functional
Eq.~(\ref{groundb}) with Eq.~(\ref{groundc}) and with the
condition $\bm{p}=0$ (Penrose-Onsager criterion for BEC):
\begin{equation}
\rho(\bm{x})=|\psi(\bm{x})|^{2}=\frac{M}{4\pi\hbar^{2}a}[\mu
-V(\bm{x})_{ext}], \label{Ground}
\end{equation}
where $\mu$ is a Lagrangian multiplier and is the chemical
potential.

 Eq.~(\ref{Ground}) is of the utmost importance in understanding
superfluidity of BEC in a trap. It tells us two things: first,
that $f(r)$ is a function of thermodynamic variable $\mu$ and the
external potential $V(\bm{x})_{ext}$, and that the density is
inhomogeneous in space and that the standard equation of motion of
a superfluid, $d{\bm v(\bm x)}/dt = d[\bm\nabla S(\bm
x)]/dt/M=-{\bm\nabla}\mu$ breaks down due to fluctuations of $
U_{eqmp}$ at the free surface, and thus defines the domain of
ODLRO; second, that the broken symmetry takes place at the free
(nodal) surface to maintain the steady state \cite{SJHan1}. {\it
This is precisely the nature of superfluid that they cannot assume
a stationary state under an external field or pressure gradient}
\cite{Anderson65}. It may be worth while to point out that Landau
\cite{Landau41} introduced the chemical potential to describe the
a superfluid as a potential flow, {\it i.e.,}
$\bm{\omega}=\bm{\nabla}\times\bm{v_{s}}=0$ and
$\frac{\partial}{\partial
t}\bm{v}_{s}+\bm{\nabla}[(1/2)\bm{v}_{s}^{2}+\mu]=0$. And hence
Landau emphasizes the superfluid velocity and its equation of
motion in his two-fluid model \cite{Landau41,Anderson65}.

 In order to obtain the explicit relation between the mean field
in ODLRO and many-body ground state wave function, we have to
invoke the phase coherence \cite{SJHan1,SJHan2,SJHan3} which
clearly shows the role of the mean field in many-body theory
\cite{Penrose56,Anderson66}. Here the phase-coherence is defined
as follows: if we write the ground state wave function in ODLRO as
$\psi(r,t)=f(\bm{r},t)exp[\frac{i}{\hbar}S(r,t)]$ where $S(r,t)$
is the action (phase), then we may state the phase coherence as
\begin{equation}
\label{coherence}
\bm{\xi}\cdot\bm{\nabla}S(\bm{x}_{0},t)
=\sum_{i}\bm{\xi}_{i}\cdot\bm{\bm{\nabla}_{i}}S_{0,i}(\bm{x}_{0,i},t),
\end{equation}
where $\bm{\xi}$ are the Lagrangian displacement vectors
\cite{BFKK58,SJHan91}. The summation of the phase change by an
individual atom on the right-hand side is not measurable as it is
hidden variable in Bohm's interpretation, but the left-hand side
in the mean-field can be measured in an experiment
\cite{Anderson64}. This is is an excellent example of a hidden
variable of a single particle phase in Bohm's quantum theory
\cite{Bohm52}. One very important point about the phase-coherence
is that it is a necessary condition for the existence of
superfluidity by ODLRO of Penrose-Onsager theory of Bose-Einstein
condensation. The phase-coherence is a striking feature of
superfluidity. Its role in a superfluid is remarkable in
connection with the spontaneously broken symmetry as discussed
below.

Because He II is a Bose system, the ground state wave function is
symmetric under particle exchange and has no nodes in the case of
a homogeneous system. Based on these properties, Feynman went on
to argue that {\it there can be no single particle excitation}
\cite{Feynman53}, which is essentially equivalent to the
phase-coherence in our analysis. Hence the collective excitations
we study in the Bose-Einstein condensation in a trap are a phase
coherent, longitudinal sound wave (a phonon) \cite{Anderson58},
the dispersion relation which is given in the form of $\omega
\simeq ck$ \cite{SJHan1,SJHan3}. In contrast to the two-fluid
model of Feynman in his study of atomic theory of liquid helium
\cite{Feynman53}, the ground state of BEC in a trap
Eq.~(\ref{Ground}) is not uniform and has a nodal surface on which
the complex order parameter $\psi(r,t)$ undergoes the second-order
phase transition and hence the symmetry is broken spontaneously as
will be shown below.

\subsection{\label{sec:level2} Landau's Two-fluid Model}

 When liquid He$^{4}$ is cooled below $T_{\lambda}=2.19^{o}$K, it
enters a new phase known as He II. This thermodynamic transition
is marked by a peak in the specific heat, which behaves like
$ln|T-T_{\lambda}|$ on both sides of the transition which
represents the onset of Bose condensation - the
$\lambda$-transition.

 The He II has remarkable hydrodynamic properties, many of which
can be explained by Tisza's phenomenological two-fluid model
\cite{Tisza47} and by Landau's two-fluid model
\cite{Landau41,Landau47}. Landau recognized that He II resembles
 a macroscopic mixture of two noninteracting components: a
 superfluid phase of zero viscosity and a normal fluid made of
 collective excitations of phonons (and rotons) at all finite
 temperature \cite{Landau41,Landau47,Mott49}. The flow of the
 superfluid component is irrotational
 $\bm{\nabla}\times\bm{v}=0$; the collective excitations are
 called the normal fluid with entropy which behaves like a
 classical fluid with viscosity. The density $\rho_{n}/\rho$
 depends on temperature, where $\rho=\rho_{s}+\rho_{n}$. The
 two-fluid model \cite{Landau41} is essentially correct
 \cite{Andro46,Feynman53}, but a question for its
 completeness was raised by Osborne in his experiment of a rotating
 He II \cite{Osborne50}, because the two-fluid model does not allow
 a physical separation of the superfluid and normal parts of the
 fluid \cite{Landau41,Mott49}. Except for the surface-layer in which
 the two-fluid model breaks down because of the boundary conditions
 \cite{Lamb45}, the model is no more than a convenient description
 of the phenomena that occur in a fluid where quantum effects are
 important.

A single particle wave function $\psi(r)$ is introduced as the
superfluid order parameter (a mean value) in the definition of an
off-diagonal long range order (ODLRO) \cite{Penrose56}. Due to a
convincing argument of Penrose and Onsager
\cite{Penrose51,Penrose56} for the theory of Bose-Einstein
condensation and liquid helium as an extension of London's theory
of the $\lambda$-transition in a strongly interacting Bose gas
\cite{London38}, ODLRO was generally accepted as the definition of
a superfluid \cite{Anderson66}. And yet it still lacks an
operational (or direct) link to the two-fluid model by ODLRO as
the way in which the BCS microscopic theory is shown to be
equivalent to the macroscopic Ginzburg-Landau theory by Gor'kov
\cite{Gorkov59}.

Although the two-fluid model was a conceptual breakthrough
\cite{Landau41,Tisza47} and was quite successful in explaining
the dynamics of He II in a homogeneous flow with emphasis on the
superfluid velocity and its equation of motion
$\bm{\omega}=\bm{\nabla}\times\bm{v_{s}}=0$, it is known as an
incomplete phenomenological theory.

 The most important equation of motion that defines a superfluid
in both the two-fluid model and the Bose-Einstein condensation
follows from a ground state wave function defined in ODLRO
\cite{Penrose51,Penrose56,Beliaev58,Yang62,Anderson07} and is
given by Bohm's second assumption in his quantum theory
\cite{Bohm52}, $\psi(r,t)=f(\bm{r},t)exp[\frac{i}{\hbar}S(r,t)]$
in which $S(r,t)$ is the action, then
\begin{equation}
\label{superfluid}
\bm{\nabla}\times\bm{v_{s}}=\bm{\nabla}\times\bm{\nabla}S(\bm{x},t)/M=0,
\end{equation}
where $\bm{p}=\bm{\nabla}S$. Hence this identity shows its
limitation as it is true only in the classical limit
($\hbar\rightarrow 0$). If we take the point of view that $S$ must
be a solution of the quantum Hamilton-Jacobi equation
Eq.~(\ref{subeq:2}), then $S$ breaks the gauge symmetry
\cite{Anderson66}; it will be shown below how the fluctuation and
dissipation driven by the effective quantum mechanical potential
Eq.~(\ref{Eqmp}) brings about the spontaneously broken gauge
symmetry at the nodal surface.

The most significant achievement of the two-fluid model
\cite{Landau41} has been in the analysis of the energy spectra of
collective excitations in He II and the experimental data by
Henshaw \cite{Henshaw61} confirmed the prediction of the model in
great detail. There are, however, many problems for which the
spatial inhomogeneity of the fluid with a boundary poses a major
stumbling block, in the study of the two-fluid model and still
remains unsolved.

More specifically, Landau \cite{Landau41,Landau47} studied the
collective excitations to explain the macroscopic properties of
He II with a particular form of the energy-momentum curve that
rises linearly for small momentum $p=\hbar k$, passes through a
maximum, falls to a local minimum, and rises again. The
excitations in the linear region are quantized sound waves
(phonons); their energy, measured relative to the ground-state
energy, is given
\begin{equation}
\label{phonon} \epsilon_{k}=\hbar c k,
\end{equation}
where at $T=1.12^{0}$K, data were obtained in the momentum range
$k=p/\hbar=0.25 \sim 2.5 \AA^{-1}$, $c \approx 238$ m/sec is the
speed of (first) sound \cite{Yarnell59,Henshaw61}.
 Near its local minimum of the energy-momentum curve, the energy
spectrum can be approximated by a parabola
\begin{equation}
\label{roton}
\epsilon_{k}=\Delta +
\frac{\hbar^{2}(k-k_{0})^{2}}{2\mu_{r}}.
\end{equation}
Landau \cite{Landau41} further proposed that the excitations in
this region represent rotons, the quantum analog of smoke rings.
In Landau's two-fluid model, the total free energy arises from the
thermally excited quasi-particles (phonons and rotons), which are
treated as an ideal degenerate Bose gas. The theoretical curve
that describe the above Eq.~(\ref{phonon}) and Eq.~(\ref{roton})
has been confirmed in great detail by inelastic neutron scattering
experiments including the roton parameters
$\Delta/k_{B}=8.6^{o}$K, $k_{0}=1.91 \AA^{-1}$, and
$\mu_{r}=0.16m_{He}$ in Eq,~(\ref{roton})
\cite{Henshaw61,Cohen56}.

However, for over a half century it has been known that the
superfluid flow can exist only below Landau's critical velocity
$[\varepsilon/p]_{min}$, where $\varepsilon$ is the energy
spectrum of a phonon and roton; the roton limited critical
velocity $v_{cr}\approx 60m/sec$ has been observed
\cite{Landau41,Rayfield66}. Although the Landau's critical
velocity is an important parameter in flow experiments, yet it
remained a puzzle why they have observed the apparent breakdown of
superfluidity at the core of a vortex line at a velocity well
below the Landau's critical velocity \cite{Rayfield66}.

 Another problem that requires a fresh examination is the experiments
on a rotating He II by Osborne and Meservey
\cite{Osborne50,Meservey64}. The measured surface curvature in
steady rotation of He II (within a thin surface layer; average
depth $5.0\times10^{-3}cm$ along the curvature) is very nearly
given by $\gamma=\omega^{2}/g$ which is independent of temperature
and is incompatible with the two-fluid model \cite{Landau41},
where $\gamma$ is the maximum surface curvature
\cite{Osborne50,Meservey64}. Both Onsager and Feynman
independently have investigated the vorticity in a superfluid and
reached the similar conclusion that the peculiarly normal fluid
like behavior of the surface of a rotating He II can be explained
by the vorticity \cite{Onsager49, Feynman55}.

A number of theories \cite{Andro66} have been developed to explain
the observed data \cite{Osborne50,Meservey64}, notably Landau and
Lifshitz's vortex sheet model \cite{Lifshitz55}, and Hall and
Vinen's vortex line model \cite{Hall56} as suggested by Onsager
\cite{Onsager49} and Feynman \cite{Feynman55}. After a careful
observation on a rotating He II, Meservey has concluded none of
these theories appeared to be completely adequate to explain his
data; He II behaved like {\it a normal fluid} in a steady
rotation. As emphasized by Mott \cite{Mott49}, Meservey speculated
that {\it the surface energy} might have played a role in a manner
analogous to the Meissner effect in a superconductor in which the
exclusion of the external magnetic field from a superconductor was
attributed to the surface energy as shown in the Ginzburg-Landau
phenomenological theory of superconductivity
\cite{Ginzburg50,Tinkham96}.

\section{\label{sec:level1} Collective Excitations in Atomic BEC
in a Trap}

First we review the Bose-Einstein condensation based on a
degenerate Bose gas derived by Einstein \cite{London38} and then
extends the theory to a weakly interacting (imperfect) Bose gas.

\subsection{\label{sec:level2} Condensation in Degenerate Bose gas}

In this section, I should like first to review the Bose-Einstein
condensation in a degenerate Bose gas as discovered by Einstein
\cite{London38} to avoid the confusion from the weakly interacting
atomic gas in a trap.

At low temperatures, a Bose gas at constant density obeys the
following equation \cite{Landau80}:

\begin{equation}
\label{Bose}
\frac{N}{V}=\frac{g(mT)^{3/2}}{2^{1/2}\pi^{2}\hbar^{3}}\int_{0}^{\infty}
\frac{\sqrt{z}dz}{e^{z-\mu/T}-1},
\end{equation}
where $g=2s+1$ with $s$ the spin of the Bose particle and
$z=\epsilon/T$.

This equation implicitly determines the chemical potential of the
gas as a function of its temperature and density (N/V). By setting
$\mu=0$ at a temperature determined by the following equation:

\begin{equation}
\label{tempzero}
\frac{N}{V}=\frac{g(mT)^{3/2}}{2^{1/2}\pi^{2}\hbar^{3}}\int_{0}^{\infty}
\frac{\sqrt{z}dz}{e^{z}-1},
\end{equation}
which can be expressed in terms of the Riemann  zeta function. The
critical temperature $T_{0}$ can be expressed then as
\begin{equation}
T_{0}=\frac{3.31}{g^{2/3}}\frac{\hbar^{2}}{m}\left(\frac{N}{V}\right)^{2/3}.
\end{equation}

The total number of particles with $\epsilon>0$  with $\mu=0$ is
given by

\begin{equation}
N_{\epsilon>0}=\frac{gV(mT)^{3/2}}{2^{1/2}\pi^{2}\hbar^{3}}\int_{0}^{\infty}
\frac{\sqrt{z}dz}{e^{z}-1}=N(T/T_{c})^{3/2}
\end{equation}

 Thus the remaining particles with $\epsilon=0$ is then
\begin{equation}
N_{\epsilon=0}=N[1-(T/T_{0}^{3/2})].
\end{equation}

The steady increase of particles in the state with $\epsilon=0$ is
called Bose-Einstein condensation. This peculiar condensation is
also known as the momentum-space condensation to emphasize that
the cause of the condensation is solely due to the symmetry of
wave function for a degenerate ideal Bose gas. This is precisely
the reason why the mechanism appeared to have little physical
significance, since all real gases are condensed at the critical
temperature in the presence of molecular forces.

The energy of the degenerate gas for $T<T_{0}$ is given by
\begin{equation}
E=\frac{gV(mT)^{3/2}T}{2^{1/2}\pi^{2}\hbar^{3}}\int_{0}^{\infty}
\frac{z^{3/2}dz}{e^{z}-1}.
\end{equation}

This integral can be tabulated by the Riemann zeta function
$\zeta(\frac{5}{2})$ \cite{Landau80}, and is given by
\begin{equation}
E=0.128g\left(m^{3/2}T^{5/2}/\hbar^{3}\right)V.
\end{equation}

The specific heat is then given by
\begin{equation}
\label{specificheat0}
C_{v}=5E/2T.
\end{equation}

Next to show the the discontinuity of the first derivative of the
specific heat $(\partial C_{v}/\partial T)$ at $T=T_{0}$, we
calculate the energy of the degenerate Bose gas for a small
$|T-T_{0}|\ll \varepsilon$ by expanding the following integrand in
terms of small $\epsilon$ near $\mu=0$:
\begin{equation}
N=N_{0}(T)+
\frac{gV(m)^{3/2}}{2^{1/2}\pi^{2}\hbar^{3}}\int_{0}^{\infty}
{\sqrt{\epsilon}d\epsilon}\left[\frac{1}{e^{(\epsilon-\mu)/T}-1} -
\frac{1}{e^{\epsilon T}-1}\right],
\end{equation}
from which we obtain after some tedious algebra the chemical
potential in terms of $N-N_{0}$.

\begin{equation}
\mu=-\frac{2\pi^{2}\hbar^{6}}{g^{2}m^{3}}\left(\frac{N-N_{0}}{TV}\right)^{2}.
\end{equation}

Thus we obtain the energy for $T>T_{0}$
\begin{equation}
\label{energyright} E=E_{0}+\frac{3}{2}N_{0}\mu=E_{0}-
\frac{3\pi^{2}\hbar^{6}}{g^{2}m^{3}}N_{0}\left(\frac{N-N_{0}}{TV}\right)^{2},
\end{equation}
where $E_{0}=E_{0}(T)$ denotes the energy for $\mu=0$.

Using Eq.~(\ref{specificheat0}) and Eq.~(\ref{energyright}), we
calculate the difference the discontinuity of the derivative
$(\partial C_{v}/\partial T)_{V}$ at $T=T_{0}$
\begin{equation}
\label{thirdorder}
 \Delta\left(\frac{\partial C_{v}}{\partial
T}\right)_{V} =
-\frac{6\pi^{2}\hbar^{6}}{g^{2}m^{3}V^{2}}\left[N_{0}\left(\frac{1}{T}
\frac{\partial N_{0}}{\partial
T}\right)^{2}\right]_{T=T_{0}}=-3.66 N/T_{0},
\end{equation}
where we have followed the approach of Landau and Lifshitz
\cite{Landau80,London382,Fetter71} to calculate the values of the
first derivative $\partial C_{v}/\partial T)$ at $T_{0} \pm
\varepsilon$.

We see therefore that the degenerate Bose gas undergoes the third
order phase-transition as shown by London \cite{London382} and is
quite different from the $\lambda$ transition in a liquid helium
which is the second-order phase transition as in a superconductor.

\subsection{\label{sec:level2} Condensation in Interacting Bose gas}

We now turn to a weakly interacting Bose gas in trap. Within the
frame work of quantum field theory, Bogoliubov has developed the
theory of superfluidity by quantizing the scalar fields for Bose
particles in a Hilbert space \cite{Bog47}. It is the only accepted
approach for the derivation of the dispersion relation
$\epsilon_{k}=\hbar ck$ (or $\omega=ck$) for the phonon in He II.
 In order to come the Bogoliubov theory of superfluidity
\cite{Bog47} to grips with experimental observations of the BEC in
a trap, we must derive a phonon dispersion relation for the
collective excitations ({\it i.e.,} a sound wave) in the BEC with
proper geometrical corrections \cite{Henshaw61}. However, the
dispersion relation must be consistent with the Bogoliubov's
results \cite{Bog47} or the Bruckner and Sawada's
\cite{Brueckner57}. In a finite space problem, we cannot quantize
the fields \cite{note1,Bog59,Brueckner57} and thus we must find an
alternative perturbation method which is applicable to a
non-uniform system, and yet yields the same dispersion relation
consistent with the geometry.

 Our first task, then, is to find the condition for the
second-order phase transition (a superfluid to a normal fluid)
\cite{Ginzburg50} in BEC in a trap. Here we carry out the first
stage of the program described in the Introduction for the
breakdown of superfluidity, a spontaneously broken symmetry in the
BEC \cite{SJHan1,SJHan2,SJHan3,Anderson58}. The theory is subtle
and complex. The broken symmetry is a quantum interference
phenomenon taking place in both a Fermion system
\cite{Goldstone61,BenLee73,Weinberg96} and a Bose system
\cite{SJHan1,SJHan2,SJHan3,Anderson84} at low temperature. Since
the spontaneously broken gauge symmetry in a superconductor has
been discussed in detail ({\it i.e.,} the Meissner effect, the
flux quantization, and the Josephson effect) in a number of papers
employing the quantum field theory
\cite{Nambu60,Nambu61,Anderson58,Weinberg96}, we will focus on
spontaneously broken symmetries in a Bose system (the
Bose-Einstein condensation and He II), which is new
\cite{SJHan1,SJHan2,SJHan3}.

 We first discuss the dynamical aspect of the broken symmetry
since detailed dynamical studies are necessary to explain why the
spontaneously broken symmetry should take place at the nodal
surface. The basic algebra of a broken symmetry in a Bose system
has been described, in essence, by the semi-classical perturbation
of particle orbits given by $\bm{x}_{i}=\bm{x}_{0,i}
+\bm{\xi}(\bm{x}_{0,i},t)$, where $\bm{\xi}$ is a Lagrangian
displacement vector \cite{BFKK58,SJHan91}, and is both tedious and
complex \cite{SJHan1,SJHan2,SJHan3}. The whole point of employing
the Lagrangian displacement vectors in the perturbation analysis
is to find an alternative method that permits one to impose
appropriate boundary conditions in a finite system for which the
conventional field theory approach entirely fails.

 It should be emphasized, however, that from now on, we will discuss less
of the quantum field theoretical background of the broken
symmetry, than a number of its most important consequences for the
BEC and He II, many of which can be understood without quantum
field theories \cite{Bog47,Brueckner57} and are therefore of
fundamental interest. Here we will describe the essential algebras
and refer the reader Refs. \cite{SJHan1,SJHan2,SJHan3} for
details.

Both He II and the BEC of an imperfect Bose gas are described by
Eq.~(\ref{Odlro}) \cite{Penrose51,Penrose56} as a basic assumption
in our study of phase-coherent collective excitations. We have
come across the analogy between the $\lambda$-transition at
$T_{\lambda}=2.17^{0}$K in liquid helium (the second-order phase
transition) and the peculiar momentum condensation of the ideal
Bose gas at $T_{0}=3.14^{0}$K [{\it i.e.,} a discontinuity of the
derivative of the specific heat (phase transition in the
third-order)] \cite{London38,London382}, which seems go much
farther than the similar condensation, {\it i.e.,} the unique
feature of Bose-Einstein condensation of charged bosons provides a
clue that exhibits the phenomenon of superconductivity - the phase
transition of the second-order at the critical temperature below
which the Meissner effect occurs for sufficiently weak magnetic
fields \cite{Schafroth54}.

Finally, it has been shown by Penrose and Onsager
\cite{Penrose51,Penrose56} that, in the absence of an external
field, the BEC is always present in a spatially uniform system
with a periodic boundary conditions whenever a finite fraction of
particles have identical momenta \cite{Bog47},
\begin{equation}
\label{condens}
n_{M}/N=e^{{\cal O}(1)}\rightleftharpoons BEC,
\end{equation}
where $n_{M}$ is the largest eigenvalue of $\sigma_{1}=Ntr_{2,3,,,
N}(\sigma)$. Here von Neunmann's statistical operator $\sigma$
known as the density matrix is defined by $<q_{1}^{\prime}\cdots
q_{N}^{\prime}|\sigma|q_{1}^{\prime \prime} \cdots q_{N}^{\prime
\prime}>$ \cite{Neunmann55,Penrose56,Yang62}. In other words, our
definition of ODLRO as defined in Eq.~(\ref{Odlro}) is equivalent
to Eq.~(\ref{condens}), and thus one can see the presence of BEC
in a uniform He II in a Hilbert space in which the Bogoliubov
theory of superfluidity remains valid \cite{Bog47,Beliaev58}.

The question naturally arises what physical significance must be
ascribed to this criteria for the presence of the BEC. The answer
to the question is obvious: A single criterion for BEC in either a
fluid or a degenerate Bose gas is also applicable to BEC in He II
although it is an approximate caculation. It is clear therefore
that the theoretical interpretation of superfluidity defined by
ODLRO with Bohm's quantum theory \cite{Bohm52} is much more
fundamental than that of mere definition of a superfluid by the
superfluid velocity and its equation of motion \cite{Landau41}. It
should be stressed that the particle field operator
$\langle\psi\rangle$ also has a macroscopic mean value in a
superfluid system (the order-disorder parameter). Especially since
Bogoliubov's study of collective excitations in He II \cite{Bog47}
establishes the superfluidity of liquid helium below the $\lambda$
point \cite{Penrose56}, it is essential in our work to demonstrate
that we obtain a correct phonon spectrum for collective
excitations in the imperfect Bose gas consistent with the
Bogoliubov spectrum \cite{Bog47} and thereby to confirm the
realization of the BEC in a trap \cite{Davis95}.

As Feynman's picture of a phonon \cite{Feynman53} is essential in
our study of phase-coherent collective excitations in a
superfluid, it will be recapitulated here. In a series of papers
\cite{Feynman53}, Feynman has laid out an elaborate, complex
theory of collective excitations and an atomic theory of the
two-fluid model based on the exact partition function with his
space-time approach to quantum mechanics \cite{Feynman42}. The
main point of his theory is that the short range interaction of
any pair of Bose particles which brings about the superfluidity of
He II (the $\lambda$-transition) also ensures that the system
possesses certain collective excitations (a phonon and a roton)
that are the consequence of the off-diagonal long range order
({\it i.e.,} the absence of a long-range order)
\cite{Penrose56,Yang62}. Moreover, Feynman has given an argument
suggesting that, as emphasized by London \cite{London38}, the
$\lambda$-transition in liquid helium  is the same process as the
Bose-Einstein condensation of an ideal Bose gas \cite{Feynman53}.
The essential point of his argument is that the strong
pair-interactions in a liquid-like quantum mechanical system do
not prevent these particles from moving freely, because the system
remains in a state of a superfluid. Also this is, in essence,
Bogoliubov's theory of superfluidity \cite{Bog47}. In addition,
Feynman has pointed out that a density fluctuation satisfies the
conservation of number density: when the density is increased by
compression in one part of the system, it is decreased by
rarefaction in the other part of the system. This property leads
to Feynman's concept of back-flow in his analysis of a sound wave
in the liquid helium \cite{Cohen56}.

Finally, the ground state wave function is assumed to have a
positive amplitude similar to that of zero-point fluctuations of
the vacuum electromagnetic field in any configuration because the
ground state has no {\it nodes} in a uniform Bose system. With
these properties Feynman \cite{Feynman53} went on to argue that
there can be no low-lying single particle excitations and that the
only low-energy excitations are long-wavelength sound waves. In
Bogoliubov's theory of superfluidity, a phonon is a quantized
sound wave (a Goldstone boson - zero spin, zero mass, scalar
particle).

 When, and only when an appropriate perturbation technique, which
reproduces the Bogoliubov's spectrum \cite{Bog47}, is found, does
the perturbation method take on a firm ground from which the
qualitative description of collective excitations by Feynman
\cite{Feynman53} and the concept of ODLRO by Penrose and Onsager
\cite{Penrose56,Yang62} can be investigated quantitatively. We
will incorporate Feynman's picture of a phonon in our study of
collective excitations by introducing the Lagrangian displacement
vector \cite{BFKK58,SJHan91} in the particle orbits defined by
Bohm's quantum theory \cite{Bohm52} in ODLRO, which yields the
Bogoliubov spectrum \cite{Bog47}. It should be emphasized,
however, that the semi-classical method by the Lagrangian
displacement vectors is not applicable to a situation for excited
states with energies above the energy gap in the roton spectrum,
{\it i.e.,} states in the roton region of excitation
\cite{Brueckner57}.

As emphasized in the Introduction, the study of collective
excitations plays a unique role in interacting many-body systems
\cite{SJHan1,SJHan2,SJHan3}. \textit {In particular, we study the
collective excitation in the Bose-Einstein condensation (BEC) to
probe the dynamics of interacting Bose particles and thereby to
provide a course of experiments to confirm the realization of the
BEC in a trap} \cite{SJHan2,SJHan3,Penrose56}. At low temperature
($T \ll T_{\lambda}$), Bogoliubov's theory of superfluidity
predicts the collective excitation energy in the form of
$\omega=ck$ in the long wavelength limit \cite{Bog47}, where
$c=[4\pi a \rho\hbar^{2}]^{1/2}/M$ and k is the wave number.
Perhaps, one of the essential points in this paper that requires a
study of collective excitations for its explanation is that of
broken symmetry in the BEC in a trap - a shell-like structure,
since the spontaneously broken gauge symmetry always accompanies a
Nambu-Goldstone boson, a mass-less, spin-less particle (phonon)
\cite{Goldstone61}.

\subsection{\label{sec:level2} General Features of the Theory}

The use of the Hamilton-Jacobi equation in solving for the motion
of a particle is only matter of convenience. Thus if we write $\bm
{v}(\bm{x}_{0},t)=\bm{\nabla}S(\bm{x}_{0},t)/M$ ({\it i.e.,} a
solution of the Hamilton-Jacobi equation), we obtain the following
three dynamical equations from Eqs.~(\ref{Mean}) and
(\ref{Ground}): the equation of motion for a single particle,
\begin{equation}
M(\frac{\partial}{\partial t}\bm{v}+\bm{v}\cdot \bm{\nabla} \bm
{v})=-\bm{\nabla}\mu, \label{Motion0}
\end{equation}
the equation of continuity,
\begin{equation}
\frac{\partial}{\partial t}\rho+\bm{\nabla}\cdot(\rho\bm{v})=0,%
\label{Cont0}
\end{equation}
where $\rho$ is henceforth interpreted as the number density, and
the equation of state,
\begin{equation}
\mu(\bm{r}, t)=\mu_{\text{loc}}[\rho(\bm{r}, t)]+V_{\text{ext}},
\label{Eqs0}
\end{equation}
where $\mu_{\text{loc}}[\rho(\bf r, t)]$ is the chemical potential
in the local density approximation and is introduced to describe a
superfluid droplet confined by the external potential with a free
surface. In order to expand the chemical potential in the
lowest-order of $a$, we assume the following conditions:
$a\rho^{1/3}\ll 1$, $ k a\ll 1$, $a/\lambda\ll 1$, and
$a\lambda^{2}\rho \ll 1$, where $a$ is an s-wave scattering length
in the pair-interaction potential, $k$ is the wave number in the
new BEC \cite{KHuang57,Anderson66, Landau80,Oliva89,Yu96}.

Keeping in mind
$\psi(r,t)=f(\bm{r},t)exp[\frac{i}{\hbar}S(r,t)]$, we expand $S$,
$\bm{v}$, and $\rho$ to the first-order,
\begin{subequations}
\label{allequations} \label{First}
\begin{eqnarray}
S(\bm{r},t)=S(\bm{r}_{0},t) +
\bm{\xi}\cdot\bm{\nabla}_{0}S(\bm{r}_{0},t)\\
\label{Firsta} \bm{v}(\bm {x},t)=\frac {\partial}{\partial
t}\bm{\xi}
\label{Firstb}\\
\rho(\bm{x},t)=\rho(\bm{x}_{0})-
\bm{\nabla}_{0}\cdot[\rho(\bm{x}_{0})\bm{\xi}]=
\rho(\bm{x}_{0})+\delta \rho \label{Firstc},
\end{eqnarray}
\end{subequations}
where $\bm{p}(\bm{x}_{0})=0$ in the BEC and $\bm{\nabla}_{0}$
denotes the partial derivative with respect to $\bm{x}_{0}$ with
$\bm{\nabla}\rightarrow
\bm{\nabla}_{0}-\bm{\nabla}_{0}\bm{\xi}\cdot\bm{\nabla}_{0}$.
Eq.~(\ref{Firstc}) was derived by substituting Eq.~(\ref{Firstb})
to Eq.~(\ref{Cont0}) with $\bm{\nabla}S(\bm{x},t)=
\bm{p(\bm{x},t)}$ without taking into account of EQMP in
Eq.~(\ref{subeq:2}), then integrating over time.

By virtue of Eq.~(\ref{Firstc}), the chemical potential may be
expanded as \cite{KHuang57,Anderson66,Oliva89,Yu96}
\begin{equation}
\mu(\bm{x}, t) = \mu_{\text{loc}}(\bm{x}_{0}) + V_{\text{
ext}}-\frac{\partial}{\partial \rho}(\mu_{\text
{loc}})[\bm{\nabla}_{0}\cdot(\rho_{0}\bm{\xi})]. \label{Eqs1}
\end{equation}

In BEC, $\bm{p}(\bm{x}_{0})=0$, and the equation of motion gives
\begin{equation}
\mu_{\text{loc0}}(\rho_{0}(\bm{r}_{0}))+V_{\text{ext}}=constant
=\mu_{\text{loc0}}[\rho_{0}(0)]=\mu_{0}. %
\label{Eqs2}
\end{equation}

If we now set $V_{\text{ext}}=\omega_{0}^{2}r^{2}/2$, the density
profile describes a spherically symmetric, nonuniform
Bose-Einstein condensation (SSNBEC). This is the problem from
which our investigation started. As a special example of
collective excitations in BEC in trap, we limit our analysis to
the problem of SSNBEC.

We proceed with the understanding that the conventional
microscopic perturbation theory based on the quantum field theory
such as employed by Bogoliubov \cite{Bog47} is not applicable to a
finite space problem with the boundary conditions
\cite{note1,Bog59}. Eq.~(\ref{Ground}) exhibits peculiar features
which show that the density of the ground state is not uniform,
but it has a nodal surface. The surface layer of the SSNBEC is
described by the distribution of a degenerate Bose gas with
$\varepsilon > 0$ by the Bose statistics with the chemical
potential $\mu=0$. It is the most convincing experimental proof of
the validity of Eq.~(\ref{Ground}) and thus Eq.~(\ref{Eqs2}) - a
second-order phase transition
\cite{Lamb41,Kothari47,Singh62,Atkins65,Landau80}.

In order to describe the collective excitations in
Eq.~(\ref{Ground}), we linearize Eq.~(\ref{Motion0}) along with
Eq.~(\ref{Firstb}), Eq.~(\ref{Firstc}), Eq.~(\ref{Eqs1}), and
Eq.~(\ref{Eqs2}). It is straightforward algebra, though somewhat
tedious, to arrive at the equation of motion in terms of
$\bm{\xi}$. This algebra is also carried out in Ref.
\cite{SJHan2}. The result is
\begin{equation}
\frac{\partial^{2}}{\partial t^{2}}\bm
{\xi}=\frac{\mu_{0}}{M}\bm{\nabla}\sigma-\omega_{0}^{2}[(\bm{
\xi}\cdot\bm{\nabla})\bm{r}+(\bm {r}
\cdot\bm{\nabla}\bm{\xi})]\\
-\omega_{0}^{2}[\sigma\bm {r}
+\frac{1}{2}r^{2}\bm{\nabla}\sigma].%
\label{Main}
\end{equation}
Here we have taken $V_{ext}(r)=M\omega_{0}^{2}r^{2}/2$, $\partial
\mu_{loc}/\partial \rho =4\pi\hbar^{2}a/M$,  $\sigma={\bf
\nabla}\cdot{\bf\xi}$, and have also dropped the subscript in
$r_{0}$. The perturbation analysis leading to Eq.~(\ref{Main})
bears considerable resemblance to Feynman's atomic theory of the
$\lambda$ transition in helium \cite{Feynman53}, and
Eq.~(\ref{Main}) leads to the Bogoliubov dispersion relation for
a longitudinal sound wave \cite{Bog47} in He II as will be shown
below. However, it is not necessary to introduce his concept of a
back-flow to derive the dispersion relation for a sound wave,
because we insist on the phase-coherence in our analysis which is
more narrowly defined concept and is also valid in quantum
phenomena.

 Although SSNBEC would not be expected to correspond to a
realistic shape of condensation in experiments, one would hope
that insight gained would be helpful in more realistic
geometrical configurations in the prolate spheroidal coordinates.
Even the simplest problems in the spherical coordinates, however,
appear to have given rise confusion in recent literature
\cite{Stringary99,Leggett01,Fetter01}. Here we wish to suggest a
straightforward approach that automatically includes all the
modes of excitations in SSNBEC. The question of collective
excitations is put in the form of an initial-boundary value
problem in Eq.~(\ref{Main}) with the boundary conditions which
play a pivotal role in this analysis.

The solutions of Eq.~(\ref{Main}) yield the dispersion relations
for the collective excitations which fall into two groups: a
phonon,  a quantized longitudinal sound wave, the spectrum of
which is $\omega\simeq ck$ for a typical sound wave; a surface
wave under the external force. It should be noted that
Eq.~(\ref{Main}) is a typical second-order (inhomogeneous) partial
differential equation from which we obtain two particular
solutions, one corresponding to surface waves and the other to
longitudinal sound waves. The general solution to
Eq.~(\ref{Main}), then, is a combination of the two particular
solutions which can be obtained by a different set of boundary
conditions.

We now make use of the unique feature of the Lagrangian
displacement vectors with the appropriate boundary conditions:
$\bm{\nabla}\times\bm{\xi}=0$ which follows from
$\bm{\nabla}\times\bm{\nabla}S=0$ with
$\bm{\nabla}S(\bm{x}_{0},t)= M \bm{v}(\bm{x}_{0})$, the
irrotational motion of a superfluid, and the other boundary
condition on the free surface is due to the incompressibility of a
fluid $\bm{\nabla}\cdot{\bm {\xi}}=0$ which follows from the
equation of continuity Eq.~(\ref{subeq:1}) \cite{Lamb45}.

In order to understand the physical meaning of the above boundary
conditions, let us take a quick look at the similarity between the
phenomenological theory of the Meissner effect \cite{London35} and
the surface phenomena in He II, before the introduction of the
specific features in Nambu's gauge invariant calculation of
Meissner effect. Just as in a superconductor, the surface layer of
He II behaves like a normal fluid. This similarity between the
surface phenomena of a superfluid in a gravitational field and the
Meissner effect in a superconductor in a static magnetic field led
us to study a possible broken symmetry in He II. It is this
uniqueness of Eq.~(\ref{Main}) that will paly the essential role
in the explanation of the broken symmetry in He II; it requires
the two-sets of boundary conditions for its solutions in a way
that is analogous to the London equations [{\it i.e.,}
$\bm{\nabla}\cdot{\bm {J}}\equiv\bm{\nabla}\cdot{\bm {v}}=0$ and
$\bm{\nabla}\times\bm{v}=e\bm{h(x)}/Mc$ with $\bm{h(x)}$ as a
magnetic field] which was the first theoretical interpretation of
the Meissner effect \cite{London35,Nambu60,Anderson58}. We shall
bring this similarity further later in this paper along the line
of Nambu's analysis \cite{Landau41,London54,Nambu60}. It should be
emphasized that these boundary conditions are also consistent with
the extended two-fluid model of Landau \cite{Mott49}.

\subsection{\label{sec:level2} Surface Waves}

We now solve Eq.~(\ref{Main}) with the boundary conditions to
show that the symmetry is broken at the free surface. We begin
with the boundary conditions $\bm{\nabla}\cdot\bm{\xi}=0$ and
$\bm{\nabla}\times\bm{\xi}=0$ on the free surface \cite{Lamb45},
which simplifies the algebra considerably. The boundary
conditions suggest that we may solve Eq.~(\ref{Main}) as a
potential flow problem, since $\nabla^{2}\chi=0$ and $\bm
{\xi}_{\text{s}}=-\bm {\nabla}\chi$. Here the subscript $s$
stands for the surface waves.  There are a number of different
surface waves which show different characteristics driven by a
different force. We discuss only gravity and capillary waves in
SSNBEC.

\subsubsection{\label{sec:level3}  Gravity Waves}

We discuss first the gravity wave driven the external trapping
force only. The solution of Eq.~(\ref{Main}) as a potential flow
$\chi$ is then given by
\begin{equation}
\chi(r, t)=
\sum_{\ell,m}[\chi_{+}^{\ell}(t)r^{\ell}+\chi_{-}^{\ell}(t)r^{-(\ell+1)}]
 Y_{\ell,m}(\theta, \phi),
\end{equation}
where we may set $\chi_{-}^{\ell}(t)=0$ for a quantum liquid
droplet. We now expand $\bm{\xi}_{s}$ in terms of three
orthogonal vector spherical harmonics \cite{SJHan1},
\begin {equation}
\bm{\xi}_{s}=\sum_{\ell,m}[\xi_{1s}^{\ell,m}(r,t)\bm{
a}_{1}+\xi_{2s}^{\ell,m}(r,t)\bm{a}_{2}+\xi_{3s}^{\ell,m}(r,t)\bm{
a}_{3}]. \label{Surf1}
\end {equation}
Here we have defined the three vector spherical harmonics as
\[\bm {a}_{1}=\bm {e}_{r}Y_{\ell,m}(\theta,\phi),~ \bm{a}_{2}=r \bm
{\nabla}Y_{\ell,m}(\theta,\phi), ~\bm {a}_{3}=\bm {r}\times \bm
{\nabla}Y_{\ell,m}(\theta,\phi).\] It is now only a matter of
elementary algebra to obtain the dispersion relation by
substituting Eq.~(\ref{Surf1}) into Eq.~(\ref{Main}); the result
is
\begin {equation}
\omega^{2}_{surf}=\ell\,\omega_{0}^{2}. \label{Surf2}
\end{equation}

This dispersion relation gives the eigenfrequencies of gravity waves
under the external trapping force. The frequency for $l=0$ vanishes,
since it corresponds a uniform radial oscillation which is not allowed
for an incompressible fluid. The $l=1$ mode corresponds to a translational
motion of a fluid droplet without any deform of a shape.

Our entire argument critically depends on the interpretation of
the dispersion relation Eq.~(\ref{Surf2}). We see at once that the
dispersion relation Eq.~(\ref{Surf2}) is independent of the
internal dynamics of the imperfect Bose gas described by the
interaction term in the nonlinear Schr\"{o}dinger equation
Eq.~(\ref{Cat2}) that yields the quantum ground state
Eq.~(\ref{Ground}) which depends on $\hbar$.

And yet, Eq.~(\ref{Surf2}) implies that the surface wave is a
classical wave and thus a fluid that supports the surface wave (a
gravity wave) is a normal fluid; these results are consistent with
Mott's analysis of the two-fluid model, and with Lamb and
Nordsiek's observation \cite{Mott49,Lamb41}. More importantly, the
dispersion relation Eq.~(\ref{Surf2}) shows that \textit{the
two-fluid model of Landau \cite{Landau41} breaks down in the
surface layer}, since it implies a separation of the superfluid
and normal parts of a quantum fluid \cite{Osborne50,Meservey64}.

Before we proceed further, we ask why do we have a classical form
of dispersion relation in a quantum mechanical analysis? This
question has been studied previously by Anderson, {\it et al.},
\cite{Anderson65}. They have shown that it is precisely the nature
of a superfluid which cannot assume a stationary state,
Eq.~(\ref{Ground}), under an external field or pressure gradient,
but will have a time-dependent order parameter in a superfluid
({\it i.e.,} a presence of dissipation) \cite{Anderson65}. Hence
the surface layer must be a classical fluid to maintain a
stationary state under the trapping potential and it shows a
shell-like structure of BEC in a trap \cite{Lamb41,Mott49}.
Similarly we obtain the dispersion relation for a surface wave in
a cylindrically symmetric condensate as $\omega^{2}_{surf}=
m\omega_{0}^{2}$ where $m$ is a mode number \cite{SJHan3}.

\subsubsection{\label{sec:level2} Capillary Waves}

So far we have paid no attention on the surface tension on the BEC
droplet. In particular, in deriving Eq.~(\ref{Main}) we have taken
no account of processes of energy dissipation, which occur during
the course of initiation and propagation of sound waves in BEC as
a result of viscosity of surface layer and heat exchange between
different parts of it. Through these processes, however, we are
able to conserve the energy of an isolated system. Hence it is
this unique characteristic of the surface layer that explains the
essence of the broken symmetry - a breakdown of superfluidity.

In order to observe the surface free energy in an experiment, we
must study the capillary waves that depend on the surface tension
which is expected to be roughly in the order of ($\alpha \propto
0.35 \; erg/cm^{2}$) \cite{Allen38,Singh62,Atkins65} on the
surface of BEC. The surface tension in He$^{4}$ has been studied
extensively since the initial analysis of a surface phenomenon by
Mott \cite{Mott49,Allen38}. It will be discussed fully later in
the study of surface phenomena in He II.

For the derivation of dispersion relations for the capillary
waves, we follow Landau and Lifshitz \cite{Landau59} for the sake
of self-contained presentation. However, we shall present it in
the framework of our semi-classical approach based on the
Lagrangian displacement vectors. And yet we must explain why the
presence of the capillary wave is a natural consequence of the
surface tension driven by the sound waves and is a necessary
consequence of the conservation of energy in an isolated system.
It is further suggested that the basis of our justification is
provided by Kapitza's conjecture that the capillary waves are
driven by the instability of the fluid \cite{Kapitza48}.

To describe capillary waves on a spherical droplet driven by the
surface-tension, it is necessary to introduce the Laplace formula
that gives the equilibrium condition for the surface layer under
the action of the external pressure and the surface-tension
\cite{Sommerfeld50,Landau59}.

This formula for a spherical droplet is obtained from the
thermodynamic equilibrium condition by
\begin{equation}
\delta W = -\int(p - p_{0})\delta\zeta df + \alpha\delta f,
\label{work}
\end{equation}
where $W$ is the work necessary to bring about the volume change,
$\alpha$ is the surface-tension. For example, $\alpha$ between
liquid helium and its vapor is very small $\alpha=0.35\;
erg/cm^{2}$ at $0^{\circ}K$; $p_{0}$ the constant external
pressure to maintain the mechanical equilibrium, and $f$ is the
area of a surface.

Next the change of the surface area of separation is given by
\cite{Sommerfeld50},
\begin{equation}
\label{areachange} \delta
f=\int\delta\zeta(\frac{1}{R_{1}}+\frac{1}{R_{2}})df,
\end{equation}
where $R_{1}$ and $R_{2}$ are the principal radii of curvature at
a given point of a surface.

Substituting this expression in Eq.~(\ref{work}), we obtain the
Laplace formula,
\begin{equation}
p_{1}-p_{0} = \alpha (\frac{1}{R_{1}}-\frac{1}{R_{2}}).
\label{Laplace1}
\end{equation}

Since the Laplace formula was derived from the Euler equation, it
is convenient to apply the boundary conditions derived from the
velocity in potential flow of an incompressible fluid
$\bm{\nabla}\cdot\bm{v}=0$ and $\bm{\nabla}\times\bm{v}=0$
\cite{Lamb45}, which are equivalent to our boundary conditions
$\bm{\nabla}\cdot\bm{\xi}=0$ and $\bm{\nabla}\times\bm{\xi}=0$
provided that the wave amplitude is smaller than the wavelength
$a\ll \lambda$, so that we can safely neglect the convective term
$(\bm{v}\cdot\bm{\nabla})\bm{v}\ll
\partial\bm{v}/\partial t$ in the Euler equation. The velocity in
potential flow may be expressed as the gradient of scalar
function, $\bm{v}=\bm{\nabla}\phi$, the velocity potential. Thus
the boundary conditions imply that $\phi$ is a solution of the
Laplace equation $\nabla^{2}\phi=0$.

 We shall next show how the surface area is calculated in
differential geometry. Let us take up the cartesian coordinate
problem first. Then a typical surface equation is given by $z =
\zeta(x,y)$. And in differential geometry, $d\sigma=|\sec\gamma|
dxdy$, with $\cos\gamma = \bm{n}\cdot\bm{k}$ where $\bm{n}$ is
the normal vector to the surface, and $\bm{k}$ a unit vector in
the z-direction, it is now a simple exercise to derive the area
of a surface in cartesian coordinates \cite{Jeff53} as

\begin{equation}
 f =\int[1+(\frac{\partial \zeta}{\partial
x})^{2}+(\frac{\partial \zeta}{\partial y})^{2}]^{1/2}dxdy,
\label{cartesian}
\end{equation}
where $\zeta \ll 1$.

Having thus shown how the surface area in the cartesian
coordinates is calculated in differential geometry, let us next
calculate the surface area $\delta f$ for a spherical droplet in
spherical coordinates with $r=h(\theta,\varphi)$
\begin{equation}
f = \int_{0}^{2\pi}\int_{0}^{\pi}[1+\frac{1}{r^{2}}(\frac{\partial
h}{\partial
\theta})^{2}+\frac{1}{r^{2}\sin^{2}\theta}(\frac{\partial
h}{\partial \varphi})^{2}]^{1/2}r^{2} \sin\theta d\theta
d\varphi. \label{surface1}
\end{equation}

We now expand the integrand of Eq.~(\ref{surface1}) in terms of
$\zeta$ which is defined by $r=r_{0}+\zeta$, where $r_{0}=b$.
\begin{equation}
 f \approx \int_{0}^{2\pi}\int_{0}^{\pi} \{(r_{0}+ \zeta)^{2}
+\frac{1}{2}[(\frac{\partial \zeta}{\partial
\theta})^{2}+\frac{1}{\sin^{2}\theta}(\frac{\partial
\zeta}{\partial \varphi})^{2}]\}\sin\theta d\theta d\varphi.
\label{surface2}
\end{equation}

Here Eq.~(\ref{surface2}) is obtained by expanding the integrand
of Eq.~(\ref{surface1}) with $\zeta\ll 1$.

It is a simple algebra to calculate the variation of $f$ with
respect to $\zeta$,
\begin{equation}
\delta f =
\int_{0}^{2\pi}\int_{0}^{\pi}\{2(r_{0}+\zeta)\delta\zeta+\frac{\partial\zeta}
{\partial\theta}\frac{\partial\delta\zeta}{\partial\theta}+\frac{1}{\sin^{2}\theta}
\frac{\partial\zeta}{\partial\varphi}\frac{\partial\delta\zeta}{\partial
\varphi}\}\sin\theta d\theta d\varphi.
\end{equation}

Integrating the second term by parts with respect to $\theta$, the
third term with respect to $\varphi$, and using
Eq.~(\ref{areachange}), we find
\begin{equation}
\delta f =
\int_{0}^{2\pi}\int_{0}^{\pi}\{2(r_{0}+\zeta)-
[\frac{1}{\sin\theta}\frac{\partial}{\partial\theta}(\sin\theta
\frac{\partial\zeta}{\partial\theta})+\frac{1}{\sin^{2}\theta}\frac{\partial^{2}\zeta}
{\partial \varphi^{2}}]\}\delta\zeta\sin\theta d\theta d\varphi.
\end{equation}

If we divide the integrand by $(r_{0}+\zeta)^{2}\approx r_{0}(r_{0}+2\zeta)$
and  compare the result with Eq.~(\ref{areachange}), we find the following
formula correct to the first order in $\zeta$
\begin{equation}
\frac{1}{R_{1}}+\frac{1}{R_{2}}=\frac{2}{r_{0}}-\frac{2\zeta}{r_{0}^{2}}
-\frac{1}{r_{0}^{2}}[\frac{1}{\sin\theta}\frac{\partial}{\partial\theta}
(\sin \theta\frac{\partial\zeta}{\partial
\theta})+\frac{1}{\sin\theta^{2}}\frac{\partial^{2}\zeta}{\partial
\varphi ^{2}}].
\label{Laplace2}
\end{equation}

Since the condition $(\bm{v}\cdot\bm{\nabla})\bm{v}\ll \partial\bm{v}/\partial t$
is assumed in the Euler equation ({\it i.e.,} for a wave whose amplitude is much
smaller than the wavelength), we may then write the equilibrium condition
for the surface layer,
\begin{equation}
\rho_{0}\frac{\partial\phi}{\partial t}+\alpha\{\frac{2}{r_{0}}-\frac{2\zeta}{r_{0}^{2}}
-\frac{1}{r_{0}^{2}}[\frac{1}{\sin\theta}\frac{\partial}{\partial\theta}
(\sin \theta\frac{\partial\zeta}{\partial
\theta})+\frac{1}{\sin\theta^{2}}\frac{\partial^{2}\zeta}{\partial
\varphi ^{2}}]\} + p_{0}=0,
\label{equilibrium}
\end{equation}
where $\rho_{0}$ is the surface density $(gr/cm^{2})$ which is finite
in the presence of sound waves.

Since $v_{r}=\partial\zeta/\partial t=\partial\phi/\partial r$, we obtain
the equilibrium condition on $\phi$ by differentiating with respect to time
and omitting the constant terms
\begin{equation}
\label{Balance}
\rho_{0}\frac{\partial^{2}\phi}{\partial t^{2}}-\frac{\alpha}{r_{0}^{2}}
\{2\frac{\partial\phi}{\partial r}+\frac{\partial}{\partial r}
[\frac{1}{\sin\theta}\frac{\partial}{\partial\phi}
(\sin \theta\frac{\partial\phi}{\partial
\theta})+\frac{1}{\sin\theta^{2}}\frac{\partial^{2}\phi}{\partial
\varphi ^{2}}]\}=0,
\end{equation}
as an equilibrium condition for the surface layer at $r=r_{0}$.

Next we shall seek a solution to Eq.~(\ref{Balance}) in the form of
a stationary wave: $\phi=
e^{-i\omega t}\eta(r,\theta,\varphi)$, where $\eta$ satisfies the
Laplace equation $\nabla^{2}\eta=0$.

Since
\begin{subequations}
\label{Eigenvalues}
\begin{eqnarray}
\nabla^{2} & = &\frac{1}{r}\frac{\partial^{2}}{\partial r^{2}}(r)-
\frac{L^{2}}{r^{2}},\\
L^{2} & = &-[\frac{1}{\sin\theta}\frac{\partial}{\partial\phi}
(\sin \theta\frac{\partial\phi}{\partial
\theta})+\frac{1}{\sin\theta^{2}}\frac{\partial^{2}\phi}{\partial
\varphi ^{2}}],\\
L^{2}Y_{lm} & = &l(l+1)Y_{lm},
\end{eqnarray}
\end{subequations} and further assume the spatial variation in the
form $\phi=Ae^{i\omega t}r^{l}Y_{lm}(\theta,\vartheta)$, and thus
with Eq.~(\ref{Eigenvalues}), we readily obtain the dispersion
relation from Eq.~(\ref{Balance}) for the capillary waves

\begin{subequations}
\begin{eqnarray}
\rho_{0}\omega^{2}+\alpha l[2-l(l+1)]/r_{0}^{3}  & = & 0,
\end{eqnarray}
\begin{eqnarray}
 or \;\;\;\;\;    \omega^{2} & = & \alpha l(l-1)(l+2)/(\rho_{0} r_{0}^{3}).
\end{eqnarray}
\label{capillary1}
\end{subequations}

As in the gravity wave problem Eq.~(\ref{Surf2}), we see from
Eq.~(\ref{capillary1}) that $l=0$ is not allowed, since the mode
is a uniform radial oscillation, which is impossible for an
incompressible fluid; the $l=1$ mode is a translational motion of
a spherical drop ({\i.e.,} a displacement of the center). Since
$\omega^{2}\approx\alpha k_{\theta}^{3}/\rho_{0}$, where
$k_{\theta}\approx l/r_{0}$,  we may neglect the capillary waves
in the long wavelength limit $k_{theta}\ll a_{1}$, where
$a_{1}=[2\alpha/(g \rho_{0})]^{1/2}$ is a capillary constant
\cite{Landau59}. On the other hand, in the short wavelength limit,
the gravity wave can be neglected.

A question then arises naturally: how do we observe the capillary
wave alone in BEC in a trap, especially since the gravity waves
and the capillary waves can be observed simultaneously by the
combined action of capillarity and the external trapping force?
This is precisely the reason why it is difficult to observe the
capillary waves in a fluid with small viscosity
\cite{Kapitza48,Vinen00}. In the course of our work, we became
aware of the difficulty of studying sound wave propagation in BEC
in a trap, which is essential to the realization of BEC in an
atomic trap experiment. Our original motivation for studying the
capillary wave was to provide an indirect means of testing the
extended two-fluid model, since it may be considered as evidence
for the presence of the (first) sound wave dissipation process at
the surface layer. We shall next concentrate on (first) sound
waves in BEC in a trap.

\subsection{\label{sec:level2}  Compressional Waves - Sound Waves}

We now solve the equation Eq.~(\ref{Main}) with the boundary
conditions inside the nodal surface $\bm{\nabla}\cdot\bm{\xi}\neq
0$ but $\bm{\nabla}\times\bm{\xi}=0$, {\i.e.,} the fluid is now
compressible, but it remains irrotational,
$\bm{\omega}=\bm{\nabla}\times\bm{v_{s}}=\bm{\nabla}\times\bm{\nabla}S/M=0$
by the second assumption of Bohm's quantum theory \cite{Bohm52},
so that the superfluid supports the phase coherent sound waves in
BEC. One point we would like to emphasize in this paper is that
the boundary conditions in terms of the displacement vector
$\bf{\xi}$ are unique in that they can be used to define the
surface layer (normal fluid) and the bulk fluid (superfluid)
simultaneously in the discussion of collective excitations for a
system of Bose-particles described by the single Schr\"{o}dinger
equation (Gross-Pitaevskii) Eq.~(\ref{Cat2}).

It is also noteworthy that the above boundary conditions are very
similar to those of London Equations ({\it i.e.,}
$\nabla\cdot\bm{v}=0$ and $\bm{\nabla}\times{\bm p}=0$) for the
Meissner effect \cite{London35}, which explains why we shall see
the similarity between the Meissner effect and the peculiar
phenomenon of a superfluid in the gravitational field later in our
discussion of He II leading to the theory of spontaneously broken
gauge symmetry \cite{Nambu60,Nambu61,Goldstone61}.

As the mathematical detail may obscure the essentially simple
steps involved, it will be helpful to give a brief outline of the
mathematical analysis. First we define the compressibility of the
fluid as $\sigma=\bm{\nabla}\cdot{\bm {\xi}}$ from
Eq.~(\ref{Firstc}), and then its spectrum can be obtained from
Eq.~(\ref{Main}) by taking the divergence on both sides. After
straightforward algebra with the help of the vector identity
${\bm\nabla}\cdot[({\bm r}\cdot{\bm\nabla}){\bm\xi}]={\bm
r}\cdot{\bm \nabla}\sigma+\sigma$ and the condition of
superfluidity, ${\bm\nabla}\times{\bm\xi}=0$, we obtain
\cite{SJHan2}
\begin{equation}
\frac{\partial^{2}}{\partial t^{2}}
\sigma(r,t)=\frac{1}{2}\omega_{0}^{2}(\alpha^{2}-r^{2})\nabla^{2}\sigma-
3\omega_{0}^{2}r\frac{\partial}{\partial
r}\sigma-5\omega_{0}^{2}\sigma,
\end{equation}
where $\alpha^{2}=8\pi\hbar^{2}a \rho_{0}(0)/(M\omega_{0})^{2}$.

The solution of the equation is not entirely trivial. Writing
$\sigma(\bf r, t)$= $S(t)W(r)Y_{\ell,m}(\theta, \phi)$, we obtain
the variable separated equations:
\begin{equation}
\frac{d^{2}}{dt^{2}}S(t)+\lambda_{n} S(t)=0%
\label{Time}
\end{equation}
and
\begin{eqnarray}
(\alpha^{2}-r^{2})[\frac{1}{r}\frac{d^{2}}{dr^{2}}(rW_{n}(r))-
\frac{\ell(\ell+1)}{r^{2}}W_{n}(r)]\nonumber\\
-6r\frac{d}{dr}W_{n}(r)\nonumber\\
+(2\lambda_{n}/\omega_{0}^{2}-10)W_{n}(r)=0,%
\label{Space}
\end{eqnarray}
where $\lambda_{n}$ is a constant of separation. The eigenvalues
are determined by Eq.~(\ref{Space}), and by the boundary
conditions on $\sigma(r,t)$, {\it i.e.,} $\sigma(r,t)=0$ at the
free surface and the origin. Here we tacitly assume a small point
source at the origin to drive an outgoing spherical sound wave
\cite{Landau59}. The speed of first sound $c(r)=[4\pi a
\rho(r)\hbar^{2}]^{1/2}/M $ is the maximum at the origin and
approaches zero at the free surface. What's more, we recall that
the fluid of the surface layer, where the two-fluid model of
Landau \cite{Landau41,Tisza47,Mott49} breaks down, is no longer a
superfluid but is a normal fluid as shown above. Hence the sound
wave propagates to the surface layer and interacts with the
normal fluid there giving rise to a surface energy by dissipation
- a manifestation of the surface tension.

This is precisely the reason why we investigate the sound wave
propagation for the dynamical study of an imperfect Bose gas in
the trap. We see a self-consistent picture emerge from the study
of a sound wave propagation which conserves the energy of an
isolated system. It is indeed remarkable to see that Nature
provides such an elegant self-consistent picture through
Eq.~(\ref{Main}).

The solution of Eq.~(\ref{Time}), then, yields the dispersion
relation for the collective excitations with the eigenvalues that
are a function of the s-wave scattering length, the radial trap
frequency, the peak density at the center of the trap, {\it i.e.,}
the speed of first sound $c_{ctr}=[4\pi a
\rho(0)\hbar^{2}]^{1/2}/M$, the trapping frequency $\omega_{0}$,
and the wave number with $k_{\theta}\simeq \ell/r$. To show this
explicitly, we transform Eq.~(\ref{Space}) to the Sturm-Liouville
problem and obtain the eigenvalues in terms of a complete set of
functions with the orthogonal properties:
\begin{equation}
\acute{\lambda_{n}}=\int_{0}^{b}r^{2}(\alpha^{2}-r^{2})^{3}
[(\frac{d}{dr}W_{n})^{2}+
\frac{\ell(\ell+1)}{r^{2}}W_{n}^{2}(r)]dr %
\label{Eigen}
\end{equation}
 and with
\begin{equation}
\int_{0}^{b}r^{2}(\alpha^{2}-r^{2})^{2}W_{m}(r)W_{n}(r)dr
=\delta_{m,n}. \label{Norm}
\end{equation}
Here $\acute{\lambda_{n}}=(2\lambda_{n}/\omega_{0}^{2}-10)$ and
$b=\alpha=[8\pi
a\hbar^{2}\rho(0)]^{1/2}/(M\omega_{0})=\sqrt{2}c(0)/\omega_{0}$ is
the radius of the condensate. It is also well worth of pointing
out that the upper-limit of the integral in Eq.~(\ref{Eigen}) is a
function of speed of first sound $c_{ctr}=[4\pi a
\rho(0)\hbar^{2}]^{1/2}/M$  at the center.

It should be noted that the integral equation Eq.~(\ref{Eigen})
for $\sigma(\bf r, t)$ remains valid only for the collective
excitation energy in the phonon regime. Higher excitation energy
spectrum such as a roton excitation cannot studied by the
perturbation method described in this paper; there is an energy
gap in the roton excitation spectrum, separating the ground state
from the excited state for which the Green's function approach
based on the quantum field theory in a spatially homogeneous
system is better suited \cite{Beliaev58,Cohen57}. However, we will
discuss later a mechanism by which the rotons are created in the
fluctuation-dissipation process that leads to a broken symmetry.

Let us now analyze this integral in some detail since it shows the
nature of collective excitations. Equations
~(\ref{Eigen})-(\ref{Norm}) exhibit the essence of our results:
that the eigenvalues are, indeed, a function of the speed of
(first) sound, $b=[\sqrt{2}/\omega_{0}]\,c_{ctr}$, where the sound
speed $c_{ctr}= c(0)=[4\pi a \rho(0)\hbar^{2}]^{1/2}/M$, and the
trapping frequency $\omega_{0}$, for all values of the angular
momentum $\ell$. The functional relation of the speed of (first)
sound and the wave number in the excitation spectrum is also clear
in this integral representation. Perhaps more important is that
the mean field $\psi(r,t)$ in ODLRO can be perturbed in the
particle orbits by the semi-classical method
\cite{BFKK58,Low58,SJHan91} which yields a quantum mechanical
result by means of Bohm's quantum theory \cite{Bohm52}. It is also
worth while pointing out that the above semi-classical method is
actually a well-defined quantum mechanical perturbation method by
the phase-coherence since it involves atomic displacements as in
Feynman's qualitative picture of a phonon in his analysis on
two-fluid model \cite{Feynman53}.

Another important point is that the eigenvalue is a function of
the speed of first sound not at the free surface but at the center
of SSNBEC. This has a simple physical interpretation: an outgoing
spherical sound wave can travel with little energy loss toward the
free surface where the sound wave (a phonon) interacts with the
normal fluid of the surface layer and completely dissipates at the
surface, giving rise to a surface energy. It is precisely the
nature of the extended two-fluid model of Landau
\cite{Landau41,Tisza47,Mott49} that a phonon cannot interact with
a superfluid component as emphasized by Mott \cite{Mott49}, but it
will, however, interact with a normal fluid. There has been little
investigation as yet of sound wave propagation from the deep
inside of liquid helium to a surface area under the gravitational
field although we can infer the surface wave study by Vinen's
group \cite{Vinen00}, which would provide a simple demonstration
of the spontaneously broken symmetry in He II. The effect due to
the interaction between a phonon and the surface layer would be
very small since the phonons have a small amplitudes; nonetheless
we are only interested in the question of basic principle.

The analytical expression Eq.~(\ref{Eigen}) gives in principle a
complete solution to the eigenvalue problem. Unfortunately, the
eigenvalue equation Eq.~(\ref{Eigen}) for the phonon is not
analytically tractable and one must resort to numerical
integration. Here we take a different but equivalent approach to
the eigenvalue problem; that is, we shall solve Eq.~(\ref{Space})
as an eigenvalue equation in a differential equation. This simple
but powerful mathematical technique will permit us a detail study
of the eigenvalue problem. With the substitution
$W_{n}(r)=r^{\pm(\ell+1/2)-1/2}Z_{n}(r)$ together with
$x=r^{2}/\alpha^{2}$, we obtain
\begin{equation}
x(1-x)\frac{d^{2}}{dx^{2}}Z_{n}+[c-(a+b+1)x] \frac{d}{dx}Z_{n}
-abZ_{n}=0 \label{Last}
\end{equation}
This is just the Gauss differential equation \cite{WW63} with
$c=\pm(\ell+1/2)+1$, $a+b=\pm(\ell+1/2)+3$, and
$ab=(1/4)\,\acute{\lambda}_{n}-6\,[\pm(\ell+1/2)-1/2]$.

To obtain explicit solutions of Eq.~(\ref{Last}), one has to
resort to a numerical method. Fortunately a number of
simplifications can be made by studying analytic structure of the
equation. Eq.~(\ref{Last}) has regular singular points at $x=0$,
$x=1$ and $x=\infty$. Its solution is the hypergeometric
function, which is analytic in the complex plane with a cut from
$1$ to $\infty$ along the real axis \cite{WW63}. A simple, but
accurate, numerical method \cite{Hildebrand74} has been employed
to evaluate the eigenvalues in the domain $[0,1]$ in which the
solutions are analytic. Since we are interested in the low-lying
excited states (a phonon), it is only necessary to find the
smallest eigenvalues in the differential equation. This is
consistent with Feynman's picture of a phonon - a sound
(longitudinal) wave with a small amplitude \cite{Feynman53}.

Now returning to Eq.~(\ref{Time}) and taking $S(t)=e^{i\omega t}$,
we obtain the dispersion relation as
\begin{equation}
\omega_{ph}=\pm[\acute\lambda_{s}/2]^{1/2}\omega_{0},
\label{Sound}
\end{equation}
where $\acute\lambda_{s}$ is the smallest eigenvalues from
Eq.~(\ref{Last}).

The principal result of this section is  Eq.~(\ref{Sound}); it is
given in Fig.1, where the ratio $\omega_{ph}/\omega_{0}$ is
plotted against the angular momentum $\ell$. The pertinent
question, "How do we interpret the Fig. 1" ?, remains.
Eq.~(\ref{Sound}) is a statement of a dispersion law, an energy
spectrum ($\varepsilon =\hbar\omega$) relating to the wave number
$k_{\theta}$ of a longitudinal spherical wave with the speed of
the (first) sound $c_{ctr}=[4\pi a \rho(0)\hbar^{2}]^{1/2}/M$, in
the BEC in a trap. A close examination of the dispersion curve
shows that in the phonon regime the energy spectrum of a
longitudinal spherical wave is nearly linear with respect to
$\ell$ (\textit{i.e.,} the wave number $k_{\theta}\approx
\ell/r)$. \textit{"The existence of the phonon spectrum is a
consequence in part of the finite energy gap for particle
excitation and hence intimately associated with the statistics of
the particles, and that there is no single particle excitation"}
\cite{Bog47,Brueckner57,Feynman53}. A more detailed study of the
excitation spectrum Eq.~(\ref{Sound}) is given later by deriving
the Bogoliubov spectrum in the phonon regime $\omega= ck$
\cite{Bog47} in a homogeneous He II.

Collective excitations in SSNBEC may be defined as solutions for
which the dispersion curve Fig. 1 is a part of the complete
solution to Eq.~(\ref{Main}). In an inhomogeneous medium, the
Bogoliubov dispersion relation $\omega_{ph}=ck$ with $c(r)=[4\pi a
\rho(r)\hbar^{2}]^{1/2}/M$ must be studied approximately with
$k_{\theta}\simeq\ell/r$ at a given radius in SSNBEC. In
particular $\omega_{ph}/\omega_{0}= 1.5972$ for $\ell=0$ is a
unique value in a finite space problem, which corresponds to a
uniform radial perturbation. Finally, since the eigenvalues are
positive for all values of $\ell$, the Landau critical velocity
$v_{c}=\varepsilon_{ph}/p $ is also finite. Hence the imperfect
Bose gas in the trap is a superfluid, whereas a degenerate Bose
for which $c$ is independent of the scattering length is not,
because $c$ and $v_{c}$ vanish identically, which is consistent
with Singh's analysis \cite{Singh62}.

Here the behavior of the dispersion curve as a function of $\ell$
is of particular interest, since the azimuthal wave number
$k_{\theta}$ is given by $k_{\theta}\approx \ell/r$ in the
spherical geometry and the slope of the curve can represent the
average speed of a spherical longitudinal sound wave travelling
toward the free surface. Moreover, the dispersion relation, Fig.1,
also shows how the presence of the repulsive pair-interaction
determines the excitation spectrum of a phonon, which is nearly
linear with respect to $k_{\theta}$ as expected from the
Bogoliubov spectrum $\omega=ck $, where $c= [4\pi a
\rho(r)\hbar^{2}]^{1/2}/M$, since the slope is the average speed
of sound wave \cite{Bog47, Fetter71}. For a cylindrically
symmetric long condensate \cite{SJHan3}, we obtain a similar
dispersion curve in terms of mode number $m$ with the wave number
$k_{\theta}=m/r$.

To implement the Feynman picture of a phonon \cite{Feynman53} in
our microscopic study of collective excitations in BEC in a trap,
we have made some basic assumptions. One of which, as discussed
above, is the semi-classical perturbation theory by the
Lagrangian displacement vector $\bm{\xi}$ in a particle orbit in
Bohm's quantum theory \cite{Bohm52} along with ODLRO as an
alternative perturbation method, but not a standard quantum field
theoretical approach with an interaction Hamiltonian in terms of
creation and annihilation operators \cite{Bog47}. It is indeed
remarkable to see that the dispersion curve is the anticipated
form from the Bogoliubov dispersion relation $\omega=\;ck$ for a
phonon \cite{Bog47}.

Next we have defined the compressibility of a fluid as
$\sigma=\bm{\nabla}\cdot{\bm {\xi}}$ from Eq.~(\ref{Firstc}),
which was derived from the equation of continuity. Hence there
was no need to introduce the concept of the back-flow for a
propagation of the sound wave, since the conservation of the
number density and the phase-coherence automatically enforced.

It should be stressed, however, that {\it the above results
contradict the recent theoretical and experimental work in BEC in
a trap \cite{Stringary99,Leggett01,Fetter01}, but agree well with
the Bogoliubov's result with a proper geometrical correction
\cite{Bog47,Feynman53}}. The errors made in the previous work
\cite{Stringary99,Leggett01,Fetter01,Anderson07} are obvious,
because the dispersion relation must be a function of the speed of
(first) sound $c= [4\pi a \rho(r)\hbar^{2}]^{1/2}/M$ in the phonon
regime at low temperature $T \ll T_{\lambda}$. Hence much of the
previous work on the BEC in an atomic trap, {\it i.e.,} either
experimental or theoretical work,
\cite{Stringary99,Leggett01,Fetter01} is, to put it mildly,
questionable \cite{Anderson07}. Moreover, a finite space problem
in quantum fluids require a new mathematical approach as
emphasized in the Introduction \cite{note1,Bog59}. It has been
shown that a direct application of the perturbation method to the
phase-coherent collective excitations in He II and BEC through the
Lagrangian displacement vectors in particle orbits is
straightforward and free of ambiguity.

\subsection{\label{sec:level2} Spontaneously Broken Symmetries
in Atomic BEC in a Trap}

An alternative perturbation method is developed to deal with the
regime of low energy excitation in the BEC  where the usual
quantum field theoretical perturbation theory \cite{Bog47}. The
effectiveness of the above elaborate algebra with the Lagrangian
displacement vectors can be appreciated by the simplicity of the
dispersion relations Eq.~(\ref{Sound}) and Eq.~(\ref{Surf2}), the
first of which predicts a phonon spectrum and the second shows the
presence of surface waves. The existence of the surface layer is a
consequence in part of the finite energy gap for particle
excitation predicted by Mott \cite{Mott49,Landau80}. The ground
state wave function, $\psi(r,t)$, which is equivalent to the order
parameter $\eta$ of Ginzburg and Landau \cite{Ginzburg50},
undergoes the second-order phase transition, {\it i.e.,} from a
superfluid to a normal fluid at the free surface. A model that can
capture the essence of the broken symmetry is a broken $U(1)$
global symmetry that accompanies a phonon as a mass-less and
spin-less boson, similar to that of lattice gauge due to a
dislocation in the translational symmetry in a lattice. It is
therefore natural to identify the underlying basic mechanism for
the symmetry breaking as \textit{a spontaneously broken symmetry}
at the free surface which accompanies a phonon as Nambu-Goldstone
boson \cite{Nambu60,Goldstone61,BenLee73,Weinberg96,Anderson07}.
The difference between the macroscopic quantum phenomena (a
quantum fluid) and those of the classical phenomena (a classical
fluid) is so clear ({\it i.e.,} $\hbar$ difference) that it is
easy to see how the surface phenomena are intimately associated
with the statistics of particles \cite{Landau80}.

 Finally if we introduce a small disturbance adiabatically at the
center of SSNBEC to initiate an outgoing longitudinal spherical
sound wave, it will travel toward its surface and will be
dissipated at the (nodal) surface by the interaction with the
normal fluid there, giving rise a surface energy. Thus the
conservation of energy of an isolated Bose system is maintained.
It is therefore clear that the spontaneously broken symmetry is
both a necessary and a sufficient condition to hold the law of
conservation of energy. The broken symmetry also explains why it
has been exceedingly difficult to confirm the realization of the
BEC in a trap in an experiment, and it also shows a uniqueness of
shell-like structure of the BEC in equilibrium in a trap.

The sole purpose of the above discussion of the collective
excitations is an attempt to investigate the superfluidity of Bose
particles in a trap and the spontaneously broken symmetry from a
microscopic point of view.  By virtue of superfluidity
\cite{Bog47,Penrose56}, one expects the dispersion relation
Eq.~(\ref{Sound}) should be satisfied in the trapped gas as
evidence for the Bose condensation. On the other hand, the case of
broken symmetry can be shown by Eq.~(\ref{Surf2}) which is
independent of internal dynamics of the BEC. It should be
emphasized that, in a finite, inhomogeneous system, it is the only
way one can show explicitly the second-order phase transition,
{\it i.e.,} from a superfluid to a normal fluid - a spontaneously
broken symmetry. The discussion here is intentionally somewhat
sketchy, but it is sufficient to explain the main points. We also
note that, with a few minor changes in the formalism, the method
also yields a theory of collective excitations in a long
cylindrical condensation \cite{SJHan1,SJHan2,SJHan3}.

In passing we remark that a similar analysis can be performed for
a charge neutral Fermi gas in a trap to study the phase coherent
collective excitations \cite{Landau57} with strong repulsive
interactions \cite{Gottfried59,Abrikosov59}. There is, however, a
fundamental difference between a Bose gas and a neutral Fermi gas
because of Pauli's exclusion principle which limits the possible
inter-particle interactions in a neutral Fermi system at low
temperature. In 1957 Landau \cite{Landau57} predicted that at
sufficiently low temperature a new type of sound, which he calls
zero sound, propagates freely in He$^{3}$. Based on Landau's
idea, a more detailed theory of sound propagation has been worked
out by Abrikosov and Khalatinikov \cite{Abrikosov59}. At
sufficiently low temperature, it is predicted that the
attenuation of zero sound is proportional to $T^{2}$ and
independent of frequency. But in the phonon regime, the first
sound attenuates as $\omega^{2}/T^{2}$ corresponding to classical
viscous attenuation. Both of these temperature and frequency
dependence are observed in experiments with He$^{3}$. The data
indeed show that $(c_{0}-c_{1})/c_{1}\approx0.040$
\cite{Keen65,Wheatley66}.

A detailed study of phase coherent collective excitations and
attenuation of sounds would provide the direct experimental
confirmation of Landau's theory for an imperfect Fermi gas in a
trap. Theory, in the absence of direct measurements of the sound
(first and zero) propagations
\cite{Landau57,Gottfried59,Abrikosov59}, often could go off down
blind alley and it is not meaningful to speak of a Fermi liquid
with interactions in a trap as reported in a recent paper
\cite{Fermigas1}, since the observation on the sound propagations
is the essential requirement for the dynamical study of the
interacting many-body system. Only upon making careful
observations of (first and zero) sound propagations for the
collective modes do we understand the dynamics of the many-body
system \cite{Keen65,Wheatley66}, and perhaps, even provide the
qualitative understanding of an anisotropic BCS-type superfluid
state with triplet pairing. The experiments
\cite{Osheroff72,DMLee72,Wheatley73,Leggett73,Osheroff74}
demonstrated that the liquid He$^{3}$ which undergoes the
second-order phase transition below $2.7 mK$ and approximately
correct specific heat jump suggested by BCS exhibits the two
phases: A-phase as an $l=1$ Anderson-Morel phase \cite{AM61}, and
the B-phase as Balian-Werthamme state \cite{BW63}.

\section{\label{sec:level1} Collective Excitations in He II}

The above analysis may also be applied to He II with the short
range interaction with off-diagonal long range order. It should be
emphasized, however, that the basic equations
Eqs.~(\ref{Odlro})-(\ref{Cat2}) are far less restrictive than
those of the previous microscopic theory \cite{Bog47}. To a great
extent, the present study with the Lagrangian displacement vectors
in ODLRO \cite{Penrose56}, like the Bogoliubov theory of
superfluidity \cite{Bog47}, can serve as a model even when it is
not a perfect microscopic theory. Yet the present approach is a
satisfactory model for finite space problems that require the
apparatus of a new perturbation method with which one can
formulate a quantitative analysis of Feynman's theory of two-fluid
model \cite{Mott49, Landau41,Feynman53}.

One shortcoming in the present model is that the two-body
interaction is sufficiently small, so that the nonlinear
Schr\"{o}dinger equation (Gross-Pitaevskii) Eq.~(\ref{Cat2}) is
assumed to be applicable to He II. Hence one might raise an
objection against the argument of Bogoliubov \cite{Bog47} on the
grounds that his theory is limited to a weakly interacting Bose
gas and that the inclusion of a strong interatomic force is
essential to the treatment of collective excitations in He II. We
shall argue, however, following Feynman \cite{Feynman53} that
London's view on the BEC of an ideal Bose gas is essentially
correct and that the inclusion of a strong force between He atoms
will not alter the central features of Bose condensation and the
superfluidity of He II \cite{Feynman53} as the  strong interaction
forces in a liquid-like quantum fluid would not prevent these
particles from behaving like a free particles, a degenerate Bose
gas \cite{Landau80}.

In this section we are concerned with the problems in a uniform
superfluid in which we show the similarities between the Meissner
effect in a uniform superconductor and the surface phenomena in He
II in a gravitational field. In the case of He II which is
stationary under the gravitational field, some care is necessary
in separating surface waves from the longitudinal sound waves in
the bulk fluid. The entire argument is based on the the boundary
conditions in terms of the Lagrangian displacement vectors by
which we can construct the boundary conditions that are consistent
with the basic equations Eqs.~(\ref{Mean}). The physical boundary
conditions allow a separation of a domain in which a dynamical
calculation shows different properties of fluids; the surface
layer is a normal fluid (and solenoidal) and the bulk fluid is a
superfluid (and irrotational) \cite{Lamb41}.

The physical principles underlying the mathematical technique are
so clear that it is easy to see how the present results can be
applied to the explanation of spontaneously broken symmetry in He
II. As a simple model which retains the main feature of the
problem, we consider the free surface of a superfluid in a
gravitational field. The simplest correct procedure, then, is to
consider the gravity wave first and then extend the dispersion
relation for both the gravity and capillarity by using the Laplace
formula \cite{Landau59}.

\subsection{\label{sec:level2}  Surface Waves}

In the surface layer of He II, there are several different surface
waves of which a transverse wave plays an important role in
explaining the broken symmetry just as the transverse surface
current in Nambu's theory of gauge invariant explanation for the
Meissner effect in a superconductor.

\subsubsection{\label{sec:level3} Gravity Waves}

 We shall first consider gravity waves on a free surface of a
uniform superfluid in which the velocity of moving particles is so
small that we may neglect the convective term
$(\bm{v}\cdot\bm{\nabla})\bm{v}$ in comparison with $\partial
\bm{v}/\partial t$ in the Euler equation. As before we use the
boundary conditions $\bm{\nabla} \cdot\bm{v}=0$ and
$\bm{\nabla}\times\bm{v}=0$ which can be written
$\bm{v}=\bm{\nabla}\phi$ and $\nabla^{2}\phi=0$ for a potential
flow. Again our boundary conditions are equivalent to the original
form $\bm{\nabla}\cdot\bm{\xi}=0$ and
$\bm{\nabla}\times\bm{\xi}=0$ in terms of the displacement vector
\cite{Lamb45,Landau59}.

For an incompressible fluid under the gravitational field, we may write the
Euler equation
\begin{equation}
(\frac{\partial\bm{v}}{\partial t})
= -\bm{\nabla}(\frac{p}{\rho_{0}})+\bm{g},
\label{Motiong}
\end{equation}
where $\bm{g}$ is the gravitational acceleration.

Eq.~(\ref{Motiong}) can be rewritten in terms of a potential
\begin{equation}
\bm{\nabla}\frac{\partial\phi}{\partial t}= -\bm{\nabla}(\frac{p}{\rho_{0}}+gz),
\label{Motiong2}
\end{equation}
where we have taken the z-axis vertically upward from the $x-y$ plane of
the equilibrium surface of the fluid under the gravitational field.

Let $\eta$ be the vertical displacement of the free surface in its
oscillation; it is a function of x, y, and t on the surface.
Since $p_{0}$ is the atmospheric pressure on the free surface, we have
\begin{eqnarray}
p = p_{0} + g\rho_{0}(\eta - z),
\label{pressure}
\end{eqnarray}

Substituting $p$ in Eq.~(\ref{Motiong2}), we obtain the equilibrium
condition
\begin{equation}
g\eta + (\frac{\partial\phi}{\partial t})_{z=\eta}=f(t).
\label{Eta1}
\end{equation}

Since $\eta$ is small, we may write $v_{z}=\partial \eta/\partial t$ to the
same degree of accuracy as in Eq.~(\ref{Motiong}). But
$v_{z}=\partial\phi/\partial z$, so that
\begin{equation}
\label{Eta2}
(\partial\phi/\partial z)_{z=\eta}=\partial\eta/\partial t.
\end{equation}

Without losing a generality, we may take $f(t)=0$ in Eq.~(\ref{Eta1})
and differentiate it with respect to time to obtain the equation of
the free surface in oscillations along with Eq.~(\ref{Eta2}).

\begin{equation}
\label{freesurface}
(\frac{\partial\phi}{\partial z}+
\frac{1}{g}\frac{\partial^{2}\phi}{\partial t^{2}})_{z=\eta}=0.
\end{equation}

Since $\eta\ll 1$, we may take it as zero and solve the equation
\begin{subequations}
\label{gravitywaves}
\begin{eqnarray}
(\frac{\partial\phi}{\partial z}+
\frac{1}{g}\frac{\partial^{2}\phi}{\partial t^{2}})_{z=0}  & =  &0,\\
\;\; with\;\; the\;\; boundary\;\; condition \;\;\;    \nabla^{2}\phi=
\frac{\partial^{2}\phi}{\partial x^{2}}
+\frac{\partial^{2}\phi}{\partial z^{2}}   & =  & 0.
\end{eqnarray}
\end{subequations}

We notice that the above derivation of the equilibrium condition
Eq.~(\ref{freesurface}) is simpler and it does not lead to any contradiction.

Next we consider surface waves propagating along the x-axis and uniform
in the y-direction. We shall then look for a solution in the form
\begin{equation}
\phi=f(z)\cos(kx-\omega t),
\end{equation}
where $k=2\pi/\lambda$ is the wave number. After a brief algebra with the
boundary condition $\nabla^{2}\phi= \frac{\partial^{2}\phi}{\partial x^{2}}
+\frac{\partial^{2}\phi}{\partial z^{2}}=0$ , we obtain the dispersion
relation,
\begin{equation}
\omega^{2}=kg, \label{gravity}
\end{equation}
which shows the relation between the wave number and the frequency
of a gravity wave \cite{Sommerfeld50}. The gravity waves are often
studied in the context of hydrodynamic instability (the
Rayleigh-Taylor instability) in a fluid under the gravitational
field, and the implosion hydrodynamics in ICF program
\cite{Taylor50,SJHan82}. Just like a spherical droplet, the
surface waves are independent of $\hbar$ and are classical waves
in He II.

\subsubsection{\label{sec:level3}  Capillary Waves}

In this section we shall consider the free surface of He II that
undergoes an infinitesimal displacement under the combined action
of a gravitational force and the surface tension. A study of the
surface phenomena in He II is important in demonstrating the
broken symmetry in quantum fluids, especially since the capillary
waves have been observed recently in liquid helium by Vinen's
group \cite{Vinen00}. In this connection, it should be pointed out
that the surface tension of liquid He$^{4}$ has been studied
extensively by Atkins and others \cite{Allen38,Atkins53,Atkins65}
and will be reviewed in the following.

The surface tension of a fluid is a measure of the free energy per
unit area in He II which is in stationary state under the
gravitational field \cite{note3}. The calculation of the surface
tension based on a degenerate Bose gas model by Singh yields
\begin{subequations}
\label{Singh}
\begin{eqnarray}
\label{Singh1}
\sigma &= &\sigma_{0} -\pi
mk^{2}T^{2}\zeta(2)/2h^{2}
 = \sigma_{0}-0.0075T^{2},\\
\sigma_{AN} & = & 0.3745-0.0096T^{2},
\label{Singh2}
\end{eqnarray}
\end{subequations}
where Eq.~(\ref{Singh1}) is a modified to Eq.~(\ref{Singh2}) to
fit to their data by Atkins and Narahara \cite{Atkins65},
$\sigma_{0}= 0.352$ erg/cm$^{2}$ is the surface tension at $0^{0}$
K \cite{Singh62}, and $\zeta(2)=\pi^{2}/6$. In an effort to obtain
further information on $\sigma_{0}$, they have made extensive
measurements down to 0.35$^{0}$K.

And yet Atkins and Narahara \cite{Atkins65} have also provided a
different theoretical interpretation with the assumption similar
to that of the Debye theory of the specific heat of solids based
on the lattice vibrations \cite{Debye12,Atkins53,Atkins65}, which
yields the surface tension as
\begin{equation}
\label{Atkins} \sigma=\sigma_{0}-
1.55(\rho/\sigma_{0})^{2/3}2\pi\hbar(k_{B}/2\pi\hbar)^{7/3}T^{7/3},
\end{equation}
where $\rho$ is the density of liquid with the cutoff frequency
$\nu_{c}=1.5\times10^{11} sec^{-1}$ and characteristic temperature
$\theta_{c}=2\pi\hbar/k_{B}\approx 7^{0}$ K.

However, the following empirical curve of $\sigma$ obtained by the
least squares fit of the equation $\sigma=\sigma_{0}+AT^{n}$,
\begin{equation}
\label{Least}
\sigma = 0.3729-0.0081 T^{2.5},
\end{equation}
shows a better agreement with the data \cite{Atkins65}.

Although Eq.~(\ref{Singh}) and Eq.~(\ref{Atkins}) give similar
curves (see Fig. 7 and Fig. 5 of Ref. \cite{Atkins65}),  it is
rather remarkable to observe that Singh's analysis $\sigma_{AN}$
\cite{Singh62} shows the best agreement with the experimental data
obtained by Atkins and Narahara \cite{Allen38,Atkins65}. The most
compelling argument for this unexpected agreement here is that
made above in the case of the BEC in the trap; the surface layer
is a normal fluid, and that there is no interaction between the
normal fluid in the surface layer and the superfluid beneath the
layer. Thus the surface layer is composed of the atoms in the
excited states as emphasized by Mott \cite{Mott49}, which is
consistent with Singh's calculation of the surface tension
\cite{Singh62,note4}.

Since we have already discussed the Laplace formula in the study
of capillary waves in a spherical droplet of BEC, we begin with
Eq.~(\ref{Laplace1}) in which $p$ is the pressure inside the fluid
in the surface layer, and $p_{0}$ is a constant atmospheric
pressure on the free surface. As before, we assume a potential
flow ({\it i.e.,} $\bm{v}=\bm{\nabla}\phi$) of an incompressible
fluid for the surface layer and take $p$ from
Eq.~(\ref{pressure}),
\begin{equation}
p=-\rho_{0}g\zeta -\rho_{0}\partial\phi/\partial t,
\end{equation}
where $\zeta$ is the displacement of the free surface in the
z-direction and $\rho_{0}$ is the surface density of an incompressible fluid
in the surface layer.

Taking $p_{0}=0$, we can write down the Laplace equation
\begin{equation}
\rho_{0}g\zeta + \rho_{0}\frac{\partial \phi}{\partial t}
-\alpha(\frac{\partial^{2}\zeta}{\partial x^{2}} +
\frac{\partial^{2}\zeta}{\partial y^{2}})=0.
\end{equation}

Differentiating the equation with respect to $t$ and replacing
$\partial \zeta/\partial t$ with $\partial \phi/\partial z$ since
$v_{z}= \partial \zeta/\partial t=\partial \phi/\partial z$, we obtain
the equilibrium condition on $\phi$ on the free surface $z=0$,

\begin{equation}
g \frac{\partial \phi}{\partial z} +\frac{\partial^{2} \phi}{\partial t^{2}}
-\frac{\alpha}{\rho_{0}}\frac{\partial}{\partial z}
(\frac{\partial^{2}\phi}{\partial x^{2}} +
\frac{\partial^{2}\phi}{\partial y^{2}})=0.
\label{Equi2}
\end{equation}

As in the gravitational wave, we consider a plane wave propagating
in the direction of the x-axis and look for a solution in the form
$\phi=Ae^{kz}\cos(kx-\omega t)$, which yields the dispersion
relation as
\begin{equation}
\omega^{2}=gk+\alpha k^{3}/\rho_{0}.
\label{capillary2}
\end{equation}

It should be stressed that, in the long wavelength limit $k \ll 1$, the
gravity wave dominates over the capillary wave. A peculiar wave motion
of the free surface due to surface tension in He II has been observed
recently by Vinen's group \cite{Vinen00}. Kapitza has argued that these
waves were driven by the instability of the fluid at comparatively
small Reynolds numbers \cite{Kapitza48}. This conjecture can be shown
to be true by studying the propagation of sound waves in He II.

We also note that the dispersion relation Eq.~(\ref{capillary2})
clearly suggests that the fluid on the nodal surface behaves like
a normal fluid, because Eq.~(\ref{capillary2}) is independent of
$\hbar$ as in Eq.~(\ref{Surf2}) for the BEC in a trap. Moreover,
the observation of the capillary wave implies that the surface
energy is another manifestation of a broken symmetry in the
observable domain, where $\alpha\cong
\sigma=0.3352-7.5\times10^{-3} T^{2} erg/cm^{2}$ \cite{Singh62}.
Since the two-fluid model does not permit a physical separation of
the two-fluids (the superfluid and normal parts of the fluid)
\cite{Landau41}, the two-fluid model breaks down at the surface
layer. It also demonstrates that the law of conservation of energy
in the system is maintained, since the sound wave initiated deep
inside is dissipated at the surface layer of He II as in the BEC
in a trap.

The wave lengths of these capillary waves are actually extremely
short, far too short to be observed in any ordinary experiment. It
is indeed a challenging experiment \cite{Vinen00} in which the
presence of capillary waves on the surface of a superfluid layer
has been observed with a strong external electric field that
mimics the gravitational field. It should be pointed out that this
experimental observation supports the theoretical explanation of
the spontaneously broken symmetry at the nodal surface of a
superfluid as discussed below.

\subsubsection{\label{sec:level3} Transverse Waves}

We have already discussed the surface waves in the surface layer
in both BEC and He II. And yet there exists still another type of
oscillations inside the surface layer; it is similar to a gravity
wave, but is driven by a entirely different mechanism, thermal
expansion of an incompressible fluid, since the free surface
cannot be a mathematical surface, but it will be a surface layer
with finite thickness \cite{Meservey64}. This new collective
excitation is a transverse wave, defined by $\bm{\nabla}\cdot
\bm{\xi}=0$ (solenoidal, that is,  free of sources and sinks). And
therefore the surface waves are essentially transverse in nature,
on the other hand the longitudinal excitation is defined by
$\bm{\nabla}\times\bm{\xi}=0$, but not solenoidal, {\it i.e.,}
$\bm{\nabla}\cdot \bm{\xi}\neq 0$.  It is this unique
characteristic of the displacement vectors that allow separation
the surface phenomena from the bulk phenomena in a fluid.

Since a transverse excitation involves breaking up the phase
coherence of collective excitations in the surface layer and plays
an important role as shown by Nambu in his study of a Meissner
effect in a superconductor \cite{Nambu60}, we show here how it
comes about in the surface layer as in a superconductor
\cite{London35,Landau59}.

Since transverse excitation in the surface layer is driven by thermal
expansion, we need a new set of hydrodynamic equations. First, we write
down the equation of conservation of the entropy:
\begin{equation}
\partial s'/\partial t + \bm{v}\cdot\bm{\nabla} s_{0}=0,
\label{entropy3}
\end{equation}
where $s'$ is defined by $s=s_{0}+s'$, and $s_{0}$ is an
equilibrium value and is a function of the vertical coordinate
$z$.

Next neglecting the convective term $\bm{v}\cdot\bm{\nabla}\bm{v}$
and defining the change in density by $\rho=
\rho_{0}+\rho'$, we write Euler's equation for a fluid element in the
surface layer under a gravitational filed,
\begin{equation}
\frac{\partial\bm{v}}{\partial t}=-\frac{\bm{\nabla} p}{\rho} +
\bm{g}
=-\frac{\bm{\nabla}p'}{\rho_{0}}+\frac{\bm{\nabla}p_{0}}{\rho_{0}^{2}}\;\rho',
\label{motion3}
\end{equation}
where the expansion of the pressure $p=p_{0}+p'$ and the
equilibrium pressure in the gravitational field
$\bm{\nabla}p_{0}=\rho_{0}\bm{g}$ are used in the derivation.

Since the variation of density is due only to the change in
entropy, and not due to the pressure change, we may write
\begin{equation}
\rho'=(\frac{\partial\rho_{0}}{\partial s_{0}})_{p}\;s',
\label{density3}
\end{equation}
then we obtain the Euler equation in the form
\begin{equation}
\frac{\partial\bm{v}}{\partial t}=\frac{\bm{g}}{\rho_{0}}
(\frac{\partial \rho_{0}}{\partial s_{0}})_{p}\;s'-\bm{\nabla}
\frac{p'}{\rho_{0}}.
\end{equation}

In the limit $(\bm{v}\cdot)\bm{v} \ll \partial \bm{v}/\partial t$,
we may neglect the change in the equilibrium density over
distances of the order of a wavelength. Hence we solve the
equations (\ref{entropy3}), (\ref{motion3}), and (\ref{density3})
with the boundary condition $\bm{\nabla}\cdot\bm{v}=0$ which is
equivalent to $\bm{\nabla}\cdot\bm{\xi}=0$ in Eq.~(\ref{Firstc})
as discussed earlier. We shall look for a solution in the form of
a plane wave $\bm{v}= A\exp[i(\bm{k}\cdot\bm{r}-\omega t)]$. Then
the boundary conditions give
\begin{equation}
\bm{v}\cdot\bm{k}=0,
\label{transverse}
\end{equation}
which implies that the fluid motion is perpendicular to the wave
vector, {\it i.e.,} a transverse wave. Eq.~(\ref{transverse}) is
equivalent to London's  assumption that the super-electrons are
assumed to be an incompressible charged fluid
$\bm{\nabla}\cdot\bm{j(x,t)}=\bm{\nabla}\cdot \bm{v}=0$
\cite{London35}.

As a detailed derivation is available in the literature
\cite{Landau59}, we skip the algebra and give the dispersion
relation for the transverse wave in the surface layer:
\begin{equation}
\omega^{2}=-\frac{1}{\rho}(\frac{\partial \rho}{\partial
s})_{p}\;\; g\frac{ds}{dz}\sin^{2}\theta, \label{transwave1}
\end{equation}
where $\theta$ is the angle between the z-axis and the wave
vector; it is defined by $\bm{g}\cdot\bm{k}=gk\cos\theta$. Here we
have also dropped the subscript zero to the equilibrium
quantities.
 We now use the thermodynamic identities \cite{Landau80}
\begin{subequations}
\begin{eqnarray}
\frac{ds}{dz}=\left(\frac{\partial s}{\partial
p}\right)_{T}\frac{dp}{dz} =-\rho g \left(\frac{\partial
s}{\partial p}\right)_{T},
\end{eqnarray}
\begin{eqnarray}
 \left(\frac{\partial s}{\partial p}\right)_{T} =
\frac{1}{\rho^{2}}\left(\frac{\partial\rho}{\partial
T}\right)_{P},
\end{eqnarray}
\begin{eqnarray}
and \;\;\;  \left(\frac{\partial \rho}{\partial s}\right) =
\frac{T}{c_{p}}\left(\frac{\partial \rho}{\partial T}\right)_{p}
\end{eqnarray}
\end{subequations}

to rewrite the dispersion relation Eq.~(\ref{transwave1}) in an
experimentally accessible form as

\begin{equation}
\omega =[T/c_{p}]^{1/2}\;\;\frac{g}{\rho} \left(\frac{\partial
\rho}{\partial T}\right)_{p}\sin\theta =[T/c_{p}]^{1/2}\;\;
\frac{g}{\rho}\left(\frac{\partial \rho}{\partial
T}\right)_{p}\sin\theta, \label{transwave2}
\end{equation}
where we have taken $z$-axis upward from the $x-y$ plane and thus
$\theta=\pi/2$ is the condition for a transverse wave in the
$x-y$ plane and $c_{p}$ is the specific heat per unit mass. For
$\theta=0$, the transverse wave cannot exist in the surface
layer. The frequency $\omega$ depends only on the direction of
the wave vector.

 The presence of the transverse wave demonstrates that the
positive surface energy associated with a normal and superfluid
interface is indeed a correct remedy for the serious flaw of
London's original theory of the Meissner effect \cite{London35}.
In fact, this point of view turns out to be more fundamental than
that of Ginzburg-Landau theory as far as the Meissner effect is
concerned, although it gives the surface energy in terms of the
coherence length, but it cannot address the nature of a transverse
wave on the surface layer which is essential to the gauge
invariant explanation of the Meissner effect in the microscopic
theory of BCS theory \cite{Ginzburg50,Nambu60}.

 Moreover, the surface energy is necessary to maintain the law of
conservation of energy associated with the dissipation of sound
waves in the surface layer of He II. This description of the
surface energy is also consistent with the positive surface
energy associated with the normal state joined onto the
super-conducting state in the second-order phase transition (the
Meissner state) of the Ginzburg-Landau's theory of a
superconductor \cite{Ginzburg50,Nambu60}.

The dispersion relation Eq.~(\ref{transwave2}) is different from
that of the (second) sound for which the velocity of sound depends
markedly on temperature and vanishes at the $\lambda$ - point
\cite{Landau59}; the second sound is of course a longitudinal
wave. The transverse wave in the surface layer is another
manifestation of broken symmetry and breaks up the phase-coherence
in the collective excitations just as the transverse current on
the surface of a superconductor in the Meissner effect
\cite{Nambu61}. This difference is the basis for the broken
symmetry on the free surface of a superfluid - a break-down of the
two-fluid model. The dispersion relation Eq.~(\ref{transwave2})
clearly shows that it is driven by thermal expansion of the
density in the layer. It should be emphasized that all the surface
waves in the surface layer are transverse in nature
\cite{Sommerfeld50}.

\subsection{\label{sec:level2}  Compressional waves in He II}

Let us now return to the main question on the superfluidity posed
in the Introduction, namely the Bogoliubov spectrum for a phonon
\cite{Bog47} below the $\lambda$-point in He$^{4}$. Inside the
free surface, the fluid is compressible
$\bm{\nabla}\cdot\bm{\xi}\neq 0$, but it remains irrotational
$\bm{\nabla}\times\bm{\xi}=0$ as it is a superfluid. To obtain
the phonon excitation spectrum, first note that the vortices are
in an isolated region, so that we may assume a uniform superfluid
flow $\bm{v}(\bm{x}_{0})$, and take the first-order terms
$\rho_{1}$ and $S_{1}$ varying as
$C\exp[i(\bm{k}\cdot\bm{x}-\omega t)]$.

It is then straightforward algebra to obtain the first-order
linearized equations of motion  from Eq.~(\ref{subeq:2}) and
Eq.~(\ref{First}),
\begin{subequations}
\label{whole}
\begin{eqnarray}
-i\omega\rho_{1}-\frac{\rho_{0}}{M}k^{2}S_{1}=0 \\
\label{lasta} -i\omega S_{1}+\frac{4\pi \hbar^{2} a}{M}\rho_{1}
+\frac{\hbar^{2}}{4M}\frac{1}{\rho_{0}}k^{2}\rho_{1}=0,
\label{lastb}
\end{eqnarray}
\end{subequations}
where we have neglected the external potential $V_{ext}$ in
Eq.~(\ref{subeq:2}), since the gravitational force is too small
compared to the short range force between the helium atoms.

We at once obtain the dispersion relation for a longitudinal
sound wave. The result is:
\begin{equation}
\omega^{2}=\frac{4\pi a\rho_{0}\hbar^{2}k^{2}}{M^{2}}+
\frac{\hbar^{2}k^{4}}{4M^{2}}. \label{Phonon}
\end{equation}

We notice that the difference between the phonon spectrum
Eq.~(\ref{Phonon}) and the spectrum of the gravity-capillary wave
Eq.~(\ref{capillary2}) in the normal fluid is striking
\cite{Fetter71}. It is also surprising, but pleasant to see that
we have obtained the exactly same dispersion relation
Eq.~(\ref{Phonon}) as Bogoliubov's microscopic results, for which,
of course, calculations were performed with his well-known
interaction Hamiltonian for an imperfect Bose gas in terms of the
creation and annihilation operators in quantum field theory
\cite{Bog47}.

In the phonon regime $k\rightarrow 0$, $\omega=c k$, where $c=
[4\pi a \rho(r)\hbar^{2}]^{1/2}/M$. The energy spectrum is
characteristic of a sound wave with the speed of first sound
$c=[4\pi a \rho(r)\hbar^{2}]^{1/2}/M$ in an imperfect Bose gas.
One may conclude, then, immediately from the above analysis that
our approach with the mean field by ODLRO \cite{Penrose56} and
Bohm's quantum theory \cite{Bohm52} is far more powerful and
simpler in the study of superfluid dynamics than that of the
quantum field theory.

Both the transverse excitations Eq.~(\ref{transwave2}) and the
longitudinal excitations Eq.~(\ref{Phonon}) are essential to the
gauge invariant explanation of the Meissner effect
\cite{Nambu60}; the transverse excitation is viewed as a
necessary consequence of the longitudinal excitation in a
superconductor.

On the other hand the longitudinal excitations, such as those
generated by the density compression Eq.~(\ref{Firstc}), do not
break up phase-coherence in a superfluid He II, that is, the phase
coherence $\bm{\xi}\cdot\bm{\nabla}S(\bm{x}_{0},t)
=\sum_{i}\bm{\xi}_{i}\cdot\bm{\bm{\nabla}_{i}}S_{0,i}(\bm{x}_{0,i},t)$
is preserved, where the ground state wave function is defined in
ODLRD as $\psi(r,t)=f(\bm{r},t)exp[\frac{i}{\hbar}S(r,t)]$ with
the action $S(r,t)$ and $\bm{\xi}$ the Lagrangian displacement
vector. Unlike a superconductor, the transverse excitations do not
take place in a superfluid but they do occur only in the surface
layer which is a normal fluid. This difference is the basis for a
spontaneously broken symmetry in a Bose quantum fluid, He II.

\subsection{\label{sec:level2} Spontaneously Broken Symmetries in He II}

The surface layer is the boundary in which the waves show a marked
change in the property of the fluid, indicating a second-order
phase transition from a superfluid to a normal fluid. The entire
argument is based on this characteristic difference between the
two dispersion relations: Eqs.~(\ref{capillary2}) and
(\ref{Phonon}). It shows clearly that the fluid in the surface
layer behaves like a normal fluid, whereas the compressible fluid
inside the free surface behaves like a superfluid supporting the
sound wave with its excitation energy $\omega=c k$, where $c=
[4\pi a \rho(r)\hbar^{2}]^{1/2}/M$ in the phonon regime. Moreover,
the free surface cannot be a mathematical surface, but it will be
a surface layer with finite thickness {\it} e.g., $\lambda \approx
5.0\; 10^{-3}cm$ \cite{Meservey64}, in which the transverse wave
Eq.~(\ref{transwave2}) is a direct reflection of the break-down of
the two-fluid model.

It is therefore natural to identify the underlying basic mechanism
for the surface phenomena as \textit{a spontaneously broken
symmetry} at the free surface in a Bose system as in Nambu's
argument on the spontaneously broken gauge symmetry in a
superconductor \cite{Nambu61}. Here we have extended the two-fluid
model of Landau \cite{Landau41} in such a manner that, in addition
to phonons and rotons, the normal fluid now includes the
degenerate, excited Bose particles in the Bose system from the
ground state of the condensation \cite{Mott49,Singh62}. The phonon
may be interpreted as the mass-less, spin-less boson,
Nambu-Goldstone boson, as $k\rightarrow 0$ in the broken symmetry
\cite{Nambu60,Anderson58,Nambu61,Goldstone61,Levy60,Goldstone62,BenLee73}.

The picture we have presented above also explains how the
superfluidity in a vortex core breaks down and why the classical
form of the Magnus force in the study of a vortex quantization is
indeed correct one \cite{Vinen61,Reif64}. This result may also
solve one of the most important and long-standing puzzle in low
temperature physics, the parabolic surface of a rotating He II
which has troubled the intuition of many experimental physicists
over the half century \cite{Osborne50,Meservey64,Andro66}.

The surface energy increment due to the interactions with a phonon
would be extremely small, however, far too small to be observed in
any ordinary experiment except for the indirect observation of a
capillary wave. And yet, the above quantitative analysis has been
in agreement with its most consistent and systematic analysis of
the surface phenomena by Mott \cite{Mott49,Singh62} and Meservey's
precise observation of the curvature of a rotating He II
\cite{Meservey64}.

In spite of a complicated mathematical analysis, it is indeed
remarkable that, from just a simple set of boundary conditions in
terms of the Lagrangian displacement vectors, one can establish
the similarity between the Meissner effect in a superconductor in
an external static magnetic field and the surface phenomena of a
superfluid in a gravitational field, which leads to the
spontaneously broken gauge symmetry in a Bose system. What's more,
the broken symmetry provides an explanation not only of how the
superfluidity is broken at the core of a vortex, but of why a new
BEC in a trap has {\it a shell-like structure} to hold the law of
conservation of energy \cite{SJHan2}.

\section{\label{sec:level1} Broken Symmetries in a Bose
System}

What has been described in the previous sections is the first
step based on the detailed analysis of dynamical equations toward
obtaining a more general theory which might offer hopes of
understanding the broken symmetries that would be of fundamental
importance in a Bose system as the Meissner effect in a
superconductor.

Although the broken symmetry has been demonstrated by the
dynamical calculations alone as shown above, we pause here,
however, to comment on the broken symmetry by Nambu
\cite{Nambu60,Nambu61}, partly because of the inherent elegance
of his method, and partly because incidental points of interest
which has led Goldstone \cite{Goldstone61} to discover the
general theorem of spontaneously broken gauge symmetry.

It is a mathematical theorem that explains the observations over a
wide range of many-body problems. The beauty of the theorem is its
simplicity: the symmetry is said to be broken spontaneously when
the ground state (the vacuum in the field theory) does not share
the same symmetry group with the Lagrangian that describe the
dynamics of the system, and that, in the absence of a long-range
order \cite{Anderson63}, accompanies a mass-less boson (a phonon).
Hence we gain a new perspective on the broken symmetry in a Bose
system.

To see whether the theorem which is the outcome of the study of
the Meissner effect in a superconductor (a Fermion system), can be
applied to problems in an imperfect Bose system, a careful
analysis must be made for the consequences of the theorem. With
these preliminary remarks, let us turn to the study of broken
symmetry in a Bose system and ask the questions: what is then the
spontaneously broken symmetry that accompanies a longitudinal
phonon in both He II and BEC? and why?

\section{\label{sec:level1} Surface Phenomena in He II and
Meissner Effect in Superconductors}

So far the analysis has not deviated significantly from that of
phase-coherent collective excitations.  At this point, however, we
shall have to digress to show how the Goldstone theorem
\cite{Goldstone61} came into being in particle physics. In the
early 60s, the question of gauge invariance in the study of the
Meissner effect has been addressed by a number of authors
\cite{Rick59,Anderson58,Nambu60}, a proof of gauge invariance in
the derivation of the Meissner effect in BCS theory lies at the
core of the problem of superconductivity, since the Hamiltonian
used in the BCS model is not gauge invariant.

\subsection{\label{sec:level2} Nambu's Gauge Invariant Theory of
Meissner Effect and the Goldstone Theorem}

In a series of papers, Nambu \cite{Nambu60,Nambu61} has laid out
an elegantly simple picture of gauge invariant explanation for the
Meissner effect. In the BCS-Bogoliubov theory of superconductivity
\cite{Nambu61}, the energy gap parameter $\phi$ is obtained by a
generalized Hartree-Fock calculation of the electron-hole
interactions $\phi\approx\omega_{D}\, exp[-1/\varpi]$, where
$\omega_{D}$ is the Debye frequency around the Fermi surface and
$\varpi=N(0)g \approx 0.2-0.3$ is the average interaction energy
of an electron with unit energy shell of electrons on the Fermi
surface, which depends on the interaction strength
\cite{Daunt46,Glover57,Bog58,note2}. Based on the same mechanism
for the appearance of energy gap as in the theory of
superconductivity in which elementary excitations can be described
by a coherent mixture of electrons and holes, real nucleons are
regarded as quasi-particle excitations in the mathematical
formulation of a dynamical model of binding nucleon-pair into a
pion, which is, in essence, a compound model of an idealized pion
as a pseudoscalar zero-mass bound state of nucleon-antinucleon
pair \cite{Nambu61}.

 In the explanation of gauge invariance of the Meissner effect,
 Nambu \cite{Nambu60} studies a linear relation between the
 Fourier components of the induced currents $J(q)$ in the ground
 state of a superconductor and an external vector potential ${\bm
 A(q)}$ in the limit $q \rightarrow 0$. It has been emphasized
 that there is a fundamental difference between the transverse
 and longitudinal current operators in the calculation of matrix
 elements in the BCS model \cite{BCS57}. Bogoliubov \cite{Bog58}
 was the first to point out that there exist the collective
 excited states of quasi-particle pairs, a coherent mixture of
 electrons and holes driven by the longitudinal current. The pair
 of quasi-particle is a linear combination of an electron and a
 hole; it becomes an electron above the Fermi surface, and a hole
 below the Fermi surface. It was pointed out that the collective
 excited states such as quasi-particles are essential to the
 gauge invariance in the Meissner effect. Moreover, the
 longitudinal current satisfies a sum rule [see Eqs. (1.1), (5.1),
 and (5.2) of Nambu] from which one concludes that
 the longitudinal current produces no physical effect - the gauge
 invariant Meissner effect, {\it whereas any transverse
 excitation involves breaking up the phase coherence over the
 Fermi surface of at least one pair (equivalent to a Cooper bound
 state) in the super-conducting ground state} - keeping the BCS
 result intact. Eq. (5.2) of Nambu takes the form of the London
 equations, since the intermediate pair formation
 (quasi-particles) is suppressed due to the finite energy gap in
 the limit $q \rightarrow 0$. In the previous chapter, we have
 also derived the dispersion relation for the transverse wave in
 the surface layer Eq.~(\ref{transwave2}), which breaks
 phase-coherence in He II. It is a dynamical calculation with
 \textit{the boundary conditions} for the surface phenomena in He
 II and is in remarkably consistent with Nambu's picture of gauge
 invariant explanation of the Meissner effect
 \cite{Nambu60,Anderson58,Nambu61}.

The models studied by Nambu \cite{Nambu60,Nambu61} involve the
basic four-fermion interactions. The boson excitations appear as
collective modes of the fermion system. Goldstone
\cite{Goldstone61,Goldstone62} has, on the other hand, studied
the theories in which bosons are taken as elementary fields.
These elementary bosons transform by an irreducible
representation of a continuous group under which the Lagrangian
remains invariant. From these models, Goldstone conjectures that
\textit
 {whenever the Lagrangian admits a continuous symmetry group, but
the vacuum is not invariant under the same group (whose
expectation value of boson fields is not zero), then massless
boson states must appear} - broken symmetry. The broken symmetry,
the energy gap, and the collective excitations are thus all
closely related phenomena in a superconductor - the
BCS-Bogoliubov theory \cite{note2}.

On the other hand, the Meissner effect in a superconductor is
explained by the gauge invariant theory
\cite{Nambu60,Anderson58,Anderson84}, since the Hamiltonian used
in the BCS calculation is not gauge invariant. However, as we have
seen in the previous discussion, the phase-coherence is broken by
transverse excitations in both He II and the Meissner effect in a
superconductor. Hence we can clearly see that the phase-coherence
is a more fundamental concept in the study of the broken symmetry.

Although much of what we have discussed on the Goldstone theorem
\cite{Goldstone61,Goldstone62} may be regarded as something of a
digression from the main object of the paper and is not new
either, the discussion of underlying physics is a pedagogical
contribution at the very most to our understanding of the broken
symmetries which have played an important role in our
understanding of superconductivity - the gauge invariant
explanation of the Meissner effect in the theory of
superconductivity. Hence it is helpful to discuss the broken gauge
symmetry in a liquid helium in some detail, both for its intrinsic
mathematical interest and as a means of solving long standing
problems in low temperature physics.

In order to show why the Goldstone theorem plays an important role
in the broken symmetry, we must cast it in a mathematical form so
as to yield a precise mathematical theory. It can be shown by the
$\sigma$-model of Gell-Mann and Levy within the framework of
quantum field theory \cite{Bjorken65,BenLee73, Levy60}: the
Lagrangian density for two scalar fields $(\sigma, \pi)$ are given
by
\begin{equation}
{\cal L} =\frac{1}{2}[\partial_{\mu}\sigma\partial^{\mu}\sigma+
\partial_{\mu}\pi\partial^{\mu}\pi]
-V(\sigma^{2}+\pi^{2}), \label{Lagrangian}
\end{equation}
where the potential
\begin{equation}
V(\sigma^{2}+\pi^{2})=\frac{1}{2}\mu^{2}(\sigma^{2}+\pi^{2})+
\frac{1}{4}\lambda(\sigma^{2}+\pi^{2})^{2}.
\label{potential}
\end{equation}

One can easily show that the lagrangian density
Eq.~(\ref{Lagrangian}) is invariant under $O(2)$ (or $U(1)$)
group by:
\begin{eqnarray}
\left(
\begin{array}{c}
\tilde{\sigma}\\ \tilde{\pi}
\end{array}
\right) =\left(
\begin{array}{c}
\cos\theta\quad\sin\theta\\ -\sin\theta\quad \cos\theta
\end{array}
\right) \left(
\begin{array}{c}
\sigma\\ \pi
\end{array}
\right)
\end{eqnarray}
In order to carry out a perturbation analysis, it is necessary to
find the minimum values for the potential. This self-consistent
condition yields the minimum conditions if both
\begin{subequations}
\label{minimum}
\begin{eqnarray}
\partial V/\partial \sigma &=&\sigma[\mu^{2}\;+\;\lambda(\sigma^{2}+ \pi^{2})]\\
\partial V/\partial \pi &=&\pi[\mu^{2}\;+\;\lambda(\sigma^{2}+ \pi^{2})]
\end{eqnarray}
\end{subequations}
satisfy simultaneously
\begin{equation}
\partial V/\partial\sigma=\partial V/\partial\pi=0.
\end{equation}

If $\mu^{2}< 0$ in Eq.~(\ref{minimum}), the minimum occurs on the
circle $[\sigma^{2}+\pi^{2}]^{1/2}=[-\mu^{2}/\lambda]^{1/2}$.
Thus in the $\sigma-\pi$ plane, we may assign the vacuum
expectation values as
\begin{equation*}
\langle\sigma\rangle_{0}=[-\mu^{2}/\lambda]^{1/2} \quad and \quad
\langle\pi\rangle_{0}=0.
\end{equation*}
Hence the symmetry of ground state (vacuum in quantum field
theory) is not invariant under the same group $O(2)$ and
therefore the symmetry is broken {\it spontaneously}.

In the $\sigma$ - model of Gell-Mann and Levy \cite{Levy60}, they
add a small symmetry breaking term ($c\sigma$) to the potential
$V$.  Then the minimum occurs if
\begin{equation*}
\sigma[\mu^{2}+\lambda(\sigma^{2}+\pi^{2})]=c  \quad and \quad
\pi[\mu^{2}+\lambda(\sigma^{2}+\pi^{2}0]=0
\end{equation*}
The equations show that $c\sigma$ breaks the rotational symmetry
in the $\sigma-\pi$ plane around the third axis and hence there
is no solution for the equations except for $\pi=0$ and
$\sigma(\mu^{2}+\lambda\sigma^{2})=c$. In the limit $c\rightarrow
0$, either $\sigma=0$ or $\sigma=[-\mu^{2}/\lambda]^{1/2}$ as
before. Hence the solution takes the minimum value if
$\mu^{2}<0$. The nature of symmetry is made much clearer if we
perform translation $\tilde{\sigma}=\sigma-<\sigma>_{0}$ with
$<\sigma>^{2}=-\mu^{2}/\lambda$ and rewrite the Lagrangian
density Eq.~(\ref{Lagrangian}) in terms of
($\tilde{\sigma},\tilde{\pi}$):
\begin{equation}
{\cal L}
=\frac{1}{2}[\partial_{\mu}\tilde{\sigma}\partial^{\mu}\tilde{\sigma}+
\partial_{\mu}\pi\partial^{\mu}\pi]+\mu^{2}\tilde{\sigma}^{2}-
\lambda<\sigma>_{0}\tilde{\sigma}(\tilde{\sigma}^{2}+\pi^{2})
-\frac{1}{4}(\pi^{2}+\tilde{\sigma}^{2})^{2}, \label{Lag2}
\end{equation}
which shows that $\pi$ field is for a massless Goldstone mode and
$\tilde{\sigma}$ field is for a particle with a positive mass
$-\mu^{2}$ \cite{Bjorken65}. As explained above the presence of a
massless particle is necessary in the spontaneously broken
symmetry. From this example, a conceptually very simple
interpretation of the Goldstone theorem emerges from the
mathematics of a simplified  $\sigma$ - model of Gell-mann and
Levy \cite{Levy60}.

It will be shown below that there is a unique aspect about the way
the spontaneously broken symmetry manifests itself in a Bose
system, but it is consistent with the conservation of energy as
shown in the previous Gedanken experiment in BEC. Since the above
mathematical derivation for the Goldstone theorem in the quantum
field theory is valid only in the Hilbert space \cite{Bjorken65},
we now see why it was necessary to show by a semi-classical method
that a phonon, whose spectrum is $\omega=c k$, is present in the
spontaneously broken symmetry in the Bose system by
Eq.~(\ref{Sound}) and Eq.~(\ref{Phonon}), where $c= [4\pi a
\rho(r)\hbar^{2}]^{1/2}/M$ in the phonon regime.  And thus it only
remains to justify on physical grounds that the ground state is
not invariant under the global symmetry $U(1)$ for the
spontaneously broken symmetry to take place in the liquid helium
under the gravitational field.

In the following we shall consider a formal similarity between the
surface phenomena in He II in a gravitational field and the
Meissner effect in a superconductor in weak external magnetic
fields. There is, however, a significant difference between the
ways in which the symmetry is broken in He II under the
gravitational field and a superconductor in an external static
magnetics field. In He II, the Hamiltonian Eq.~(\ref{grounda}) is
invariant under $U(1)$, but the symmetry is broken at the surface
layer as the profile of ground state function Eq.~(\ref{Eqs2}) is
not invariant and looks like a Mexican hat just as a vacuum
expectation (ground state) Eq.~(\ref{potential}) in quantum field
theory \cite{Goldstone61,BenLee73}.

\subsection{\label{sec:level3} London Equations and Surface
phenomena in He II}

We proceed by first noting that Landau's two-fluid model which
describes beautifully many aspects of He II, in particular, the
excitation spectra of phonons and rotons, seems a well-grounded
theory despite the fact that underneath all lies the peculiar
absence of physical boundaries which brings about the broken
symmetries in the liquid helium \cite{Landau59}.

Much of the similarities between the surface phenomena in He II
and the Meissner effect in a superconductor has been due to the
London equations for the Meissner effect in terms of the classical
magnetohydrodynamics, ($\bm{\nabla}\cdot\bm{J}=
-n_{s}\bm{\nabla}\cdot\bm{v}=0$ and
$\bm{Q}=\bm{\nabla}\times\bm{v}-e\bm{h}/Mc=0)$. If we set
$\bm{h}=0$, the London equations are essentially  equivalent to
our boundary conditions ($\bm{\nabla}\cdot\bm{\xi}=0$ and
$\bm{\nabla}\times\bm{\xi}=0$) with the gravitational acceleration
$\bm{g}=0$ \cite{London38,London54,Gorkov59} along with Bohm's
second assumption $\bm{p}=\bm{\nabla}S$ which implies
$\bm{\nabla}\times\bm{p}=0$ for a free surface of He II in the
absence of gravitational field \cite{Lamb45}.

It may seem strange that the characteristics of the surface layer
of He II under the gravitational field are precisely analogous to
those of a superconductor in a static magnetic field (Meissner
effect). Yet the similarity is indeed remarkable. It is this
critical observation of the surface phenomena in He II that led us
to study the broken symmetries. Thus the spontaneously broken
symmetry on the surface layer in He II is not quite accidental.
However, unlike a superconductor in a uniform magnetic field for
which the penetration depth is finite $\lambda_{L}=[mc^{2}/(4\pi
n_{s}e^{2})]^{1/2}$ due to the expulsion of the magnetic flux, the
penetration depth of the gravitational field in He II is infinite
$\lambda\rightarrow \infty$ but the gravitational field strength
is negligible compared to the molecular forces; and the thickness
of a surface layer is only $\lambda_{d} \approx 5.0 \times
10^{-3}cm$. Thus we see that He II is not ever be free of
vorticity under the gravitational field, which allows one to
assume the upward diffusion of vortices in a rotating He II as
will be shown below.

 Since the ground state wave function in ODLRO can be written as
$\psi(r,t)=f(\bm{r},t)exp[\frac{i}{\hbar}S(r,t)]$
\cite{Bohm52,Aharonov63}, we may interpret the broken symmetry of
the ground state wave function at the nodal surface, the point at
which $S(r,t)$ must satisfy the prescribed boundary conditions by
which we have demonstrated in section IV that a degenerate Bose
system in an external field exhibits the second-order phase
transition at the nodal surface - a normal fluid with the surface
energy onto a superfluid just as the Meissner effect in a
superconductor in an external magnetic field, {\it i.e.,} {\it a
spontaneously broken symmetry}
\cite{London38,Singh62,Anderson84,BenLee73,Tinkham96}.

This is analogous to that of the second-order phase transition
between the normal state and the super-conducting state based on
the free energy function in terms of order parameter in the
macroscopic Ginzburg-Landau theory of a superconductor in which
the energy gap function $\Delta$ shows the existence of spatial
inhomogeneity, {\it i.e.,} the normal state joined onto the
super-conducting state near the second-order phase transition (the
Meissner effect) \cite{Tinkham96}. What's more, the second-order
phase transition in a superconductor with
$\kappa=\lambda(T)/\xi(T)\ll 1$, where $\kappa$ is the
Ginzburg-Landau parameter, is almost identical to our problems -
the superfluid joined on to surface layer (a normal fluid) in He
II under the gravitational field or the shell-like structure of
BEC in a trap.

 It is also well-known that a lattice possesses a translational
symmetry, {\it i.e.,} invariant under $U(1)$ transformation and
that the symmetry can be broken by dislocation \cite{Landau70}. In
this sense the spontaneously broken symmetry in He II is similar
to what a lattice breaks the translational symmetry by dislocation
or by impurity - a broken $U(1)$ lattice gauge
\cite{Anderson63,Anderson84}, since the Bose particles in He II
are charge neutral which can be described by the global symmetry
without a local gauge field like Fermions. Our discussion on the
spontaneously broken symmetry in He II should be, therefore,
justified by the Goldstone theorem alone
\cite{Goldstone62,BenLee73,Weinberg96}.

It should be emphasized, however, that Goldstone studied a model
that has  a single fermion interacting with a single pseudoscalar
boson (quantized) field in the long wavelength limit $\bf{k}=0$
\cite{Goldstone62}; the model thus has a boson field at the
outset in contrast to Nambu's dynamical model of a pseudoscalar
zero-mass bound states of nucleon-antinucleon pair (soft pions)
\cite{Nambu61}. Therefore we see that our dynamical study of
phonons as collective excitations in He II is a semi-classical
approach (a compressional wave) in which we confined our
attention to verifying that the phonons (Goldstone modes) are
present with a correct dispersion relation for the traveling
sound waves in He II and hence our approach is quite different
from those of both Nambu's  and Goldstone models. Yet we can
still apply the Goldstone theorem for the spontaneously broken
symmetry in our problem, because we can construct the Lagrangian
from the Hamiltonian Eq.~(\ref{grounda}) that possesses a
continuous symmetry group $U(1)$ under which the ground state
Eq.~(\ref{Ground}) is not invariant, for it has a nodal surface.
What is important here is that the broken symmetry always
accompanies \textit{a phonon} (Nambu-Goldstone boson or Goldstone
mode) \cite{Anderson07}.

In the case of a vortex, the breakdown of superfluidity, however,
accompanies a roton whose effective mass is $\mu_{ro}=0.16
m_{He}$. In fact there is also a spontaneously broken approximate
symmetry for which we add a small symmetry breaking term in the
action for the derivation of the Goldstone theorem and from which
we see the appearance of low mass, spin-zero bosons,
pseudo-Goldstone boson as shown above in the $\sigma$ model of
Gell-Mann and Levy \cite{Levy60,Weinberg72}.

\section{\label{sec:level1} Broken Symmetries by
the Fluctuation-Dissipation}

Another approach to the study of broken symmetry is to ask a
question whether the mean-field approach can explain the
spontaneously broken symmetry at the nodal surface in He II based
on Bohm's interpretation of quantum theory \cite{Bohm52}. The
affirmative answer to the question is essential to the basic
argument of this paper. Here we discuss the broken symmetry based
on the following picture: He II is described by the mean field
$\psi(r,t)$ in ODLRO and has the hidden symmetry which is given in
terms of the action $S(r,t)$. He II breaks this symmetry at the
nodal surface by the effective quantum mechanical potential
Eq.~(\ref{Eqmp}), which drives the particles to fluctuations ({\it
i.e.,} Bohm's irreducible disturbance) and thereby breaks the
phase coherence - a necessary condition for superfluidity. Thus
the spontaneously broken symmetry can be readily explained by
Bohm's quantum theory and the mean field defined in ODLRO.

An important step in our argument is based on the phase-coherence
of the collective excitations in ODLRO. Since the
phase-coherence, $\bm{\xi}\cdot\bm{\nabla}S(\bm{x}_{0},t)
=\sum_{i}\bm{\xi}_{i}\cdot\bm{\bm{\nabla}_{i}}S_{0,i}(\bm{x}_{0,i},t)$,
is a necessary condition for a superfluid to exist, if it is
broken due to the quantum fluctuation \cite{Bohm52} at the nodal
surface where $\bm{\nabla}\rho(\bm{x})$ is discontinuous and the
effective quantum mechanical potential Eq.~(\ref{Eqmp}) drives
particles to fluctuate rapidly at the nodal surface of He II,
thus the fluid is no longer a superfluid due to breakdown of the
phase-coherence at the surface where the boundary conditions play
an important role just as the London equations for the Meissner
effect in a superconductor. On the other hand, we note that,
inside of He II away from the surface layer where the boundary
conditions are unimportant, the action $S(r,t)$ is analytic and
the longitudinal sound waves remain phase coherent
\cite{Anderson58,Nambu61}.

If, for the purpose of illustration, we assume that a superfluid
droplet, which is suspended by an external force and in which the
symmetry is not broken, and that a sound wave was initiated
adiabatically at the center of the droplet, then the phonons must
bounce back to the center. Hence we see immediately that we cannot
maintain {\it the law of conservation of energy} in an isolated
system
\cite{Nambu60,Nambu61,Goldstone61,Levy60,Goldstone62,Anderson63,BenLee73},
because the sound wave initiated at the center of a superfluid
droplet cannot be dissipated at the surface layer. However, if the
symmetry is broken, the sound wave dissipates by the interaction
with the normal fluid at the surface layer, which would give rise
a surface energy manifested by a surface tension -
Eq.~(\ref{Singh2}). This is because acoustic phonons, which
possess many attributes of particle, cannot interact with a
superfluid component, as Mott emphasized in his study of the
two-fluid model \cite{Mott49}, it does, however, interact with a
normal fluid, giving rise to a surface energy manifested by the
capillary wave Eq.~(\ref{capillary2}). This dissipation process of
the sound wave is the same as in the case of the BEC in a trap. It
should be made clear that \textit{this Gedanken experiment} of
(first) sound wave propagation in a superfluid droplet described
above is different from the experimental study of the propagation
of second sound by Hall and Vinen \cite{Landau41,Hall56}. In the
simple example described above, the energy supplied from an
external source is to drive the phonons to form a degenerate Bose
gas with {\it an energy gap} out of Bose particles near the
surface layer in a trap \cite{London38,Landau80}. Thus the above
picture provides a simple model to help understand the
mathematical treaties of broken symmetry.

It is obvious that the above discussion is not a standard method
to describe a dissipation process. In this we have departed from
the Kubo's approach for the electrical conductivity in a
dissipative medium \cite{Kubo57,Mori58,Luttinger64-1}, which would
be a more familiar approach to the traditional many-body theory.
The broken symmetry in the Bose system follows, however, from the
statistics and simplicity of the ground state defined in ODLRO,
and explains a number of unsolved problems in He II. The
discussion on Luttinger's analysis on the fluctuation and
dissipation in a superconductor is given below to show that the
general features of the broken symmetry by the fluctuation and
dissipation presented here are similar to those employed in his
analysis, in particular his basic assumption on the chemical
potential in the equation of motion, {\it i.e.,} Eq.(2.4) of
Luttinger \cite{Luttinger64-2}.

Since there is no Hamiltonian that describes a thermal gradient
and the temperature is a statistical property of the system, the
Kubo's formula has been derived by using local variables
\cite{Mori58} assuming the local equilibrium. In the absence of
an external magnetic field, the dissipation mechanism must be
then temperature dependent.

Indeed, the equations with phenomenological transport coefficients
for a superconductor has been derived in the presence of
temperature gradient by Luttinger \cite{Luttinger64-2} with
additional assumption on the system. From these equations, he
derives an expression for the the thermal conductivity as a
correlation function of the Kubo-type based on the two-fluid model
and the  basic equation [Eq. 2.4 of Luttinger] which is similar to
our Eq.~(\ref{Motion0}) [see also Eq.~(\ref{Eqs0})]. The
fundamental assumption on the presence of the chemical potential
in the equations of motion implies that a system of steady-state
is made up with a free surface and thus the two-fluids in which
heat and particle flows are necessary to maintain
quasi-equilibrium in the presence of temperature gradient
\cite{Luttinger64-2,Anderson65,Gorkov58}.

In other words, a superfluid alone cannot maintain a steady state
without the broken symmetry at the free surface. In our study of
BEC, the chemical potential $\mu$ is introduced in
Eq.~(\ref{Ground}) as a Lagrangian undetermined multiplier, a
condition to maintain the number of particles constant in an
isolated system. A more elaborate explanation was offered by
Luttinger using the mean-occupation number formalism in
statistical mechanics \cite{Luttinger68}, and he gives a full
justification for his model of thermal transport coefficients of a
superconductor \cite{Luttinger64-2}. Precisely, it is also the
reason why we have seen the broken symmetry in an isolated system
of the BEC at the free surface - a shell-like structure.

The dissipation process of sound waves in the surface layer of He
II is essentially similar to that of Luttinger's model of thermal
transport coefficients of a superconductor
\cite{Luttinger64-1,Luttinger64-2}, since his derivation of
transport coefficients are appropriate to a normal conductor. Thus
there is a discontinuity in the free energy which leads to the
second-order phase transition ({\it i.e.,} a superconductor
transforms to a normal metal).

In a simple system of a charged boson gas as a superconductor in
equilibrium, Schafroth showed the system exhibits a phase
transition of the second kind at a critical temperature $T_{e}$
\cite{Schafroth54}. It was further emphasized that, in the
presence of the vector potential of an  applied weak inhomogeneous
magnetic field, the penetration depth of a superconductor is not
determined solely by the density of condensed boson $\rho_{s}$
[see Eq. 4 of Schafroth], but it also depends on the normal fluid
$\rho_{n}$ \cite{Schafroth54,BCS57}. Thus the charged boson gas
$\rho_{s}$ in the surface layer may be considered as the bound
two-electron state due to interactions mediated by phonons in the
lattice. Also, it may be interpreted as another manifestation of
Bose-Einstein condensation below $T_{e}$ \cite{Landau80}. Thus the
second-order phase-transition at the surface layer in a
superconductor in an external magnetic field \cite{Nambu60} is
similar to that of a shell-like structure of Bose-Einstein
condensation in an external potential. The model of a
superconductor by Schafroth gives further evidence of the bound
two-electron state in a superconductor \cite{Onsager61,Yang61},
{\it i.e.,} a correct flux quantization constant $\Phi=nhc/2e$
realized in experiments by Deaver and Fairbank and also by Doll
and Nabauer \cite{Ring61}.

 In conclusion we note that {\it the broken symmetry at the
surface layer is both a necessary and a sufficient condition for
an isolated Bose system to conserve the energy at low temperature
and to conserve the number of particles in a trap}. Although the
effect on the surface energy is very small in practice, we are
only interested in the question of basic principles, and an
arbitrarily small effect on the surface tension is just as good as
a large one. Perhaps, I should remind the reader of how a neutrino
has been discovered in the $\beta$ decay in nuclear physics ({\it
Fermi theory of $\beta$ decay}) \cite{Holliday55}. The discovery
of the neutrino was in a similar circumstance as in the
second-order phase transition in He II
\cite{London38,Singh62,Ginzburg50,Gorkov58}. So far we have
outlined only a number of the salient features in theoretical
developments for the broken symmetry in a Bose system and we now
turn to experimental aspects of the broken symmetry in He II which
provide a complete picture of the broken symmetry.

\section{\label{sec:level1} Basic Experiments}

Based on the phenomenological theory of broken symmetry presented
above and Anderson's concept of phase slippage we now discuss the
four puzzling problems in He II and BEC: 1) the vortex
quantization, 2) Magnus force, 3) the curvature of a rotating He
II, and 4) a question of possible formation of a super-lattice in
He II by compression.

\subsection{\label{sec:level2} Vortices in He II}

It is useful to recall that phonons and rotons are not the only
collective excitations in He II. There also exist macroscopic
excitations, {\it vortices}, which involve the flow of large
amount of fluid in considerably higher energy. The study of
vortex motion in He II is of particular interest, because it helps
not only understand the decay of vorticity in a homogeneous
turbulence state which is created in a steady state of He II and
the subsequent decay of vortices \cite{Vinen93}, but also the
role played by the current-induced vortex motion ({\it i.e.,} the
motion of Abrikosov flux lines \cite{Abrikosov57}) for
dissipative processes in type-II superconductors
\cite{Brandt01,Kim64,AndersonKim64}.

In a multiply connected region in He II the condition that
$\psi(r,t)$ be single-valued leads to the requirement that the
phase $S/\hbar$ need not be single-valued but merely need return
to its original value $\pm 2n\pi$ on traversing the nodal surface
of a vortex. Hence the Onsager-Feynman quantization
\cite{Onsager49,Feynman55} can be stated in a more precise manner,

\begin{equation}
\sum_{cir} \bm{\nabla}S\cdot\bm{\xi}= \oint\bm{\nabla}S\cdot
d\bm{l}= nh. \label{vortex}
\end{equation}

 Here the summation is taken around the nodal surface and
$\bm{\xi}$ is taken to be small since it is an atomic
displacement. In He II, if one takes $\bm{\nabla}S = \bm{p}=M
\bm{v}$ by Bohm's second assumption which states that $S$ is a
solution of the Hamilton-Jacobi equation, then
$\kappa=\oint\bm{v}_{s}\cdot d\bm{l}$, where
$\kappa_{0}=2\pi\hbar/M=0.997\times10^{-3}cm^{2}/sec$
\cite{Bohm52,Landau77}.

As we have discussed in VIII that Eq.~(\ref{vortex}) breaks up the
phase-coherence $\bm{\xi}\cdot\bm{\nabla}S(\bm{x}_{0},t)
=\sum_{i}\bm{\xi}_{i}\cdot\bm{\bm{\nabla}_{i}}S_{0,i}(\bm{x}_{0,i},t)$.
Here $\bm {p}=\bm{\nabla}S(\bm{x},t)$, where $S(\bm{x},t)$ is
assumed to be analytic and is also independent of time. The
phase-coherence is broken spontaneously due to fluctuations in
particle orbits driven by the EQMP in the phase-space ({\it i.e.,}
Bohm's irreducible disturbance) and is also quantized by Planck's
quantum condition. Thus this can be interpreted as the
spontaneously broken symmetry manifested by breaking up the
phase-coherence. Better yet is the conceptual difference that the
spontaneously broken symmetry is more fundamental for the creation
of a vortex line \cite{Bohm52,Anderson66} than that of the
Bohr-Sommerfeld quantization of the linear momentum
\cite{Onsager49,Feynman55}, for which there is no underlying
mechanism by which a spontaneous quantum jump can be explained
other than the statistical law. The absence of an identifiable
cause for the spontaneous quantization troubled Einstein whose
objection to Bohr's atomic model is well known
\cite{Bohm52,Bohm51,Pais91}.

This quantized circulation has been confirmed in a superfluid
helium at low temperature, He II. In an ingenious experiment by
Rayfield and Reif \cite{Reif64} in which they have used ions as a
microscopic probe to study the motion of vortices in He II. The
motion of ions creates a flow favorable for creating a vortex
around the probe which is trapped at the core of the vortex. The
probes are interpreted as an ion-structure clustered by van der
Waals forces \cite{Lifshitz56,Sabisky73}. Moreover the motion of
the ion-cluster is accurately measured by externally applied
electromagnetic forces. An interpretation of the ion mobility in
He II involves the nature of collective excitations and their
interaction with the ionic probe particles.

Rayfield and Reif also proposed a mechanism for the vortex-ring
formation in He II by ion probes. But there was a difficulty in
explaining how the ion probes have been trapped in the vortex-ring
\cite{Reif64,KHuang65,Careri65}. Rayfield
\cite{Reif64,Rayfield67,Rayfield68} repeated the same experiment
with impurity of He$^{3}$ ($1$ part of He$^{3}$ per
$5.35\times10^{3}$ parts of He$^{4}$) \cite{Zharkov57} and at the
temperature below $0.30^{\circ}$K to limit the scattering of
rotons with the ion probes which complicates the study of
vortex-ring formation in He II \cite{Careri65}. Based on the new
data, Rayfield has proposed a mechanism
\cite{Rayfield67,Rayfield68} by which a vortex-ring is being
formed in He II with ion probes at the center and which overcomes
a logical inconsistency in the previous work \cite{Reif64}.

More specifically the new experimental data indicate that "{\it
the vortex lines associated with the formation of the vortex-ring
is slowly peeled away from the ion complex in the form of a
steadily growing loop as the electric field increased}". This
picture explains why the ion complex is not required to hop into
the ring and there was no discontinuous change in the drift
velocity after the ion reaches a critical velocity. Thus the
peeling out of vortex line begins when the velocity field of the
superfluid in the neighborhood $\oint \bm{v}\cdot d\bm{l}=h/M$
(see Fig.2 of Ref.\cite{Rayfield67}). If the process is completed
at the ion-critical velocity, the discontinuity results in the
curve of velocity vs electric field which shows a spontaneous
symmetry breaking in He II \cite{Rayfield67}.

Based on the concept of Anderson's phase slippage
\cite{Anderson66}, this process can be described more precisely by
$2\pi$ phase slip of a vortex line at the inner surface of the
macroscopic vortex-ring and the momentum exchange as the ion
cluster begins to slow down, since the microscopic vortex line
attached to the ion complex can be detached through the
interaction with the macroscopic vortex-ring at its inner surface.
As emphasized by Rayfield \cite{Rayfield67}, it is a difficult
task to obtain the dispersion relation for the ion-vortex-ring
experimentally, for the velocity field is a superposition of the
flow field of ion and vortex-ring. As will be shown below, the
vortex-ring formation is very similar to that of a parabolic
surface of a rotating He II.

It is noteworthy, however, that we have applied a spontaneously
broken symmetry as a mechanism to explain the breakdown of
superfluidity at the vortex core instead of the  Bogoliubov-de
Gennes equations which were applied to the study of the energy gap
parameter $\Delta$ to explain the breakdown of superconductivity
in the Abrikosov vortex line \cite{Caroli64, Leggett01}. The
reason for this is that the Bogoliubov equations are valid in a
Hilbert space, but not to a problem with a boundary, because they
are derived for a uniform system \cite{Bog59}. Hence this proves
our approach is indeed consistent with the quantum field theory,
as it gives us a correct concept on the breakdown of superfluidity
from which we describe the basic notion of broken symmetry.

\subsection{\label{sec:level2} Circulation Constants $(\kappa)$
and Magnus Force}

As we have discussed above, it is well known that, in addition to
phonons, rotons, and a vortex-line, there also exist macroscopic
quantum excitations involving the flow of much larger amount of
liquid at much higher energy - a vortex-ring in a superfluid. It
is also known that the quantized vortices in a neutral superfluid
play an important role in the decay of superfluid currents.
Likewise in a superconductor in a high magnetic field, the motion
of quantized magnetic flux lines is the main mechanism for
electrical resistance in superconductors \cite{Kim64}.

 It is therefore important to understand the dynamics of
quantized vortices for a study of dissipation mechanism in both a
neutral superfluid and the superconductors. Rayfield and Reif
\cite{Reif64} have presented the vortex formation by the exact
expressions which are derived from classical hydrodynamics for the
energy and translational velocity of a vortex ring, moving in an
{\it incompressible} fluid of density $\rho$. We follow this
procedure since we wish to maintain a close contact with the
previous theoretical work, but we wish to offer an entirely
different interpretation of the results. The following equations
describe the motion of the surface of a vortex-ring in a
superfluid \cite{Lamb45}.
\begin{subequations}
\label{hydrodynamics}
\begin{eqnarray}
E &=&(1/2)\rho\kappa^{2}R[\eta-(7/4)]\\
\label{Energy} v &=&(\kappa/4\pi R)(\eta-1/4),\label{velocity}
\end{eqnarray}
\end{subequations}
where $\eta=ln(8R/a)$ and $a$ is the core radius of a vortex. The
experimental data were, then, analyzed by eliminating
$R$.

Furthermore Eq.~(\ref{Energy}) and $\eta=(vE/B)^{1/2}+1$ with
$\eta\gg 1$ yield
\begin{eqnarray*}
vE & = &B[\eta-(7/4)](\eta-1/4)\\
\eta & = &ln\{16E[\rho\kappa^{2}a(\eta-7/4)]^{-1}\}
\end{eqnarray*}
where $B=\rho\kappa^{2}/8\pi$.

With the condition $\eta\gg 1$, which is equivalent to taking the
hydrodynamic equations to a quantum domain in which the
uncertainty principle limits the precise measurements of the
position and momentum of a particle in the system. This is
consistent with Bohm's interpretation of quantum theory
\cite{Bohm52} and Eq.~(\ref{vortex}). Perhaps more important,
Bohm's irreducible disturbance at the core of a vortex ({\it
i.e.,} fluctuations of particle orbits due to $U_{eqmp}$) is
precisely the cause of breakdown of superfluidity at the core of a
vortex-line. A combination of the above equations gives an
approximate equation for $[vE]^{1/2}$,

\begin{equation}
[vE]^{1/2}=B^{1/2}\{lnE-ln[(\frac{vE}{B})^{1/2}-\frac{3}{4}]\}
+B^{1/2}[ln(\frac{16}{\rho\kappa^{2}a})-1]. \label{vortexmotion}
\end{equation}

Since $vE$ is a slowly varying function of $E$, Rayfield and Reif
\cite{Reif64} were able to use a graphic technique to obtain the
values for $\kappa$:

\begin{eqnarray}
\label{Kappa2}
\kappa &=&(1.00 \pm 0.03)\times10^{-3}cm^{2}sec^{-1}\\
a &=&(1.28\pm0.13)\AA,
\end{eqnarray}
where $\rho=0.1454\,g\,cm^{-3}$ for liquid helium has been used.
It is indeed remarkable that the experimental value $\kappa$ with
$n=1$ is, within the limits of estimated error due to a
discontinuity, equal to $\kappa_{0}=h/M$ where $M$ is the mass of
He$^{4}$ - the second-order phase transition
\cite{Ginzburg50,Schafroth54}; it is another manifestation of a
spontaneously broken symmetry in a superfluid.

The first experiment designed to investigate the question of
quantized circulation was Vinen's closed cylindrical vessel of
superfluid in which he measured the transverse vibrations of a
fine wire along the axis of rotation from which the Magnus force
on the wire was deduced \cite{Vinen61}. The experiment, however,
encountered some difficulties in establishing steady state and in
avoiding partial attachment of vortex lines to the wire. The data
showed, therefore, a wide range of uncertainty in the observed
values of $\kappa$, although a pronounced peak was found near the
value of $h/M$.

Although Rayfield and Reif \cite{Reif64} did not directly measure
the Magnus force to study the vortex quantization, it is
essential to understand the dynamics of a charged vortex ring
from the point of view of hydrodynamics which provides a greater
physical insight into the question of breakdown of superfluidity
at the vortex core. In spite of its importance in vortex
dynamics, there is still no agreement on a correct form of the
Magnus force \cite{Iordanskii65,Sonin97}; Thouless, {et al.},
\cite{Thouless99} pointed out that \textit {there are many,
conflicting results in the literature, as many as there are
theorists active in the field}. Perhaps it is not surprising that
there should have been such a diverse disagreement, for the
Landau's two-fluid model was confirmed in an experiment by
Andronikashvili \cite{Andro46} and was never before suspected for
its breakdown at the nodal surface, although there was clear
experimental evidence for the break-down of superfluidity in a
rotating He II as early as 1950 in Osborne's experiment
\cite{Osborne50}.

 Out of such seemingly hopeless confusion that has plagued our
intuition since the first experiment by Vinen \cite{Vinen61}, a
systematic study of Thouless group over the years has brought the
question of a correct form of the Magnus force as a major
stumbling block in our understanding of dissipation processes in
He II. But a general consensus on the form of Magnus force is
still lacking, because there has been no convincing theoretical
analysis to date. The failure of strenuous efforts  by Thouless
group led us to speculate at first that some novel concept is
lurking behind this pervasive confusion over the half century
\cite{Thouless99}.

Yet there is one outstanding a hint in the way of accepting any of
those previous theories: the crux of controversy on the form of
the Magnus force is the very simple observation that we have no
experimental evidence for the presence of a superfluid component
at the vortex core and the data are always analyzed based on
\textit{the classical form of the Magnus force}
\cite{Vinen61,Reif64,Lamb45}. The current state on the Magnus
force reminds us the $\tau-\theta$ puzzle in nuclear physics that
led the discovery of the parity non-conservation in the $\beta$
decay in particle physics by Lee and Yang in the 1956
\cite{Lee-Yang56,CSWu57}.

Indeed the most significant consequence of the spontaneously
broken symmetry is that we adapt the classical form of the Magnus
force, since the two-fluid model breaks down at the nodal surface.
Thus the core of charged vortex ring may be regarded as a charged
thin ring made of a normal fluid ({\it i.e.,} a smoke ring with an
ion probe) that satisfies the equation of motion for a uniform
motion:

\begin{equation}
\bm F+ \bm G=0,  \;\;\;\; with \;\;\;\;\;  \bm G=\rho\bm
{\kappa}\times\bm U, \label{Magnus}
\end{equation}
where $\bm F$ is the force by the applied electric field, and $\bm
G$ is the hydrodynamic Magnus force per unit length, where $\rho$
is the density of a normal fluid. Here $\bm U$ is the relative
velocity of the core element with respect to the fluid. Thus $\bm
U=\bm{u}-\bm{u_{f}}$, where $\bm{u_{f}}$ denotes the fluid
velocity at this position caused by all sources other than the
core element. The Magnus force is, therefore, proportional to its
velocity of circulation and perpendicular to its direction of
motion, and thus causes a precession of the plane of the
vibration.

With the empirical dispersion relation of a charged vortex, they
obtain the Magnus force:
\begin{equation}
G_{x}=-\pi\rho\kappa R^{2}\dot{\theta}=-F_{x}.\label{Magnus1}
\end{equation}

It should be emphasized at this point that the basic equations
Eqs.~(\ref{hydrodynamics}) used in their data analysis are
obtained from classical hydrodynamics \cite{Lamb45} except for
$\kappa$ value which was determined by the data - a spontaneously
broken symmetry by which it is understood that
$\kappa=\oint\bm{v}_{s}\cdot d\bm{l}$, where $\bm{v}_{s}$ is the
velocity of a normal fluid (of an inner surface of a vortex - a
surface layer) and $\kappa$ is quantized in units of
$\kappa_{0}=h/M=0.997\times10^{-3}cm^{2}/sec$ in He II. Notice
that the circulation is not only constant but also quantized on a
macroscopic scale in units of $h/M$, whereas the motion of a
vortex-line by an external force should be described by the
classical physics.

Moreover, the relation between the Magnus force and the applied
external force Eq.~(\ref{Magnus}) can be used to test the
self-consistency since the applied field is given by $\bm F=e{\cal
E}$ which can be measured precisely, although the precise radius
of a vortex core is difficult to measure in experiments. The
velocity measurements in terms of energy showed that there is a
unique relation between the velocity and the energy. Moreover, the
energy is proportional to $\kappa^{2}R$, that is
$E\propto\kappa^{2}R$. But these points are completely classical
in nature except for the $\kappa$ value by the quantization,
Eq.~(\ref{vortex}).

Once again we remind the reader that the free surface at the
vortex core is a nodal surface where the broken symmetry takes
place \cite{Hall60} and should therefore behave like a normal
fluid. This explains why Rayfield and Reif \cite{Reif64} have
observed that a vortex responds like a normal fluid to an ion
probe at the core ({\it i.e.,} a breakdown of superfluidity).  As
the superfluid density $\rho_{s}$ approaches to zero near the
nodal surface of a vortex-line which rotates approximately with
the speed of the first sound $c=\kappa/(\sqrt{8}\pi\zeta)$ with
$\zeta$ the coherence length, the effective quantum mechanical
potential (EQMP) fluctuates rapidly to excite the rotons near the
vortex line, because on the vortex line $U_{eqmp}$ is highly
singular.

The first experimental observation on the collective excitations
of rotons near the vortex line was reported in 1968
\cite{Glaberson68}. With the picture of a roton as the quantum
analog of a smoke ring whose effective mass is
$\mu_{ro}=0.16m_{He}$ and whose excitation energy is much higher
than that of a phonon, we have limited our discussion on the
mechanism of how rotons are created near the vortex line to
explain the breakdown of superfluidity at the core - a breakdown
of Landau's two-fluid model \cite{Landau41}.

The broken symmetry is demonstrated so strikingly in the
experiments by Rayfield and Reif \cite{Reif64} that it has finally
helped resolve a longstanding controversy over the Magnus force
\cite{Iordanskii65,Sonin97,Thouless99}. We have known the
erroneous (and confusing) interpretation of the Magnus force,
however, for a good many years, since the first experiment on a
vortex quantization was analyzed by the classical form of the
Magnus force that has given the correct circulation in units of
$\kappa_{0}=h/M$ \cite{Vinen61,Lamb45,Glaberson68}. The reason why
this controversy has persisted for so long may be traced to our
failure to recognize the broken symmetry at the nodal surface and
to take account the fluctuations of particle orbits ({\it i.e.,} a
break down of superfluidity by the breaking up the
phase-coherence) driven by the fluctuating EQMP with decreasing
radius of a curvature of the free surface at the core
\cite{Bohm52}. It also explains why Vinen's experimental data
\cite{Vinen61} are essentially correct, which was obtained by the
classical form of the Magnus force.

We would like to clarify about the confusing remark that one finds
in the literature on the Iordanskii's extension of the Magnus
force in the two-fluid model \cite {Thouless99}, which was
supported by experimental data \cite{Donnelly91}, but is now
questionable, because the extension motivated by the two-fluid
model and contradicts Vinen's experimental data \cite{Vinen61}. We
find the data are not credible \cite{Donnelly91} and contradict
the analysis by Rayfield and Reif [see Eq.~(\ref{Kappa2}) and
Eq.~(\ref{Magnus})].

\subsection{\label{sec:level2} A Rotating He II}

From Landau's two-fluid model \cite{Landau41}, one obtains the
equation $z=(\omega^{2}r^{2}\rho_{n})/(2g\rho)$, where
$\rho=\rho_{s}+\rho_{n}$, for the shape of the free surface of a
rotating He II, but in experiments one observes instead the
temperature independent equation for the curvature
$\gamma=\omega^{2}/g$ \cite{Osborne50,Meservey64}. This
observation presents a profound paradox \cite{Andro66}: Meservey
\cite{Meservey64} has questioned whether the two-fluid model can
explain the shape of the free surface without violating
hydrodynamics of a superfluid $\bm{\nabla}\times\bm{v_{s}}=0$,
which takes nonzero values at discrete points [see the comments
following Eq.~(\ref{vortex})] in a multiply connected region.
Moreover, Mott's analysis on the surface energy on a rotating
cylinder suggests that there might occur the Onsager-Feynman
quantization on the surface of a rotating superfluid, which is
essentially similar to that of Anderson's phase slippage, but it
cannot still explain how the energy from a rotating superfluid is
transferred to the surface \cite{Mott49}. And Reppy and Lane
\cite{Lane65} also questioned if a sufficient number of vortex
lines were generated and maintained in a rotating He II during the
course of the experiment to satisfy Feynman's criteria for the
areal vortex number density
$\Gamma_{\Omega}\geq\frac{2\Omega}{\kappa}$ with
$\kappa=\frac{h}{M}=0.997\times10^{-3}cm^{2}/sec$ \cite{Hall60}.

We approach this long standing riddle in a rotating He II from an
entirely different point of view \cite{Osborne50,Meservey64}.
Since a superfluid is described by the potential flow
$\bm{\nabla}\times\bm{v_{s}}=0$, a steady rotating superfluid flow
exerts no force on the rotating container (d'Alembert's paradox).
The normal fluid (phonon and roton), on the other hand, exerts
drag force on the wall of a rotating container. The fundamental
limitation of the two-fluid model brought about by the absence of
the surface phenomena defined by an external pressure. In 1950,
Osborne was the first to question if the superfluid component
rotates when we rotate a container of He II. To answer this
question, he observed the contour of the free surface of a
rotating He II that was independent of temperature, which is
incompatible with the two-fluid model \cite{Landau41}.

In order to overcome this difficulty, Landau and Lifshitz proposed
a vortex sheet model \cite{Lifshitz55}. On the other hand based on
the suggestion by Onsager \cite{Onsager49} and Feynman
\cite{Feynman55}, Hall and Vinen \cite{Hall56} extended London's
model \cite{London54} and showed that the macroscopic rotation of
He II can be achieved despite the condition $\bm{\nabla}\times
\bm{v}_{s}=0$ if an array of vortices is aligned along the axis of
rotation ($\Omega$) and satisfies Feynman's criteria for the areal
vortex number density. We refer the reader Ref.~\cite{Hall60} for
details of Feynman's criteria \cite{Hall60,Glaberson74,Tsubota03}.

In contrast to the above models, Lin \cite{Lin59} suggested that
the superfluid might have a small viscosity and would thus reach a
steady state of uniform rotation in which boundary slip would
cause the superfluid component to rotate more slowly than the
normal fluid at low velocities. Lin's insight into the question of
whether the superfluid component rotates was quite plausible, but
it was difficult to explain the experimental observation by Reppy
and Lane \cite{Reppy61} that a container of He II can be rotated
at a higher speed in the absence of the superfluid motion.
Moreover the notion of the boundary slip is incompatible with the
two-fluid model of Landau. In particular, the superfluid component
in a uniform motion must remain inviscid and irrotational
$\bm{\nabla}\times\bm{v}=0$, but it can be singular only at
isolated points of a rotating He II, {\it i.e.,} $\kappa=\oint
d\bm {l}\cdot\bm {v}= \frac{h}{M}=0.997\times10^{-3}cm^{2}/sec$ at
discrete points of the superfluid in motion, at which the broken
symmetry takes place to create vortices
\cite{Landau41,Mott49,Andro46,Feynman55}.

To Meservey \cite{Meservey64} who has carried out the same
experiment on a rotating He II as Osborne's \cite{Osborne50} and
whose incisive comments on the previous theoretical models
\cite{Landau41,Andro46,Feynman55} based the his data, and to Mott
\cite{Mott49,Anderson66} whose remarkable insights into the
surface energy due to velocity discontinuity with which the
present concept of broken symmetry has been deduced with
mathematical rigor, we owe them for their valuable insights into
the surface phenomena in He II.

In 1964, Meservey \cite{Meservey64} concluded that neither the
simple vortex line model \cite{Hall60}, nor the vortex sheet
model \cite{Lifshitz55} is adequate to explain all of his data.
In addition, Reppy and Lane \cite{Lane65} also pointed out that
there is no reasonable mechanism by which the vortex line model
can be realized in actual experiments, since the macroscopic
turbulence in the superfluid has been observed in a rotating He
II which implies the vortex lines were unstable long before the
rotating He II reached a steady state \cite{Glaberson74,vortex2}.
In fact so random is the upward motion of vortices, diffusion of
vortex-lines - that is, short smoke ring like objects, but they
are much longer in length than those of rotons - that the
vorticity of the system has been evolved into an infinite number
of patterns which can be described only by a homogeneous
turbulence in He II \cite{Vinen93}. Thus the experimental support
for Feynman's vortex line model \cite{Hall60} is doomed to
failure; the unusual argument for a stable vortex line model
leads to counterintuitive results that cannot still describe the
shape of curvature of a rotating He II, since a vortex line is
absolutely unstable to the $m=1$ hose mode
\cite{Glaberson74,vortex2}.

Meservey's \cite{Meservey64} argument for the role of a surface
free energy associated with velocity discontinuities suggested by
Mott \cite{Mott49} is quite plausible, yet they cannot still
explain why the curvature of the surface of rotating He II is
temperature independent and the free surface does rotate. This is
the principal difficulty of the vortex-nucleation model in a
superfluid: at no point can one really exclude the second order
phase-transition in order parameter, in view of the break-down of
the phase coherence due to fluctuation-dissipation at the free
surface.

Also, of course, it is precisely the nature of a superfluid that
it cannot have a steady state under a gravitational field whether
it is in motion or at rest \cite{Anderson65,Anderson66}, but it
can have a steady state with a free surface on which the two-fluid
model breaks down as shown in section III. We have already seen in
section IV that, in the dynamical study of the BEC in trap, from
which we have discovered a shell-like structure of the BEC in an
external field, it was necessary to introduce the chemical
potential as an undetermined Lagrangian multiplier to maintain the
equilibrium state of an imperfect Bose gas in an external
potential [see Eq.~(\ref{Motion0}) and Eq.~(\ref{Eqs0})]. This
study of the BEC in a trap has led us to study the spontaneously
broken symmetry to explain the shell-like structure.

With the spontaneously broken symmetry, which takes place only
locally at the free surface and is purely topological in nature,
and Anderson's $2\pi$ phase slippage of vortices at the free
surface \cite{Anderson66,Avenel85,Packard92,Hess95}, it is
straightforward to derive the contour of a free surface of a
rotating He II \cite{Sommerfeld50},
\begin{equation}
z=\frac{\Omega^{2}}{2g}r^{2},
\end{equation}
since no superfluid component is present at the surface layer
\cite{Singh62,Atkins65}.

For the simplest dissipation case, let us consider a small bundle
of $n$ vortices which are created at the bottom of rotating
cylinder and move upward like micro-bubbles in a beer can. The
bundle moves upward in a complicated manner, perhaps undergoing a
few twists and reconnections \cite{Schwarz88} before finally being
nucleated at the free surface by the phase slippage, imparting the
angular momentum to the surface. This will lead to a $2\pi n$
phase slip and the energy dissipated at the surface will be $n$
times greater than the $2\pi$-event. Hence the surface rotates and
reaches a steady state with the angular velocity $\Omega$ after
awhile \cite{Sommerfeld50,SJHan1}. Here the free surface being a
normal fluid plays a role similar to that of a wall in Anderson's
model - that is,  nucleation of vortices by a $2\pi n$ phase
slippage \cite{Anderson66,Avenel85,Packard92,Hess95}. Hence the
curvature must remain temperature independent, which is consistent
with the extended two-fluid model of Landau
\cite{Landau41,Mott49,Singh62,Gorkov59,Anderson65}.

It is also worth noting that the above picture of how a parabolic
free surface is formed in a rotating He II is similar, in many
respects, to that of formation of a vortex-ring by an ion complex
in He II \cite{Rayfield67}. Thus $2\pi$ phase slip is indeed a
basic concept that applies to a number of problems including the
motion of a vortex in an orifice \cite{Anderson66}. In accordance
with the broken symmetry, not only should we able to resolve the
recent controversy over the Magnus force \cite{Thouless99}, but
also explain how the superfluidity breaks down at the core of a
vortex line \cite{Tinkham96,note2}.

With the detailed dynamical calculations in terms of the action
of the mean-field defined in ODLRO \cite{Penrose56,SJHan1}, we
have been able to explain a number of problems that have been
profoundly puzzling over the last half century \cite{Andro66},
such as the  curvature of a rotating He II and the breakdown of
superfluidity at a vortex core in He II
\cite{Osborne50,Meservey64,Rayfield67}.

Before we close this discussion on the rotating He II, one more
comment on a recent paper \cite{Suzuki01} is in order: based on
the notion of a regauged space translation, Su and Suzuki have
shown a new derivation of the quantization of a vortex-line, and
have shown the similarity between the main characteristic of the
surface layer of a rotating He II and the Meissner effect in a
superconductor \cite{Nieh95,Sewell97}. When a particle
displacement is introduced into the Bose system, it invariably
perturbs the particle density of the system. Moreover, the gauge
field is a dynamical variable which must be analytic unless it is
spontaneously broken. We therefore find it difficult to believe
the idea of the regauged space translation \cite{Nieh95,Suzuki01}
is tenable. This idea has also led to the self-contradiction in
their work as pointed out by Shi \cite{Shi03}. Moreover, there was
a fundamental misunderstanding \cite{Nieh95,Sewell97} on the gauge
invariant explanation of the Meissner effect
\cite{Nambu60,Anderson58} and the BCS theory \cite{BCS57}.

\subsection{\label{sec:level2} A Super-Lattice}

 Lastly we discuss a possible formation of a close-packed lattice
as $a \rightarrow 0$ in BEC, which is equivalent to applying high
pressure to an atomic gas in a trap. Our ground state wave
function describes a degenerate Bose gas modified as little as
possible by the presence of the repulsive pair-interaction [see
Eq.~(\ref{Cat2})] \cite{SJHan2,Dyson57}. Hence the BEC in a trap
provides an excellent test for this controversial problem
\cite{Meisel92}. Moreover, it is highly unlikely that, from our
experience in a superconductor, a long range force could change
the essential picture of our discussion in the absence of the
electron-pair interactions (Cooper pairs) of electrons in the
system \cite{Ginzburg50,Schafroth54}.

It may be worth emphasizing that Gorkov's theory of
superconductivity \cite{Gorkov58} demonstrated that both the
energy-gap function $\Delta$ and the Ginzburg-Landau order
parameter $\Psi$ vary as $e^{-2i\mu t/\hbar}$, where $\mu$ is the
Fermi energy. Because of this time dependence, the Fermi level
plays an important role in the phenomenology of superconductivity
which is different from that of normal metals, and thus the
energy gap function due to a virtual electron-electron
interaction near the Fermi surface is essential to a
super-conducting state \cite{Cooper56}.

Nevertheless, a detailed analysis shows that, in the limit $a
\rightarrow 0$ [see Eq.~(\ref{Eigen})] which is equivalent to
applying huge compressional pressure to the BEC in a trap, we
obtain the dispersion relation for a close-packed lattice with
zero-point vibration from Eq.~(\ref{Eigen}) \cite{SJHan2}:

\begin{equation}
\omega_{latt}=(5)^{1/2}\omega_{0}.
\end{equation}

It is again independent of $\hbar$. This implies that if there is
actually a transition from BEC to a crystalline ground state, then
the ground state is a classical lattice. The above discussion
explains to some extent why a simple model based on the imperfect
Bose gas that does not include long range interaction predicts a
better logical structure with the classical close-packed lattice
than a super-lattice.

The theory of broken symmetry is remarkably successful to give an
unequivocal answer to this controversial question of a
super-lattice \cite{Chan04,Meisel92}. The physical mechanism by
which the close-packed lattice can be realized is again the
spontaneously broken symmetry that accompanies a longitudinal
phonon as Nambu-Goldstone boson \cite{BenLee73,Weinberg96}. Since
the spontaneously broken symmetry is the consequence of
superfluidity described by $\psi$ in ODLRO, it is highly unlikely
that we could observe a super-lattice in an laboratory experiment
by compression, since we cannot find a physical mechanism by which
we can transform He II into a superconductor
\cite{Mezhov66,Andreev69,Chan04,Meisel92}.

\section{\label{sec:level1} Future Prospects and Discussion}

In this paper, we have developed a perturbation method appropriate
to the determination of a phonon spectrum of collective
excitations in a finite Bose system. The new perturbation method
based on ODLRO and the particle orbit perturbations by the
Lagrangian displacement vectors in Bohm's quantum theory shows
that a remarkably simple algebra yields not only a correct phonon
spectrum of the sound wave but also the dispersion relations for
surface waves in a finite space problem. A detailed analysis of
the dispersion relations shows, furthermore, that a spontaneously
broken symmetry takes place in a degenerate Bose system in a
boundary layer - breakdown of the two-fluid model. Moreover, we
are able to provide the direct cause of the Bohr-Sommerfeld
quantization Eq.~(\ref{vortex}) by the spontaneously broken gauge
symmetry. Finally we have shown that the characteristics of
surface phenomena in He II in the gravitational field are similar
to those of the Meissner effect in a superconductor in a weak
external magnetic field.

It seems appropriate to point out in conclusion what was apparent
at the outset - namely, that the conventional hydrodynamic
perturbation methods to many-body Bose systems which have been
employed by many authors in the past, are extremely primitive and
are not applicable to the problems with boundary conditions
\cite{Fetter01}. On the other hand the quantum field theoretical
approaches to the problems would face the same criticism; in
particular the Bogoliubov-de Gennes equation should not be applied
to a finite space problem, since the quantum field theory remains
valid only in a Hilbert space
\cite{Bog47,Caroli64,note1,note2,Leggett01}.

A finite space problem is indeed unique in that a correct approach
to the problem requires a new mathematical method. Here we have
taken the entirely different approach to incorporate Feynman's
atomic picture of a phonon \cite{Feynman53}, namely the
semi-classical perturbation method with the Lagrangian
displacement vector to solve the many-boson problems. The
Lagrangian displacement vector is basically the classical
perturbation method that has been applied to particle orbits in
Bohm's interpretation of quantum mechanics \cite{Bohm52} in ODLRO
\cite{Penrose51,Penrose56}, and yet the results are in the quantum
mechanical domain so long as we maintain the phase-coherence and
retain the effective quantum potential (EQMP). It is also
remarkable to obtain the same dispersion relation
Eq.~(\ref{Phonon}) for the collective excitations in He II as
Bogoliubov obtained in his theory of superfluidity using quantum
field theoretical technique for a uniform He II \cite{Bog47}. In
this paper I have described the essential mathematical techniques
of calculating the coherent collective excitations in finite space
problems and have shown how one can realize the spontaneously
broken symmetry in the many-boson system to explain several
puzzling experiments.

A few points that would bear a further discussion follow. The
first is the accuracy on which Bohm's interpretation of a
quantum-mechanical system depends on a precisely definable
phase-coherence by a mean field $\psi$ that also defines a
superfluid in ODLRO. Since the Goldstone theorem
\cite{Nambu60,Nambu61,Goldstone61,BenLee73} implies a zero-mass
boson, there is no difficulty of identifying a phonon as a
Nambu-Goldstone boson. However, a vortex creation by a broken
symmetry accompanies a roton whose effective mass is $0.16m_{He}$
that can be treated as a pseudo-Goldstone boson. This large mass
can be explained by the effective quantum mechanical potential by
which a particle fluctuates to a higher energy ({\it i.e.,} Bohm's
irreducible disturbance \cite{Bohm52}), because the vortex
creation involves a bulk of a large superfluid with higher energy
and the radius of curvature of a free surface is also extremely
small being in the order of many $\AA$. Hence the EQMP [see
Eq.~(\ref{Mean}) and Eq.~(\ref{Eqmp})] in the quantum
Hamilton-Jacobi equation plays a crucial role in explaining the
spontaneously broken symmetry in He II. Moreover, it is consistent
with the $\sigma$-model of Gell-Mann and Levy within the framework
of quantum field theory \cite{Bjorken65,BenLee73, Levy60}.

The second, we have introduced, in the spirit of  Feynman
\cite{Feynman53}, the Lagrangian displacement vector which
provides a new physical concept of phase-coherence in terms of the
particle displacements. More importantly, in terms of the
Lagrangian displacement vector, the boundary conditions have
defined a boundary layer in which the two-fluid model of Landau
\cite{Landau41} breaks down. One can show that Eq.~(\ref{subeq:1})
and Eq.~ (\ref{subeq:2}) with the boundary conditions give the
second-order phase transition in He II under the gravitational
field - the normal fluid joined a superfluid - just like the
Meissner effect in a superconductor in an external magnetic field.

Moreover, the phase-coherent collective excitation we have studied
is, in essence, to place Feynman's picture of a phonon in precise
mathematical formulas to obtain the excitation spectrum which
turns out to be the same as the Bogoliubov spectrum
\cite{Feynman53,Bog47}. Hence we have established the equivalence
between Feynman's atomic theory of liquid helium near absolute
zero \cite{Feynman53} and Bogoliubov's theory of superfluidity
\cite{Bog47}. More importantly, we have derived the dispersion
relation for a sound wave Eq.~(\ref{Sound}) without Feynman's
concept of a back-flow which has become an unnecessary conjecture
because of the equation of continuity Eq.~(\ref{subeq:1}).

If we are to take seriously the usual dynamical proof of BEC
\cite{Henshaw61}, we must study the dispersion relation
Eq.~(\ref{Sound}) experimentally, which will be extremely
difficult to carry out in the present geometrical configurations
for the Bose-Einstein condensation in the trap. It should be
emphasized that a typical wavelength of a sound wave
$[\lambda_{ph}=2\pi/k_{ph}=(2\pi/\omega_{ph})(4\pi a
n\hbar^{2})^{1/2}/M]$ with a small perturbation in BEC that one
can drive is comparable to the size of BEC. Moreover, the speed of
sound $c=[4\pi a \rho(r)\hbar^{2}]^{1/2}/M$ is extremely difficult
to measure \cite{Steinhauer02}, because it depends on the local
density and therefore depends on the wavelength. Its domain of
access is so small and difficult (due to a peculiar geometrical
configuration of confining magnets) in the device that, with all
our experimental efforts, we have been able to probe a small
fraction of a surface area with limited success
\cite{Onofrio00,note2}.

At the outset, the new BEC in a trap is doomed to failure, because
the wavelength $[\lambda_{ph}=2\pi/k_{ph}=(2\pi/\omega_{ph})(4\pi
a n\hbar^{2})^{1/2}/M]$ of a (first) sound is too long for the
micron-scale cloud to carry out the dynamical study of collective
excitations in BEC. [Recall the wave number $k_{ph}=p/\hbar=0.25
\sim 2.5 \AA^{-1}$ (or $\lambda_{ph}=0.06 \sim 2.51 m\mu$) used in
the study of a phonon spectrum at $T=1.12^{0}$K in He II]. It is
indeed unfortunate that the atomic Bose condensate in a trap was
once hailed as a Holy Grail in atomic physics ({\it sic})
\cite{Baym96} as an alternative to He II in testing the quantum
many-body theory of an interacting Bose gas
\cite{Fetter71,Fetter01}, and yet one must now question for its
scientific merit in the light of new developments in the
theoretical front \cite{SJHan1,SJHan2,SJHan3}.

Unfortunately the circumstance in which the new Bose-Einstein
condensation in a tap was promoted and celebrated for its
achievement suggests that the scheme is untenable in a healthy
scientific community. In spite of Bogoliubov's well-known theory
of superfluidity \cite{Bog47}, it is almost incredible that no
attention has ever been called on to study the (first) sound wave
propagation in BEC as a dynamical proof of the BEC in a trap.
Although the two-fluid model of Landau \cite{Landau41} describes
remarkably well the hydrodynamic properties of He II, it cannot,
however, explain the problems with boundary conditions, since it
is developed in a Hilbert space. On the experimental front,
Henshaw and Wood have confirmed in great detail the Landau's
prediction on the shape of collective excitation spectrum (phonon
and roton) in He II by means of the precise neutron-scattering
technique \cite{Henshaw61}. It may not be worth while, however, to
create a larger scale of a droplet of BEC for a study of the sound
wave, since {\it a strength of pair-interaction force will not
alter the central features of Bose-condensation} \cite{Feynman53},
nor will it modify London's view on the $\lambda$-transition
between He I and He II as the same process of the condensation of
an ideal Bose-Einstein gas \cite{London38}.

Experiments, of which there are many, are usually performed on a
prolate ellipsoidal condensate in a trap. One of the few
exceptions to the universal ellipsoidal condensate was a long
cylindrical condensate in which the length of the condensate can
be made much longer than the radius {\i.e.,} $L\gg r \gg \lambda$,
where $\lambda$ is the wavelength of a sound wave, and which
mimics a uniform condensate for a traveling sound wave. In an
experiment on collective excitations, Steinhauer, {\it et al.,}
\cite{Steinhauer02} have created such a cylindrically symmetric,
long Bose condensate of $^{87}$Rb atoms that satisfies the above
criteria and launched a traveling longitudinal sound wave along
the axis of symmetry. But the effort to measure the speed of the
sound wave was not successful, for they were not aware of the
breakdown of the superfluidity at the end of the condensate. The
difficulties in the measurements of the speed of sound waves are
enormous. Thus their dispersion relations are qualitatively
correct in some respects, but not in detail.

Given the present geometrical configuration of a trapping devise,
it is almost impossible to precisely measure the speed of the
(first) sound wave in any form of BEC in a trap, because, in
addition to the breakdown of superfluidity on the free surface,
the speed $c=[4\pi a \rho(r)\hbar^{2}]^{1/2}/M$ varies as the
sound wave propagates toward to the free surface in a spatially
inhomogeneous BEC \cite{Steinhauer02,SJHan2}. Moreover, the sound
wave dissipates near the surface layer as discussed above, and
thus one must take measurements deep inside of BEC where the
boundary conditions are unimportant.

The basic idea described in this paper suggests several other
lines of investigation involving the spontaneously broken symmetry
in He II: a) there is, first, a natural extension of the present
study on the collective excitations to a more general condensation
in geometry in a trap \cite{SJHan2,SJHan3}; although the present
work in a spherical and a cylindrical geometries describes the
essential physics, it is too specific in practice and must be
extended to a realistic model, a prolate ellipsoidal condensate
that has the best possible chance for a future experiment provided
the number of trapped particles increase by many times with the
much larger dimension of condensation - that is in the order of
$cm$s) \cite{SJHan4}; b) second, the stability of a vortex line
based on the broken symmetry should be carried out for a detailed
study of the upward diffusion of vortices in a rotating He II
\cite{vortex2}.

As in Anderson's argument on the singularities of Landau's order
parameter in connection with the energy-dissipation mechanism
\cite{Anderson84}, the whole points of discussion here are
essentially the emergent phenomena that follow from the
spontaneously broken symmetry. It is also clear that, for this
spontaneously broken symmetry to occur on the nodal surface, all
that is required of the analyticity of the action $S(\bm x,t)$ of
Bohm's theory of quantum mechanics (Bohm's irreducible
disturbance). The current controversy \cite{Leggett05, Laughlin05}
over the broken gauge symmetry may be be resolved simply by
identifying the Goldstone boson in the analysis of collective
excitations \cite{Anderson07}.

\section{\label{sec:level1}Acknowledgements}

I am grateful to Professor W. F. Vinen for pointing out the role
of surface tension at the free surface of He II. I am also
indebted to Dr. Robert Meservey for his helpful comments on the
upward diffusion of vortices from the solid base in his rotating
He II experiment. I also thank Professor G. B. Hess for sending me
his paper on the geometrical effect of nucleation rate of
vortices.

\newpage

\begin{figure}[tbp]
\includegraphics[totalheight=0.45\textheight,viewport=100 50 400 400,totalheight=4in,height=4in,width=6in,keepaspectratio=true]{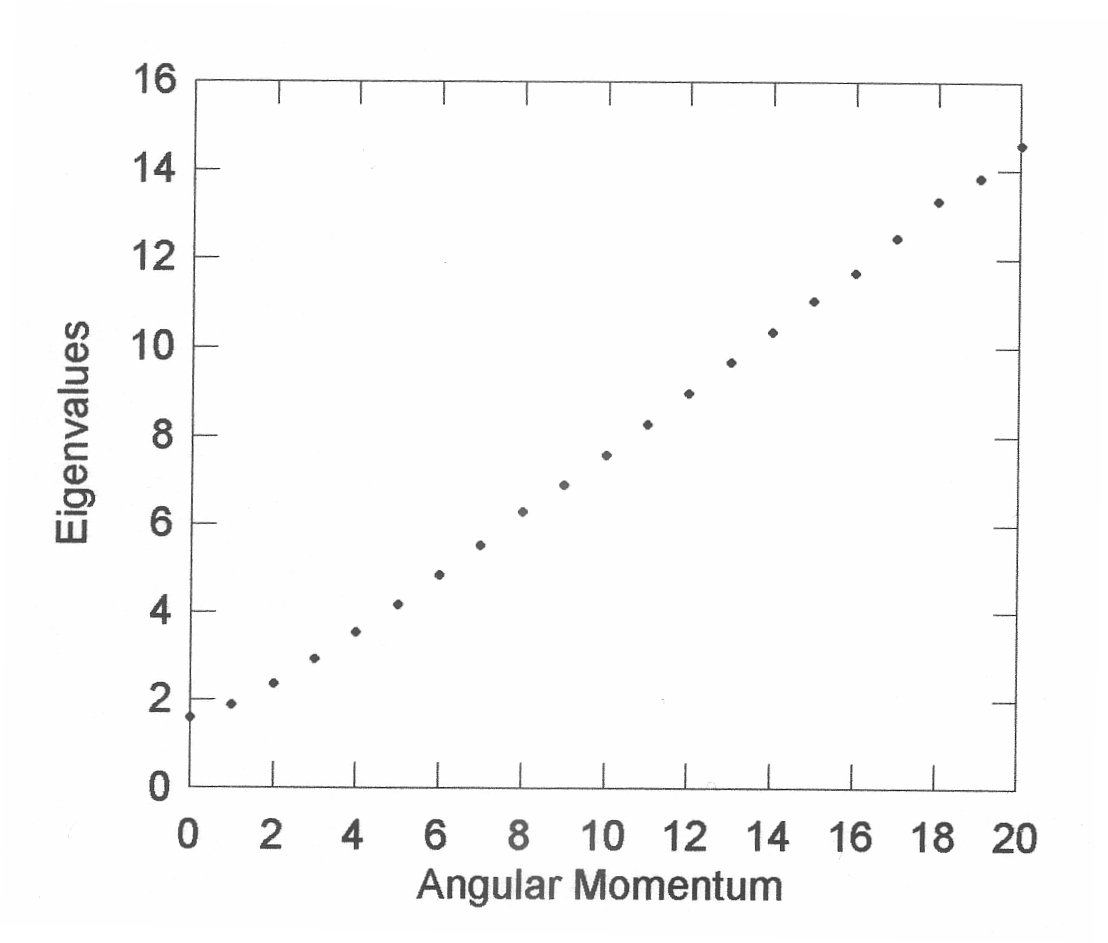}\vspace{2.0truecm}
 \caption{The ratio $\omega_{ph}/\omega_{0}=
[2\acute\lambda_{s}]^{1/2}$ is plotted against the angular
momentum $\ell$. It shows how the energy spectrum of phonons
varies with the angular momentum valid in the phonon regime
($ck_{\theta}\ll1$); and it does not increase without limit as
Landau has shown, but must approach a value $\lambda$
transition point.}%

\label{fig:fig1}
\end{figure}

\begin{thebibliography}{95}
\bibitem{note4} It should be stressed that the derivation of
        Singh's calculation is based on the ideal degerate Bose gas
        (no interaction among the He atoms) in a finite momentum
        $dp$ with the appropriate $\epsilon - p$ relation. Hence
        the surface tension deviates significantly from the experimental
        values of Allen and Misener \cite{Allen38} near the
        $\lambda$ point which is the consequence of the mutual
        interactions of He atoms.
\bibitem{note1}The conventional perturbation methods are applicable
        to a periodic or a uniform system. In particular, the
        Bogoliubov theory is applicable only to a uniform medium,
        because one can quantize the scalar field of
        quasi-particles only in a Hilbert space.
\bibitem{note3} The presence of the surface tension in the surface layer
         is an important point in the study of spontaneously broken
         symmetry, because it shows that the fluid in the surface
         layer behaves like a normal fluid, and more importantly
         it is necessary for the conservation of energy in an isolated
         sysytem as discussed below in VIII.
\bibitem{SJHan4} S. J. Han, The study of the ellipsoidal
        condensation has begun; we have only solved a surface
        wave problem in an ellipsolidal condensate.
\end{thebibliography}
\end{document}